\documentclass[a4paper,11pt]{article}
\pdfoutput=1
\usepackage{jcappub}

\usepackage{longtable}
\usepackage{booktabs}
\usepackage{xcolor}
\usepackage[normalem]{ulem}
\usepackage{caption}

\definecolor{Cerulean}{RGB}{0,123,167}

\def\QCD{{\scriptscriptstyle \rm QCD}}
\def\UV{{\scriptscriptstyle \rm UV}}

\def\W{{\scriptscriptstyle \rm W}}
\def\D{{\scriptscriptstyle \rm D}}
\def\dP{{\scriptscriptstyle \rm dP}}
\def\ALP{{\scriptscriptstyle \rm ALP}}

\newcommand\redsout{\bgroup\markoverwith{\textcolor{red}{\rule[0.5ex]{2pt}{0.4pt}}}\ULon}

\title{Axions at the meV Crossroads:\\Theory, Cosmology, Astrophysics, and Experiments}

\author[a,b]{Michele Cicoli,}
\author[c,d]{Francesco D'Eramo,}
\author[d]{Luca Di Luzio,}
\author[e]{Damiano F.G. Fiorillo,}
\author[f,g,h]{Maurizio Giannotti$^{1}$,\note{Corresponding author.}}
\author[i]{Alicia Gomez,}
\author[j]{Diego Guadagnoli,}
\author[f,g]{Mathieu Kaltschmidt,}
\author[k]{Bradley J. Kavanagh,}
\author[c,d]{Alessandro Lella,}
\author[l]{Giuseppe Lucente,}
\author[m]{David J. E. Marsh,}
\author[n]{Federico Mescia,}
\author[o,p]{Alessandro Mirizzi$^{2}$,\note{Corresponding author.}}
\author[f,g]{Javier Redondo,}
\author[w,x]{Nicole Righi,}
\author[q]{Jaime Ruz,}
\author[r]{Ken'ichi Saikawa,}
\author[s,t,u]{Elisa Todarello,}
\author[c,d]{Edoardo Vitagliano,}
\author[v]{Su-Yang Xu}

\affiliation[a]{Dipartimento di Fisica e Astronomia, Universit\`a di Bologna, via Irnerio 46, 40126 Bologna, Italy}
\affiliation[b]{INFN, Sezione di Bologna, viale Berti Pichat 6/2, 40127 Bologna, Italy}
\affiliation[c]{Dipartimento di Fisica e Astronomia, Universit\`a degli Studi di Padova, Via Marzolo 8, 35131 Padova, Italy}
\affiliation[d]{Istituto Nazionale di Fisica Nucleare (INFN), Sezione di Padova, Via Marzolo 8, 35131 Padova, Italy}
\affiliation[e]{Istituto Nazionale di Fisica Nucleare (INFN), Sezione di Napoli, Complesso Universitario di Monte Sant'Angelo, Via Cintia, 80126 Napoli, Italy}
\affiliation[f]{Departamento de F{\'i}sica Te{\'o}rica, Universidad de Zaragoza, C.\ de Pedro Cerbuna 12, 50009 Zaragoza, Spain}
\affiliation[g]{Centro de Astropart{\'i}culas y F{\'i}sica de Altas Energ{\'i}as, Universidad de Zaragoza, C.\ de Pedro Cerbuna 12, 50009 Zaragoza, Spain}
\affiliation[h]{Department of Chemistry and Physics, Barry University, 11300 NE 2nd Ave., Miami Shores, FL 33161, USA}
\affiliation[i]{ Centro de Astrobiología, INTA-CSIC, Ctra. Torrejón-Ajalvir km.4, Torrejón de Ardoz, 28850, Madrid, Spain.}
\affiliation[j]{LAPTh, Universit\'{e} Savoie Mont-Blanc et CNRS, 74941 Annecy, France}
\affiliation[k]{Instituto de F\'isica de Cantabria (IFCA, UC-CSIC), Av.~de
Los Castros s/n, 39005 Santander, Spain}
\affiliation[l]{SLAC National Accelerator Laboratory, 2575 Sand Hill Rd, Menlo Park, CA 94025}
\affiliation[m]{Physics Department, King's College London, Strand, London WC2R 2LS, UK}
\affiliation[n]{Istituto Nazionale di Fisica Nucleare, Laboratori Nazionali di Frascati, 00044 Frascati, Italy}
\affiliation[o]{Dipartimento Interuniversitario di Fisica ``Michelangelo Merlin'', Via Amendola 173, 70126 Bari, Italy}
\affiliation[p]{Istituto Nazionale di Fisica Nucleare -- Sezione di Bari, Via Orabona 4, 70126 Bari, Italy}
\affiliation[w]{Scuola Normale Superiore, Piazza dei Cavalieri 7, 56126 Pisa, Italy}
\affiliation[x]{INFN, Sezione di Pisa, Largo Bruno Pontecorvo 3, 56127 Pisa, Italy}
\affiliation[q]{Fakult{\"a}t f{\"u}r Physik, Technische Universit{\"a}t Dortmund, D-44221 Dortmund, Germany}
\affiliation[r]{Institute for Theoretical Physics, Kanazawa University, Kakuma-machi, Kanazawa, Ishikawa 920-1192, Japan}
\affiliation[s]{Leinweber Institute for Theoretical Physics, University of California, Berkeley, CA 94720, U.S.A.}
\affiliation[t]{Theoretical Physics Group, Lawrence Berkeley National Laboratory, Berkeley, CA 94720, U.S.A.}
\affiliation[u]{Universit\`a degli Studi di Torino, via P. Giuria 1, I--10125 Torino, Italy}
\affiliation[v]{Department of Chemistry and Chemical Biology, Harvard University, MA 02138, USA} 

\emailAdd{mgiannotti@unizar.es}
\emailAdd{Alessandro.Mirizzi@ba.infn.it}

\abstract{The meV mass range has emerged as a focal point in axion physics, where advances in theory, cosmology, astrophysics, and experimental techniques converge. Axions in this mass range are theoretically well motivated, can arise in ultraviolet-complete models, and can have significant cosmological impacts as dark matter or dark radiation. In parallel, their efficient production in stellar and supernova environments provides powerful astrophysical probes.
Here, we provide a comprehensive overview of meV axions across these domains, highlighting both established results and open questions. We discuss the theoretical underpinnings of meV axions, their cosmological and astrophysical signatures, and the diverse experimental strategies---ranging from helioscopes and haloscopes to quasiparticle systems and large-volume Cherenkov detectors---that aim to explore this regime. The convergence of these approaches emphasizes the pivotal role of the meV mass range for axion discovery in the coming years, identifying meV axions as a key probe for testing beyond-Standard-Model physics.
This review document is the direct outcome of the discussions at the dedicated workshop ``The meV Mass Axion Frontier: Challenges and Opportunities'', held at Laboratori Nazionali di Frascati (IT) on 27--28 October 2025, and organized by the EU funded COST Action \emph{``Cosmic WISPers in the Dark Universe: Theory, astrophysics, and experiments''} (CA21106, \url{https://www.cost.eu/actions/CA21106}). Its aim is to provide an overview of current efforts in meV axion research, their motivations, and the research goals that animate the community involved in this search.}


\begin{document}

\noindent {\sffamily\bfseries Preprints:} LAPTH-006/26, BARI-TH/785-26

\begin{table*}[ht]
\centering
\begin{minipage}[b]{0.3\linewidth}
    \centering
    \includegraphics[height=7em]{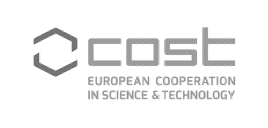}
\end{minipage}%
\hfill
\begin{minipage}[b]{0.3\linewidth}
    \centering
    \includegraphics[height=7em]{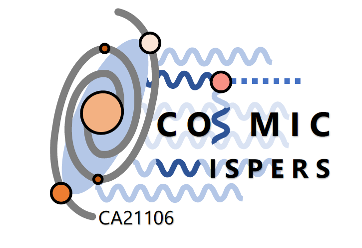}
\end{minipage}%
\hfill
\begin{minipage}[b]{0.3\linewidth}
    \centering
    \includegraphics[height=7em]{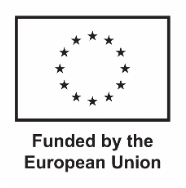}
\end{minipage}
\rule{\linewidth}{0.4pt}
\end{table*}

\maketitle

\newpage

\normalsize

\section{Introduction}
\label{sec:intro}

More than a decade ago, Ref.~\cite{Raffelt:2011ft} articulated the case for the ``meV mass frontier'' of axion physics. 
That work drew attention to the fact that axions with masses of a few meV---lying just below the astrophysical 
sensitivity threshold set by stellar cooling arguments---occupy a uniquely interesting region of parameter space for QCD axions.
The original motivation was rooted in the intriguing hint that white-dwarf (WD) cooling rates could be explained 
by axion emission with an electron coupling $g_{ae} \sim 10^{-13}$, corresponding to decay constants 
$f_a \sim 5\times 10^9$~GeV and masses $m_a \sim \mathcal{O}(\mathrm{meV})$. While the detection prospects for the 
diffuse supernova axion background appeared daunting, Ref.~\cite{Raffelt:2011ft} emphasised that 
next-generation helioscopes could realistically probe this regime, thereby establishing the meV mass range as an experimentally accessible target.
Since then, the physics case for meV axions has strengthened considerably across multiple fronts, from theoretical developments to cosmological considerations and astrophysical observations.

On the theoretical side, developments in string phenomenology have revealed that meV-scale QCD axions can emerge in large classes of ultraviolet-complete scenarios. In type IIB models with stabilised moduli, meV QCD axions can be realised as either closed string axions associated to blow-up cycles \cite{Conlon:2006tq,Cicoli:2012sz,Broeckel:2021dpz,Cicoli:2022fzy}, or as open string axions on sequestered branes at singularities \cite{Cicoli:2013cha,Cicoli:2017zbx,Petrossian-Byrne:2025mto,Loladze:2025uvf}. When the QCD axion is a closed string, its decay constant is typically of order the string scale, resulting in a pre-inflationary scenario with a non-linearly realised Peccei-Quinn (PQ) symmetry. In models with a relatively small number of K\"ahler moduli, the QCD axion decay constant can take a large range of values with a statistical logarithmic preference for higher values \cite{Broeckel:2021dpz}. Interestingly, when instead the number of K\"ahler moduli is very large, the requirement to trust the low-energy theory does not allow for large decay constants, resulting in an upper bound on the QCD axion mass around the meV range \cite{Mehta:2020kwu,Mehta:2021pwf}. On the other hand, if the QCD axion is an open string, its decay constant is suppressed with respect to the string scale, leaving the possibility to realise a post-inflationary scenario. Concurrently, field-theoretic approaches to the PQ quality 
problem---the requirement that PQ-breaking operators be highly suppressed to preserve the solution to 
the strong CP problem---favour relatively low values of $f_a$, again pointing towards meV masses as 
theoretically preferred.

From the cosmological perspective, meV axions occupy a distinctive niche. They are too heavy to constitute 
the dominant cold dark matter component via the standard misalignment mechanism with $\mathcal{O}(1)$ 
initial angles, yet they remain viable dark matter candidates under alternative production scenarios 
such as large-misalignment tuning, kinetic misalignment, or post-inflationary dynamics with domain-wall 
number $N_\mathrm{DW} > 1$. Simultaneously, thermal production in the early Universe renders meV axions 
a natural target for precision cosmology: future cosmic microwave background (CMB) experiments, including 
the Simons Observatory, will achieve sensitivity to the contribution of thermal axions to the 
effective number of relativistic species, $\Delta N_\mathrm{eff}$, precisely in the mass range where 
astrophysical probes currently set the strongest bounds.

Astrophysical constraints themselves have undergone significant refinement. Observations of neutron-star 
cooling now provide limits competitive with the classical supernova~1987A bound, placing the current 
sensitivity at $m_a \lesssim 10$--$20$~meV for KSVZ-type models. The theoretical understanding of axion 
emission from dense nuclear matter---including the role of nucleon spin fluctuations, pionic processes, 
and {in-}medium modifications---has matured, clarifying both the robustness and the residual uncertainties 
of these bounds. Furthermore, novel detection strategies have emerged: axion-to-photon conversion in 
the magnetic fields of supernova progenitor stars could yield detectable gamma-ray bursts from the 
next Galactic core-collapse event, offering a direct probe of meV-scale couplings complementary to 
laboratory experiments.

Perhaps most significantly, the experimental landscape has been transformed. The roadmap from CAST 
to BabyIAXO and ultimately IAXO represents a coherent helioscope programme capable of systematically 
scanning the meV mass range through buffer-gas techniques, with sensitivity extending to the QCD axion band. High-frequency haloscope concepts, such as CADEx, are 
pushing resonant-cavity searches towards the 100~$\mu$eV--meV regime, while broadband approaches 
offer complementary coverage. A particularly exciting development is the recent experimental 
discovery of axion quasiparticles in topological antiferromagnets, which exhibit masses 
$m_\Theta \sim 0.1$--$1$~meV set by intrinsic material scales. These condensed-matter analogues 
not only validate the underlying axion electrodynamics but also open {new} avenues for resonant 
dark-matter detection at meV frequencies through axion--polariton hybridisation, with tunability 
provided by external magnetic fields.

The convergence of these diverse approaches motivated the workshop ``The meV Mass Axion Frontier: Challenges and Opportunities'', held at Laboratori Nazionali di Frascati (IT) on 27--28 October 2025 (\url{https://agenda.infn.it/event/48801/}) and organized by the EU funded COST Action \emph{``Cosmic WISPers in the Dark Universe: Theory, astrophysics, and experiments''} (CA21106, \url{https://www.cost.eu/actions/CA21106}). The workshop played a central role in fostering interdisciplinary discussions across theory, cosmology, astrophysics, and experimental techniques, clarifying the key challenges and opportunities in this mass range, and establishing a coherent roadmap for the field. The present review emerges directly from these discussions and aims to present the meV axion as a coherent, cross-validated research programme. 
In Sec.~\ref{sec:meV_axions_model_building}, we survey the theoretical underpinnings 
that single out this mass range, including field-theoretic approaches to the PQ quality problem 
and top-down string constructions. 
We then turn to the cosmological roles of meV axions: Sec.~\ref{sec:meV_axions_as_cold_dark_matter} 
discusses their viability as cold dark matter candidates under various production scenarios, while 
Sec.~\ref{sec:dark_radiation} addresses their contribution to dark radiation, covering thermal 
production mechanisms (Sec.~\ref{sec:Thermal_production_of_axions}), observational consequences 
(Sec.~\ref{sec:Dark_radiation_as_a_cosmological_consequence}), and future prospects 
(Sec.~\ref{sec:Thermal_axions_whats_next}). 
In Sec.~\ref{sec:meV_axions_astrophysics}, we review astrophysical probes of meV-scale axions, focusing on axion production in dense nuclear media (Sec.~\ref{sec:Axion_production_in_dense_nuclear_media}), axion–photon conversion in astrophysical magnetic fields (Sec.~\ref{sec:Axion_conversion_in_astrophysical_magnetic_fields}), and searches for supernova axions in large-volume Cherenkov detectors (Sec.~\ref{sec:Supernova_axions_in_large-volume_Cherenkov_detectors}), with supernovae and neutron stars providing both stringent constraints and promising discovery opportunities. Sec.~\ref{sec:meV_axions_experiments} outlines the diverse experimental strategies 
converging on this parameter space: helioscope searches with CAST 
(Sec.~\ref{sec:Extending_axion_mass_sensitivity_in_CAST}) and the upcoming BabyIAXO 
(Sec.~\ref{sec:Axion_Mass-Scanning_Strategy_of_BabyIAXO}), post-discovery mass determination 
techniques (Sec.~\ref{sec:Post-discovery_strategies_for_axion_mass_determination}), high-frequency 
haloscope searches with CADEx (Sec.~\ref{sec:Pushing_the_haloscope_searches_towards_the_meV_range_The_CADEx_experiment}), and axion quasiparticle detection in topological materials (Sec.~\ref{sec:meV_Axion_Quasiparticles}). 
Finally, in Sec.~\ref{sec:conclusion} we present our conclusions. A comprehensive list of acronyms employed in this review can be found in App.~\ref{app:acronyms}.

The central message is that the meV frontier
has matured from a theoretically motivated target into a well-defined experimental programme. 
Multiple independent approaches---spanning particle physics, cosmology, astrophysics, and 
condensed-matter physics---now probe overlapping regions of parameter space, offering the 
prospect of robust, {cross-validated} discovery or exclusion. The coming decade promises to 
subject the meV QCD axion hypothesis to decisive tests, making this mass range one of the 
most compelling targets in the broader search for physics beyond the Standard Model (SM).

\section{meV axions model building}
\label{sec:meV_axions_model_building}


{Before discussing specific theoretical motivations that point towards the
meV range, we briefly recall the basic field-theoretic ingredients of QCD
axion model building. A QCD axion arises from a spontaneously broken global
$U(1)_{\rm PQ}$ symmetry whose associated pseudo-Nambu--Goldstone boson
couples anomalously to QCD. At energies below the PQ-breaking scale, the 
axion effective Lagrangian reads (see e.g.~\cite{DiLuzio:2020wdo})
\begin{equation}
\mathcal{L}_{a}
=
\frac{1}{2}(\partial_\mu a)^2
+
\frac{\alpha_s}{8\pi}
\frac{a}{f_a}
G^a_{\mu\nu}\tilde G^{a\mu\nu}
+
\frac{\alpha}{8\pi}
C_{a\gamma}^{0}
\frac{a}{f_a}
F_{\mu\nu}\tilde F^{\mu\nu}
+
\sum_f
\frac{\partial_\mu a}{2f_a}
C_f^0\,
\bar f\gamma^\mu\gamma_5 f
+\ldots \, ,
\label{eq:axion_UV_couplings}
\end{equation} 
where $\tilde{G^a}_{\mu \nu}=\frac{1}{2}\epsilon_{\mu \nu \rho \sigma}G^{a\, \rho \sigma}$ ($\epsilon_{0123}=+1$), 
and similarly for the electromagnetic field strength, 
$C_{a\gamma}^{0}=E/N$ is fixed by the ratio of electromagnetic and
color anomalies of the PQ current, while the coefficients $C_f^0$ depend on
the PQ charges of the SM fermions. After QCD confinement, the
axion mixes with the neutral pseudoscalar mesons and obtains a mass 
\begin{equation}
m_a
\simeq
5.7~{\rm meV}
\left(
\frac{10^9~{\rm GeV}}{f_a}
\right) ,
\label{eq:ma_fa_relation}
\end{equation}
up to small uncertainties from light-quark masses and higher-order chiral
effects \cite{Gorghetto:2018ocs}. Thus, the meV
mass range corresponds to an axion decay constant around
$f_a\sim 10^9$--$10^{10}\,{\rm GeV}$, close to the region where stellar
cooling, helioscope searches, thermal production, and several model-building
considerations become simultaneously relevant.}

{Two benchmark classes of ultraviolet (UV) completions are commonly used to
organise the phenomenology. In KSVZ models \cite{Kim:1979if,Shifman:1979if},
the PQ anomaly is generated by heavy vector-like quarks, while ordinary
SM fermions are not charged under $U(1)_{\rm PQ}$ at tree level.
As a result, the axion couplings to electrons arise from loop
effects. In DFSZ
models \cite{Zhitnitsky:1980tq,Dine:1981rt}, instead, the SM
fermions and two Higgs doublets carry PQ charges, and the axion has
tree-level derivative couplings to ordinary matter. These differences have
important phenomenological consequences. For instance, the electron coupling
is typically suppressed in hadronic KSVZ models, while it is present at tree
level in DFSZ models. This distinction is directly relevant for stellar
cooling bounds and for the interpretation of
helioscope searches. Similarly, the nucleon couplings depend on both the UV
PQ charges and the model-independent contribution induced by axion-meson
mixing, which is crucial for supernova and neutron-star constraints.} 

{At low energies, the axion-photon interaction is conventionally written as
\begin{equation}
\mathcal{L}_{a\gamma}
=
-\frac{1}{4}g_{a\gamma}aF_{\mu\nu}\tilde F^{\mu\nu} 
= g_{a\gamma} \textbf{E} \cdot \textbf{B},
\qquad
g_{a\gamma}
=
\frac{\alpha}{2\pi f_a}
\left(
\frac{E}{N}
-
1.92(4)
\right) ,
\label{eq:axion_photon_coupling}
\end{equation}
where the second term is the model-independent contribution from axion
mixing with neutral mesons \cite{Srednicki:1985xd,GrillidiCortona:2015jxo}. The ratio
$E/N$ is therefore one of the key model-building quantities controlling
the location of a QCD axion model in the $(m_a,g_{a\gamma})$ plane.
While $E/N = 0 \ (8/3)$ in KSVZ (DFSZ) models, different values of $E/N$ are possible 
in specific UV completions. Phenomenological selection
criteria have been proposed to single out ``preferred'' models \cite{DiLuzio:2016sbl,DiLuzio:2017pfr,Plakkot:2021xyx,DiLuzio:2024xnt}, 
which define the so-called QCD axion band.} 

{A similar interplay between model-independent and UV contributions 
holds for the derivative couplings to low-energy matter fields,
\begin{equation}
\mathcal{L}_{a\Psi}
=
\sum_{\Psi}
\frac{\partial_\mu a}{2f_a}
C_\Psi\,
\bar \Psi\gamma^\mu\gamma_5 \Psi ,
\label{eq:axion_fermion_couplings}
\end{equation}
where $\Psi = e, p, n,\ldots$ and the low-energy coefficients $C_\Psi$ include both UV contributions (including running effects \cite{Choi:2017gpf,Chala:2020wvs,Bauer:2020jbp,Choi:2021kuy,DiLuzio:2023tqe,DiLuzio:2022tyc})
and, in the case of nucleons, infrared corrections induced by QCD. These couplings
control many of the thermal production rates discussed in
Sec.~\ref{sec:dark_radiation}, as well as the astrophysical emission
processes reviewed in Sec.~\ref{sec:meV_axions_astrophysics}. In particular,
the axion-electron coupling drives production in electron-rich stellar
plasmas, while axion-nucleon couplings determine emission from dense nuclear
matter in supernovae and neutron stars.}

{The cosmological implications of a QCD axion also depend on the thermal
history of PQ symmetry breaking. If the PQ symmetry is broken before or
during inflation and is not restored afterwards, the observable Universe
contains a single initial value of the axion field. In this pre-inflationary
scenario, the relic abundance depends on the initial misalignment angle 
\cite{Preskill:1982cy,Abbott:1982af,Dine:1982ah}, and
isocurvature perturbations can impose strong constraints on high-scale
inflation. In contrast, if PQ breaking occurs after inflation, different
Hubble patches choose different initial axion angles and a network of global
strings and domain walls forms \cite{Davis:1986xc,Hiramatsu:2012gg,Kawasaki:2014sqa,Gorghetto:2018myk,Buschmann:2019icd,Gorghetto:2020qws,Hindmarsh:2021zkt,Buschmann:2021sdq,Saikawa:2024bta,Kim:2024wku, Benabou:2024msj,Correia:2024cpk,Correia:2025nns}. The decay of these topological defects gives
an additional contribution to the axion relic abundance, and the viability of
the scenario depends crucially on the domain-wall number $N_{\rm DW}$.
Models with $N_{\rm DW}=1$ are cosmologically safe, while stable domain walls
for $N_{\rm DW}>1$ would overclose the Universe unless the degeneracy among
vacua is lifted \cite{Sikivie:1982qv} or the cosmological history is modified.}

{These standard ingredients provide the baseline framework for axion model
building. Within this framework, the meV range is not tied to a single UV
realisation: it can arise in hadronic or DFSZ-like models, in variants with
suppressed or enhanced photon and fermion couplings (see e.g.~\cite{DiLuzio:2016sbl,DiLuzio:2017pfr,Farina:2016tgd,Agrawal:2017cmd,DiLuzio:2017ogq,Bjorkeroth:2019jtx,Darme:2020gyx,Hook:2018jle,DiLuzio:2021pxd,DiLuzio:2021gos}), and in both pre- and
post-inflationary cosmologies. In the following subsection we focus on one
specific model-building motivation for the meV range, namely the PQ quality
problem, which tends to favour relatively low values of $f_a$ in several
field-theoretic constructions.}

\subsection{Axion quality problem}

The axion solution to the strong CP problem \cite{Peccei:1977hh,Peccei:1977ur,Weinberg:1977ma,Wilczek:1977pj}, while elegant and potentially testable, raises the puzzling question 
of the origin of the PQ symmetry. In quantum field theories, global symmetries are not considered fundamental but are rather seen as accidental, as e.g.~in the familiar cases of strong isospin or baryon number. 
While some axion models \cite{Kim:1979if,Shifman:1979if,Zhitnitsky:1980tq,Dine:1981rt} impose an effective $U(1)_{\rm PQ}$ symmetry ``by hand'', a robust PQ theory should preferably generate this symmetry in an automatic way \cite{Georgi:1981pu}.
Furthermore, the PQ symmetry needs to be of very ``high quality'', 
since even a tiny 
explicit 
breaking of the $U(1)_{\rm PQ}$ 
from the ultraviolet (UV) completion 
would spoil the PQ solution to the strong CP problem, 
raising the so-called PQ quality problem \cite{Georgi:1981pu,Dine:1986bg,Barr:1992qq,Kamionkowski:1992mf,Holman:1992us,Ghigna:1992iv}.   
To quantify the latter, consider a 
PQ-breaking 
effective
operator of the type 
\begin{equation}
\label{eq:PQbreak}
V_{\text{PQ-break}}= - e^{i \delta} \frac{\Phi^n}{\Lambda_{\rm UV}^{n-4}}+\text{h.c.} ~ , 
\end{equation}
with $n>4$. Here, $\delta$ denotes a generic phase,   
$\Lambda_{\rm UV}$ is some UV scale like the Planck mass 
and $\Phi=\frac{f_a}{\sqrt{2}} e^{i a /f_a} $ is a complex scalar,  
with the radial mode integrated out and 
the angular mode $a /f_a $ given by the axion field. 
The induced axion potential then reads
\begin{gather}
V_{\text{PQ-break}}= - \frac{1}{2^{n/2-1}} \frac{f_a^n}{\Lambda_{\rm UV}^{n-4}}\cos\left(\frac{n a}{f_a}+\delta \right) ~ .\label{eq:cosine from higher dim}
\end{gather}
In general, the phase of this potential, parameterized by $\delta$, does not need to be aligned with the CP-conserving minimum of the QCD potential at $a=0$. Therefore, we 
generically expect $\delta$ to be $\mathcal{O}(1)$.
The balance between $ V_{\rm PQ-break} $ and the standard QCD axion potential, $V_{\rm QCD} = \frac{1}{2} f_a^2 m_a^2 (a/f_a)^2 + \mathcal{O}((a/f_a)^4)$, 
yields an axion vacuum expectation value (VEV) displaced from zero, that is
\begin{gather}
\label{eq:thetaeff}
	\theta_{\rm eff} \equiv \langle a / f_a \rangle 
	\simeq - \frac{n \Lambda_{\rm UV}^{4-n} f_a^n \sin \delta}{2^{\frac{n}{2}-1} f_a^2 m_a^2} ~ , 
\end{gather}
where we took into account the approximation $\theta_{\rm eff} \ll 1$. 
Taking for instance $\sin\delta = 1$, 
$\Lambda_{\rm UV} = M_{\rm Pl}$ and 
$f_a = 10^{8}$ 
GeV, 
the nEDM bound $|\theta_{\rm eff}| \lesssim 10^{-10} $ 
translates into $n\geq 9$. Therefore, the original question ``why is $\theta_{\rm QCD}$ small ?'' 
is traded for ``why PQ-breaking operators are highly suppressed ?''. 

Different approaches to the PQ quality problem have been proposed so far. 
It is useful to classify them in three categories: 
\begin{enumerate}

\item \textit{Low-scale $f_a$:}

For $f_a \sim 10^3$ GeV, only $n > 5$ is required in order not to generate a too large $\theta_{\rm eff}$ from
Eq.~(\ref{eq:thetaeff}). From this perspective, the original Weinberg-Wilczek model \cite{Weinberg:1977ma,Wilczek:1977pj} 
would have been perfectly natural,
since in the absence of SM-singlet fields, the first gauge invariant PQ-breaking operator that one can write
in the scalar potential features $n = 6$. Over the years, the increasingly lower bounds on $f_a$
let the PQ quality problem emerge and worsen. It is, however, possible to modify the standard QCD relation between 
$m_a$ and $f_a$ 
such that 
$m_a \gtrsim 10~\text{MeV}$ (see e.g.~\cite{Rubakov:1997vp,Berezhiani:2000gh}), in which case the conventional astrophysical constraints on $f_a$ 
can be avoided. 
Such heavy axion models feature an axion decay constant of the order of $10^{4 \div 5}$ GeV, 
thus improving on the PQ quality problem.

\item \textit{$U(1)_{\rm PQ}$ as an accidental symmetry:}

A prevalent strategy involves extending the Standard Model (SM) gauge symmetry, using either continuous or discrete local symmetries that are exact in the UV regime,  
so that the PQ symmetry arises as an accidental global symmetry, 
and is eventually broken by some higher-order effective operators
(see, for example, Refs.~\cite{Randall:1992ut,Dobrescu:1996jp,Babu:2002ic,Redi:2016esr,Fukuda:2017ylt,DiLuzio:2017tjx,Bonnefoy:2018ibr,Lillard:2018fdt,Gavela:2018paw,Lee:2018yak,Ardu:2020qmo,DiLuzio:2020qio,Yin:2020dfn,Contino:2021ayn,DiLuzio:2025jhv}). 

This mechanism provides a robust solution, independent of the specific UV completion. 
It requires, however, non-trivial model building, characterised by an integer parameter that 
matches the dimension $n$ of the PQ-breaking operator. Considerations of minimality and 
simplicity favour low values of $n$, which in turn imply relatively small values of $f_a$, 
close to the boundary of astrophysical limits. Consequently, meV-scale axions emerge as the 
most favourable scenarios for addressing the PQ-quality problem within field-theoretical 
four-dimensional models.

\item \textit{Extra-dimensional axions:}

Axions may also arise as zero modes of higher dimensional forms in string theory compactifications \cite{Witten:1984dg} {or, more in general, in models with extra dimensions \cite{Arkani-Hamed:2003xts,Choi:2003wr,Craig:2024dnl}}. 
In this context, the PQ shift symmetry is inherited from a higher-dimensional gauge symmetry and it is exact at perturbative level. Non-perturbative contributions to the axion potential are exponentially suppressed, and so extra-dimensional axions (when calculable) can offer a structural solution to the PQ quality problem. 
\end{enumerate}

\subsection{meV axions from string theory}

The existence of axions in field theory is motivated by the PQ solution to the strong-CP problem, but they have natural and elegant realisations in string theory {which is currently one of the most well-developed approaches to a UV-complete theory of quantum gravity}. Antisymmetric $p$-form gauge potentials in $10$ dimensions give rise, after compactification, to $4$-dimensional pseudo-scalar fields enjoying an (approximate) continuous shift symmetry, typically broken to a discrete one by non-perturbative effects. These modes are called \emph{closed string axions}. String compactifications feature also branes wrapping internal cycles and supporting QCD-like gauge theories. In these scenarios a global PQ $U(1)$ symmetry can arise at low energy as the remnant of a higher dimensional anomalous gauge symmetry, and the phases of $U(1)$ charged modes behave as axions. These are named \emph{open string axions}. The periodicity, decay constant and interactions of both closed and open string axions are set by the geometry and topology of the internal space, making axions sensitive probes of the UV completion. 

From the top-down point of view, different scenarios with meV axions seem to be possible. In fact, meV axions can arise in various corners of the string landscape, either as closed or open string modes, with an intermediate scale decay constant which can be around the string scale or much smaller, resulting in the possibility to reproduce both the pre- and post-inflationary scenario. Conversely, the potential discovery of a meV axion would provide valuable information about the ultraviolet structure of the theory, the size of extra dimensions, and the nature of moduli stabilisation, thereby offering a rare opportunity to use low-energy axion physics as a probe of string theory.

\subsubsection{meV axions as closed strings}

Axions arise ubiquitously in the closed string sector from dimensional reduction of $p$-form gauge fields, $C_p$, over internal $p$-cycles $\Sigma_p$. Their phenomenological properties are set by moduli stabilisation which determines their mass spectrum and couplings. Moduli stabilisation is therefore a crucial prerequisite for trusting any attempt to build phenomenologically viable axion models. Given that type IIB is the perturbative corner of string theory where moduli stabilisation is best understood, the majority of string axion models have been constructed in this context. In type IIB the most promising examples are $C_4$ axions:
\begin{equation}
\theta^i = \int_{\Sigma_4^{(i)}} C_4\qquad i=1,\cdots,h^{1,1}\,,
\label{eqn:p-form_axion}
\end{equation}
where $h^{1,1}$ is the number of K\"ahler moduli, $T^i=\tau^i+{\rm i}\,\theta^i$, with $\tau^i$ the volume of the $i$-th $4$-cycle $\Sigma_4^{(i)}$ in units of the string length $\ell_s=2\pi\sqrt{\alpha'}$:
\begin{equation}
\tau^i = {\rm Vol}\left(\Sigma_4^{(i)}\right)\, \ell_s^{-4}\,.   \end{equation}
The kinetic terms of the K\"ahler moduli are determined by the K\"ahler metric:
\begin{equation}
\mathcal{L}_{\rm kin} = K_{i\bar{j}}\partial_\mu T^i\partial^\mu \overline{T}^{\bar{j}}= \frac14 \frac{\partial^2 K}{\partial\tau^i\partial\tau^j} \left(\partial_\mu \tau^i\partial^\mu\tau^j +\partial_\mu \theta^i\partial^\mu\theta^j \right),
\end{equation}
where the tree-level K\"ahler potential $K$ depends on the Calabi-Yau (CY) volume $\mathcal{V}$ as:
\begin{equation}
K  = - 2\ln\mathcal{V} = -2 \ln \left(\frac16 k_{ijk} t^i t^j t^k\right) \qquad\text{}\qquad \tau_i = \frac12 k_{ijk} t^j t^k\,.
\end{equation}
In the previous expression $k_{ijk}$ are the CY intersection numbers and $t^i$ denote 2-cycle moduli.

In type IIB models, QCD-like non-Abelian gauge theories with chiral matter arise naturally on magnetised D7-branes wrapped around $4$-cycles \cite{Cicoli:2011qg, Cicoli:2016xae,Cicoli:2017axo,Cicoli:2024bxw}. In particular, the gauge kinetic function $\mathfrak{f}^a$ associated to the D7-stack wrapped around the $a$-th $4$-cycle is set by $T^a$, $\mathfrak{f}^a=T^a$. This induces a natural axion-like coupling for $\theta^a$ of the form: 
\begin{equation}
\mathcal{L}\supset \theta^a\, {\rm Tr}\left(F_a \wedge  F_a\right) + \tau^a\,{\rm Tr}\left(F_a \wedge  \star F_a\right).
\end{equation}
Note that the gauge coupling is controlled by $\tau^a$ since $\alpha_a^{-1}={\rm Re}(\mathfrak{f}^a)=\tau^a$.

The scalar potential receives D- and F-term contributions which depend on the K\"ahler potential $K$ and the superpotential $W$ as follows:
\begin{equation}
V_D = \sum_b \frac{1}{\tau_b} \left(\sum_k \hat{q}_k^{(b)} |\phi_k|^2-\xi_b\right)^2\quad V_F = e^K\left[K^{i\bar{j}}\left(W_i + W K_i\right)\left(\overline{W}_{\bar{j}}+\overline{W} K_{\bar{j}}\right) - 3 |W|^2\right].   
\end{equation}
The sum on $b$ is over all $U(1)$ factors, $\hat{q}_k^{(b)}$ is the charge under $U(1)_b$ of the $k$-th open string mode $\phi_k$, and the Fayet-Iliopoulos (FI) term $\xi_b$ reads (see App.~A of \cite{Cicoli:2011yh}):
\begin{equation}
\xi_b = -q_i^{(b)}\frac{\partial K}{\partial\tau_i}
\qquad\qquad q_i^{(b)} =  \int_{\Sigma_4^{(b)}} \hat{D}_i \wedge \mathcal{F}^{(b)}\,,
\end{equation}
where $q_i^{(b)}$ is the charge under $U(1)_b$ of the $i$-th K\"ahler modulus, $\mathcal{F}^{(b)}$ is the gauge flux on the $b$-th D7-stack, and $\hat{D}_i$ is the $i$-th element of a basis of harmonic $(1,1)$-forms.

At leading order, the D-term potential is minimised supersymmetrically at $V_D=0$. It is worth pointing out that $V_D$ does not depend on the axions but plays a crucial role in determining which axions are eaten by anomalous $U(1)$ gauge bosons. In fact, focusing for illustrative purpose on a single matter field $\phi$ and a single K\"ahler modulus $\tau$ charged under $U(1)_b$, the Abelian gauge boson becomes massive by absorbing a combination of the open string axion $\zeta$, which is the phase of $\phi=|\phi|\,e^{{\rm i}\zeta}$, and the closed string axion $\theta$, that is the supersymmetric partner of $\tau$. The mass of $U(1)_b$ can be written schematically as \cite{Cicoli:2011yh}:
\begin{equation}
m_{U(1)_b}^2 \simeq \tau_b^{-1} \left(f_{\rm op}^2+f_{\rm cl}^2\right),
\end{equation}
where $f_{\rm op}$ and $f_{\rm cl}$ denote, respectively, the open and closed string axion decay constants given by (for $q^{(b)}\simeq \hat{q}^{(b)}$):
\begin{equation}
f_{\rm op}^2=\langle|\phi|\rangle^2 \simeq \left|\frac{\partial K}{\partial\tau}\right|\qquad\text{and}\qquad 
f_{\rm cl}^2\simeq \frac{\partial^2 K}{\partial\tau^2}\,.
\label{DecConst}
\end{equation}
If $f_{\rm op} \gg f_{\rm cl}$, the combination of axions eaten by $U(1)_b$ is mostly given by $\zeta$, while if $f_{\rm op} \ll f_{\rm cl}$, the eaten axion is mostly $\theta$ \cite{Allahverdi:2014ppa}. The magnitude of $f_{\rm op}$ relative to $f_{\rm cl}$ is a function of the K\"ahler moduli that remain flat after D-term stabilisation. 

The minimisation of the F-term potential is essential to address three important issues for axion phenomenology: 
\begin{enumerate}
\item To determine the ratio $f_{\rm op}/f_{\rm cl}$ which establishes which axions are not eaten by anomalous $U(1)$'s, and so survive in the low-energy theory. If $\tau$ is fixed at large values, $\tau\gg 1$, as in the case of D7-branes wrapped around $4$-cycles in the geometric regime, the surviving axion is $\theta$. If, instead, $\tau$ shrinks to zero size, $\tau\to 0$, as for the case of D3-branes at a CY singularity, the role of the QCD axion is played by $\zeta$. 

\item To fix the ratio between the axion mass $m_\theta$ and the saxion mass $m_\tau$. Being pseudo-scalars with spin dependent couplings, the axions do not mediate any long-range force between unpolarised bodies. This is however not the case for the saxions which naturally couple with Planckian strength to ordinary matter. The non-observation of fifth-forces sets a lower bound on $m_\tau$ of the order $m_\tau\gtrsim\mathcal{O}(0.1)$ meV \cite{Hees:2018fpg}. Cosmology gives yet a much more stringent bound from the requirement of decaying before BBN, $m_\tau\gtrsim\mathcal{O}(30)$ TeV \cite{Coughlan:1983ci,Banks:1993en,deCarlos:1993wie}. Thus, to avoid the case where axions are necessarily very heavy, F-term fixing has to create a hierarchy between $m_\theta$ and $m_\tau$ of the form $m_\theta\ll m_\tau$.

\item To reproduce the value of the QCD coupling at the cut-off scale, $\alpha_\QCD^{-1}(\Lambda_\UV) = \tau_\QCD$, and, at the same time, an meV mass for the QCD axion with an intermediate scale decay constant, $f_\QCD\simeq 5\times 10^9$ GeV. Note that, in 4D string models, the UV cut-off is set by the overall volume $\mathcal{V}$ and, for purely local cycles, it corresponds to the string scale, $M_s\simeq M_p/\sqrt{\mathcal{V}}$, while for cycles with homology representatives both locally
and in the bulk, it is given by the winding scale, $M_\W\simeq M_p/\mathcal{V}^{1/3}$ \cite{Conlon:2010ji}. 
\end{enumerate}
 
The first realisation of a closed string QCD axion at the meV scale with F-term moduli stabilisation has been proposed in \cite{Conlon:2006tq} and, later on, D-term stabilisation has been included in \cite{Cicoli:2012sz}. The idea is to consider the $C_4$-axion $\theta_\QCD$ associated to a blow-up cycle with volume $\tau_\QCD$ and wrapped by D7-branes which realise QCD. Knowing that $\alpha_\QCD^{-1}(90\,{\rm GeV})\simeq 8.5$, one can work out $\tau_\QCD=\alpha_\QCD^{-1}(\Lambda_\UV)$ using RG running. Given that blow-up cycles are local, we shall consider $\Lambda_\UV\simeq M_s$. As we will see, $M_s$ sets the scale of the QCD axion decay constant, which has therefore to be intermediate to reproduce an meV axion. In turn, this fixes also the gravitino mass that gives a scale of supersymmetry breaking around a few tens of TeV. Thus, performing RG running in the SM up to a few tens of TeV, and then in the MSSM up to an intermediate scale, one finds $\tau_\QCD\sim 20$. 

The simplest CY volume including also a bulk cycle with volume $\tau_b$ looks like:
\begin{equation}
\mathcal{V}=\tau_b^{3/2}-\tau_\QCD^{3/2}\simeq \tau_b^{3/2}\qquad\text{for}\qquad \tau_b\gg \tau_\QCD\,.    
\end{equation}
Eq. (\ref{DecConst}) applied to our case would give:
\begin{equation}
\left(\frac{f_{\rm op}}{f_\QCD}\right)^2\simeq \frac{\partial K}{\partial\tau_\QCD}\left(\frac{\partial^2 K}{\partial\tau_\QCD^2}\right)^{-1} \sim \tau_\QCD\sim 20\quad\Rightarrow\quad f_{\rm op}\gg f_\QCD\,,
\end{equation}
implying that $\theta_\QCD$ is not eaten by any anomalous $U(1)$. Moreover the QCD axion decay constant is not fixed uniquely by $\alpha_\QCD^{-1}$ since:
\begin{equation}
f_\QCD^2\simeq \frac{\partial^2 K}{\partial\tau_\QCD^2}\simeq \frac{M_p^2}{\mathcal{V}\sqrt{\tau_\QCD}}\simeq \left(\frac{M_s}{\alpha_\QCD^{-1/4}}\right)^2\,.
\label{fQCD}
\end{equation}
If $M_s\sim 10^{10}$ GeV for $\mathcal{V}\sim 10^{15}$, one then obtains indeed $f_\QCD\sim 5\times 10^9$ GeV for $\alpha_\QCD^{-1}\sim 20$. Let us make two observations. The first is that $f_\QCD$ is tight to the string scale, and so any realisation of cosmic inflation in $4$ dimensions features inevitably a Hubble scale below the PQ breaking scale, $H_{\rm inf}<f_\QCD$. This leads to a typical post-inflationary scenario where $H_{\rm inf}$ is upper bounded by isocurvature constraints due to the oscillations of the light axion during inflation. The second observation is that the axion decay constant would not be decoupled from the QCD gauge coupling if one tried to realise the QCD axion with $\theta_b$. In fact, in the case of a bulk cycle, one would obtain:
\begin{equation}
f_{\theta_b}^2\simeq \frac{\partial^2 K}{\partial\tau_b^2}\simeq \left(\frac{M_p}{\tau_b}\right)^2\simeq \left(\frac{M_p}{\alpha_b^{-1}}\right)^2\,.
\end{equation}
which would give $f_{\theta_b}\sim 10^{16}$ GeV for $\alpha_\QCD^{-1}\sim 45$ which is the right QCD coupling for a higher string scale, $M_s\sim M_p/\tau_b^{3/4}\sim 5\times 10^{16}$ GeV. 

Having established that $\theta_\QCD$ is a promising meV axion candidate, it remains to show that the F-terms potential has indeed a minimum at $\mathcal{V}\sim 10^{15}$ and $\tau_\QCD\sim 20$, and that the moduli mass spectrum around this vacuum does not lead to any cosmological moduli problem (CMP) for the saxions. Such a large value of the CY volume can be achieved in LVS models where $\mathcal{V}$ is stabilised at exponentially large values in terms of the string coupling $g_s\ll 1$ as $\mathcal{V}\sim e^{c/g_s}\gg 1$ \cite{Balasubramanian:2005zx,Cicoli:2008va}, where $c$ is an $\mathcal{O}(1)$ parameter that depends on the details of the underlying microscopic construction.

Note that, contrary to closed string axions $\theta^i$ which enjoy perturbatively exact shift symmetries, the saxionic component $\tau^i$ of any K\"ahler modulus $T^i=\tau^i + {\rm i}\, \theta^i$ is not protected by any symmetry. Hence, if perturbative effects dominate over non-perturbative ones, $\tau_\QCD$ can be much heavier than $\theta_\QCD$, resulting in 
$m_{\theta_\QCD}\sim \mathcal{O}({\rm meV}) \ll m_{\tau_\QCD}\gtrsim \mathcal{O}(30)$ TeV. However in LVS models the main contribution to the potential of blow-up modes arises typically from non-perturbative corrections to $W$ due to the extended no-scale cancellation of loop corrections to $K$ \cite{Cicoli:2007xp}. This results in very heavy axions associated to blow-up cycles. The situation is, however, more subtle for blow-up modes wrapped by the QCD stack of branes due to the well-known tension between non-perturbative effects and chirality \cite{Blumenhagen:2007sm}. In fact, as we have seen, in this case $T_\QCD$ gets a non-zero charge $q_\QCD$ under an anomalous $U(1)$. Thus, any non-perturbative corrections to $W$, to be gauge invariant, has to depend also on the open string modes $\phi$ with charge $\hat{q}_\QCD$:
\begin{equation}
W\supset A\,\phi^n\,e^{-2\pi T_\QCD}\qquad\text{with}\qquad n=q_\QCD/\hat{q}_\QCD\,.
\label{Wnp}
\end{equation}
As can be seen from (\ref{DecConst}), D-term stabilisation fixes $|\phi|^2\sim \sqrt{\tau_\QCD}/\mathcal{V}\ll 1$, introducing an extra $\mathcal{V}$ suppression factor in (\ref{Wnp}) which makes this term negligible. Therefore $\tau_\QCD$ can be stabilised by contributions to the scalar potential which do not depend on $\theta_\QCD$. These are loop corrections to $K$ and supersymmetry breaking effects of order the gravitino mass $m_{3/2}\sim W_0\,M_p/\mathcal{V}$ (where $W_0$ is the $\mathcal{O}(10)$ flux superpotential \cite{Chauhan:2025rdj}) \cite{Cicoli:2017zbx,Broeckel:2021dpz}:
\begin{equation}
V (\tau_\QCD) = m_{3/2}^2 |\phi|^2 + \frac{c_{\rm loop}}{\mathcal{V}^3\sqrt{\tau_\QCD}} \simeq \frac{W_0^2}{\mathcal{V}^3}\left(c_\D\sqrt{\tau_\QCD} + \frac{c_{\rm loop}}{\sqrt{\tau_\QCD}}\right),
\end{equation}
where we have substituted $|\phi|$ in terms of $\tau_\QCD$ using D-term stabilisation, and $c_\D$ and $c_{\rm loop}$ are $\mathcal{O}(1)$ coefficients which depend, respectively, on $2$-form and $3$-form gauge flux quanta. This potential has a minimum at $\langle\tau_\QCD\rangle \simeq c_{\rm loop}/c_\D$ which can easily lie around $\langle\tau_\QCD\rangle\sim 20$. Moreover, it gives a mass to $\tau_\QCD$ of order $m_{\tau_\QCD}\sim m_{3/2}\sim \mathcal{O}(50)$ TeV for $W_0\sim\mathcal{O}(10)$, while leaving $\theta_\QCD$ as a flat direction. A similar result can be obtained also if $\langle |\phi|\rangle=0$ for geometries where $\tau_\QCD$ intersects another local blow-up cycle with volume $\tau_{\rm int}$. In that case, $V_D=0$ if the FI-term vanishes, corresponding to stabilising $\tau_{\rm int}$ in terms of $\tau_\QCD$ which, in turn, can then be fixed by string loops  at subleading order \cite{Cicoli:2022fzy, Cicoli:2011qg,Cicoli:2012sz}. 

This model seems to provide an attractive meV QCD axion with $f_\QCD\sim 5\times 10^9$ GeV, the correct QCD coupling and a saxion above the CMP bound, $m_{\tau_\QCD}\sim \mathcal{O}(50)$ TeV \cite{Conlon:2006tq}. However it faces a severe challenge associated to the decay of the volume mode after BBN due to the smallness of its mass, $m_{\tau_b}\sim M_p/\mathcal{V}^{3/2}\sim \mathcal{O}(1)$ MeV. In order to make $\tau_b$ heavy enough to avoid any CMP, one would need to focus on vacua with a smaller value of $\mathcal{V}$, $\mathcal{V}\lesssim 10^8$. This lower bound would then restrict the possible values of the QCD axion decay constant, using (\ref{fQCD}), to the window $10^{13}\,{\rm GeV}\lesssim f_\QCD\lesssim 10^{16}$ GeV, where the upper bound comes from the condition to trust the $\alpha'$ expansion that, for models with small $h^{1,1}$, can be estimated as $\mathcal{V}^{-1/3}\lesssim 0.1$. This conclusion might seem to forbid an meV-scale QCD axion, unless the CMP is solved via a late time dilution mechanism or a suppressed initial misplacement of $\tau_b$ from the minimum of its potential. In this case, the CY volume could be safely as large as $\mathcal{V}\sim 10^{15}$, resulting in $f_\QCD\sim 5\times 10^9$ GeV. 

Given that different values of $\mathcal{V}$ yield different values of $f_\QCD$, it is interesting to investigate the statistical distribution of the QCD axion decay constant in the type IIB flux landscape moving along trajectories in the field space which correspond to stabilised vacua. This analysis has been performed in \cite{Broeckel:2021dpz} by varying flux quanta at fixed $h^{1,1}$, finding a mild logarithmic preference for larger decay constants of the form $N(f_\QCD)\sim \ln\left(f_\QCD/M_p\right)$, where $N(f_\QCD)$ counts the number of flux vacua for a given $f_\QCD$. 

A complementary statistical approach has been taken in \cite{Mehta:2020kwu,Mehta:2021pwf} which derived numerically the distribution of the axion decay constants fixing a point in moduli space and varying $h^{1,1}$ by changing geometries within the largest known ensemble of CY manifolds given by the Kreuzer-Skarke database \cite{Kreuzer:2000xy}. In this construction, a CY threefold is built as a complex 3-dimensional hypersurface in a complex 4-dimensional ambient toric variety, and the largest number of K\"ahler moduli is $h^{1,1}=491$. The decay constants of all axions are estimated by the eigenvalues of the kinetic matrix using the software \textsc{CYTools}~\cite{Demirtas:2022hqf}, and their distribution turns out to be approximately log-normal with a decreasing mean value as $h^{1,1}$ increases (a similar trend has been found also in more general F-theory models \cite{Fallon:2025lvn}). In particular, $f_\QCD\lesssim \mathcal{O}(10^{10})$ GeV for $h^{1,1}\gtrsim \mathcal{O}(400)$, showing that, at large $h^{1,1}$, the lower bound on the mass of a closed string QCD axion lies exactly around the meV scale. 

\begin{figure}
\centering
\includegraphics[width=0.49\linewidth]{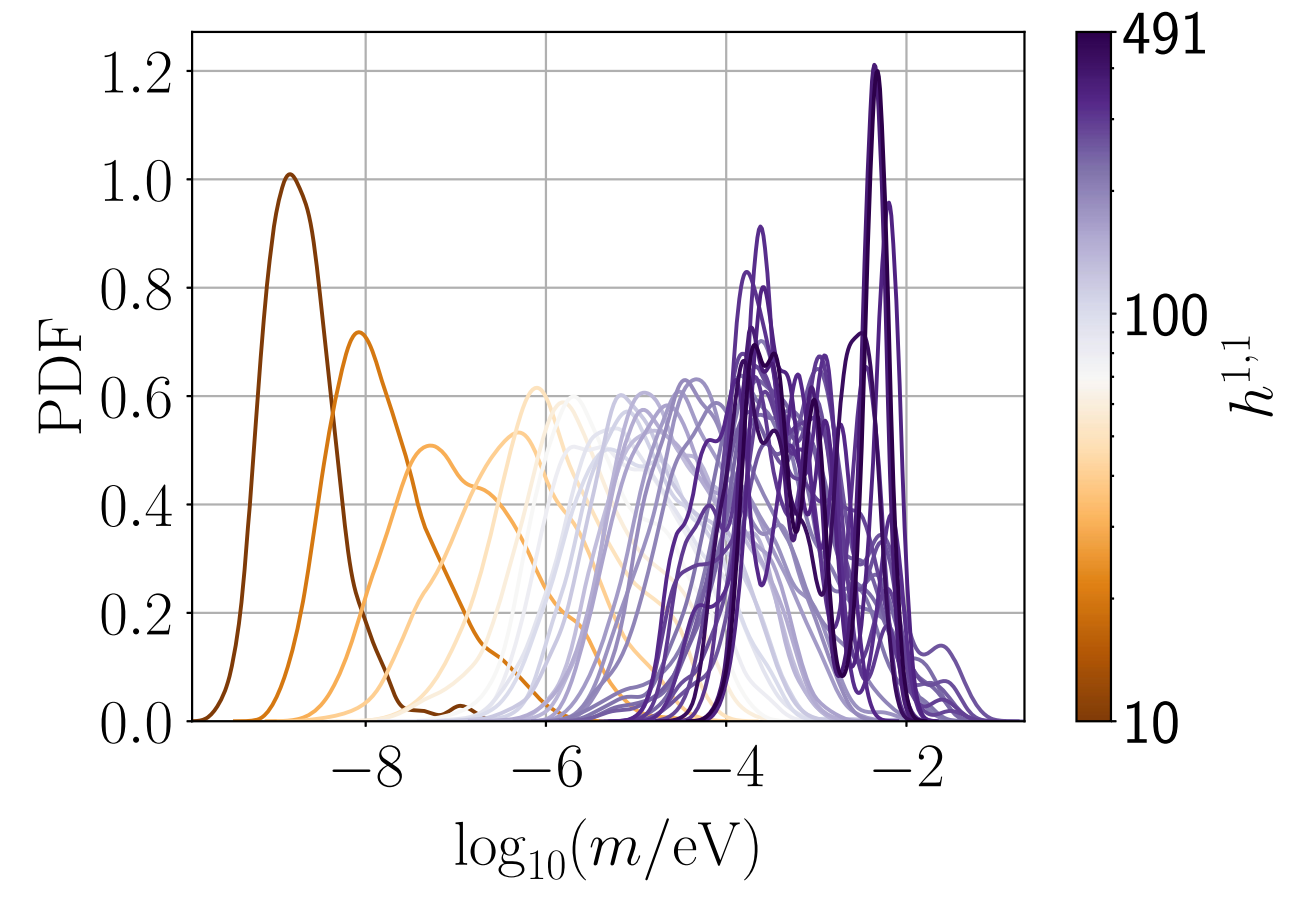}
\includegraphics[width=0.49\linewidth]{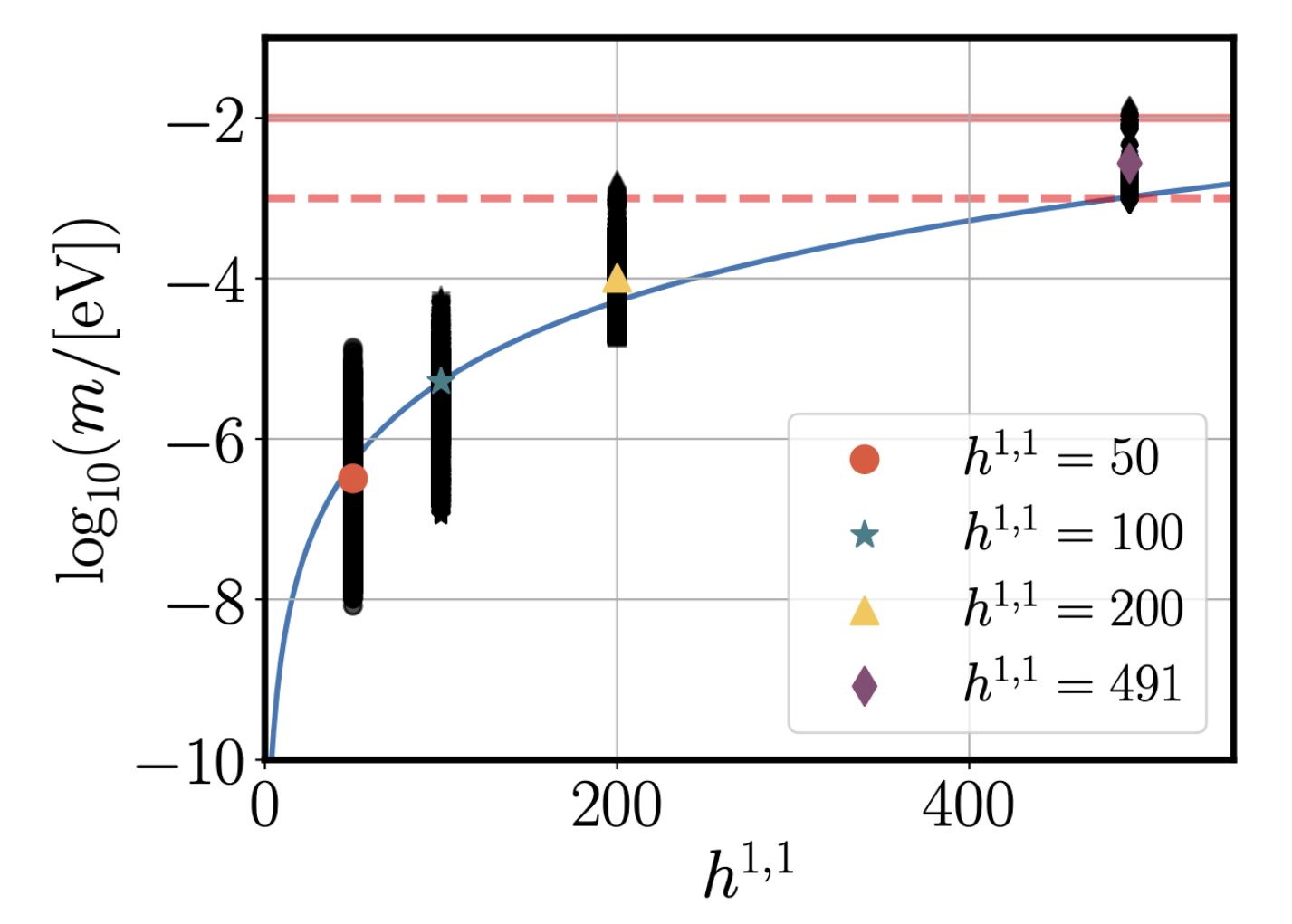}
\caption{QCD axion mass, at random points along a ray from the tip of the SKC, for different values of $h^{1,1}$ in type IIB CY models.  \emph{Left.} Probability distribution on the QCD axion mass colour coded by $h^{1,1}$. Distributions move towards meV for $h^{1,1}\gtrsim 200$ (reproduced from \cite{Gendler:2024adn}). \emph{Right.} QCD axion mass as a function of $h^{1,1}$. Coloured dots denote the average in the bin, the line denotes a power law assuming $\mathcal{V}\sim (h^{1,1})^5$. The horizontal solid line is the current limit from SN1987A, and the dotted line is the projection for the next galactic supernova (reproduced from \cite{Gendler:2023kjt}).}
    \label{fig:kreuzer-skarke_scan}
\end{figure}

The point in moduli space considered in \cite{Mehta:2020kwu,Mehta:2021pwf} is the tip of the stretched K\"ahler cone (SKC), corresponding to the minimal value of all $2$-cycle moduli $t^i$ which allows to control $\alpha'$ corrections to the EFT. Ref. \cite{Demirtas:2018akl} found that this requirement can be translated into an $h^{1,1}$-dependent lower bound on the overall volume of the form $\mathcal{V}\gtrsim 10^{-3.4} (h^{1,1})^{7.2}$. Intuitively, this is due to the fact that, as $h^{1,1}$ increases, the conditions which define the K\"ahler cone also increase, making the cone narrower, and so moving the tip of the SKC farther from the origin. For $h^{1,1}\gtrsim 400$, one would indeed obtain $\mathcal{V}\gtrsim 10^{15}$ which, as we have seen above, is in the right ballpark to realise an meV-scale QCD axion using a $C_4$-axion associated to a rigid blow-up mode. Fig.~\ref{fig:kreuzer-skarke_scan} shows the distribution of the QCD axion mass evaluated at points that reproduce the correct UV QCD coupling and are randomly selected along a ray originating from the tip of the SKC. The QCD axion mass increases as a power law with $h^{1,1}$, and starts to have support near the meV scale for $h^{1,1}\gtrsim 200$. At the largest value of $h^{1,1}=491$, the QCD axion mass appears to be distributed around $1$ meV. It is interesting to observe that the distribution barely goes above $10$ meV: these string theory models appear largely unconstrained by the SN1987A neutrino burst limits~\cite{Raffelt:2006cw,Caputo:2024oqc}, but are within reach of other meV phenomenology. 

Let us stress that these results are valid at the tip of the SKC (or along a ray generated from the tip) but one could also analyse the distribution of the axion decay constants when moving deeper inside the K\"ahler cone. An indication of the outcome of this study comes from the analysis of \cite{Broeckel:2021dpz} which implies a mild statistical preference for an meV-scale QCD axion compared to larger masses, when considering all allowed values of $f_\QCD$ at very large $h^{1,1}$. To have a complete understanding of the statistical distribution of $f_\QCD$ in the entire flux landscape, one should combine this result with the distribution at small $h^{1,1}$, counting the number of different vacua which depends on $h^{1,2}$ and the number of topologically inequivalent CY geometries. This is still an open question, due also to the difficulty to stabilise explicitly all the moduli at large $h^{1,1}$ without a CMP (see however \cite{Cicoli:2016chb} for an attempt to fix an arbitrary number of moduli perturbatively at exponentially large $\mathcal{V}$). The PQ quality is, however, in general exponentially good at large $h^{1,1}$ due to the large number of shift symmetries and the presence of axions much lighter than the QCD axion~\cite{Demirtas:2021gsq}. Different phenomenological aspects of axion physics at large $h^{1,1}$ have been studied in~\cite{Benabou:2025kgx,Sheridan:2024vtt}. 

Let us finally comment that for axion-like particles (ALP) not tied to QCD, the relation between mass and decay constant is less constrained. For $C_4$-axions associated to bulk cycles $\tau_b$, ALP decay constants and masses exhibit parametric scalings of the schematic form $f_\ALP\sim M_p/\tau_b$ and 
$m_\ALP \sim e^{-k M_p/f_\ALP}\,M_p$, where $k$ is an $\mathcal{O}(1)$ coefficient which depends on the instanton order at which the ALP becomes massive. These relations indicate that meV ALPs tend to be associated with very large decay constants, potentially as high as $f_a \sim 10^{16}\,\text{GeV}$. Conversely ALPs with intermediate scale decay constants (as would be the case for an meV-scale QCD axion) are effectively massless. Moreover, when considering mass eigenstates, these ALPs tend to couple to the visible sector extremely weakly \cite{Broeckel:2021dpz}, and so might be effectively invisible to many experimental probes. A more promising case is the one of $C_4$-axions associated to small moduli $\tau_s$ which intersect the rigid $4$-cycle supporting the QCD stack of branes \cite{Cicoli:2012sz}. In this case, the ALP can couple to the visible sector more strongly with decay constant $f_\ALP\sim M_s\sim M_p/\sqrt{\mathcal{V}}$ which can be at an intermediate scale for the same value of $\mathcal{V}$ that yields an meV-scale QCD axion. On the other hand, the ALP mass, scaling as $m_\ALP \sim e^{-k \tau_s}\,M_p$, can take a large range of values since it is very sensitive to the specific values of $k$ and $\tau_s$.

\subsubsection{meV axions as open strings}

D3-branes at del Pezzo singularities are a common way to realise chiral non-Abelian gauge theories like QCD \cite{Cicoli:2012vw,Cicoli:2013cha,Cicoli:2013mpa,Cicoli:2017shd,Cicoli:2021dhg}. The singularity is obtained via the collapse of a del Pezzo divisor, $\tau_\dP\to 0$, induced dynamically by moduli stabilisation. In these models there is always an anomalous $U(1)$ factor that acquires a St\"uckelberg mass by absorbing the closed string axion $\theta_\dP$ that is the supersymmetric partner of $\tau_\dP$. The role of the QCD axion is therefore played by an open string mode \cite{Choi:2010gm,Cicoli:2013cha,Allahverdi:2014ppa,Cicoli:2017zbx,
Petrossian-Byrne:2025mto, Loladze:2025uvf}. Being localised at a singularity, the visible sector is in general sequestered from supersymmetry breaking sourced by fluxes in the bulk \cite{Blumenhagen:2009gk,Aparicio:2014wxa}, resulting in an axion decay constant which is suppressed with respect to the string scale. Thanks to this property, one can realise an meV axion without a CMP for any saxion. Moreover, one can also reproduce, at the same time, the correct QCD coupling which for D3-branes is set by the dilaton $S$, $\alpha_\QCD^{-1}={\rm Re}(S) = g_s^{-1}$.

Let us see this picture more in detail. The K\"ahler potential expanded around the del Pezzo singularity reads:
\begin{equation}
K=-2\ln\mathcal{V} + \frac{\tau_\dP^2}{\mathcal{V}}\,.   
\end{equation}
Thus, D-term fixing leads to:
\begin{equation}
|\phi|^2\simeq \xi\simeq \frac{\partial K}{\partial\tau_\dP} \simeq \frac{\tau_\dP}{\mathcal{V}} \simeq \tau_\dP\,M_s^2\,.   
\end{equation}
This relation can be interpreted as fixing $\tau_\dP$ in terms of the open string mode $|\phi|$ whose F-term potential receives supersymmetry breaking contributions of the form:
\begin{equation}
V_F(|\phi|) = c_2\,m_0^2\, |\phi|^2 + c_3\,A\, |\phi|^3 + \mathcal{O}(|\phi|^4)\,,
\end{equation}
where $c_2$ and $c_3$ are $\mathcal{O}(1)$ coefficients while the soft scalar mass $m_0$ and the trilinear coupling $A$ scale with the CY volume, respectively, as $m_0^2 \sim M_p^2/\mathcal{V}^{p_2}$ and $A\sim M_p/\mathcal{V}^{p_3}$. If $c_2>0$, the minimum lies at $\langle|\phi|\rangle=0$ which implies $\tau_\dP=0$, ensuring that the del Pezzo divisor shrinks down to zero size dynamically. However there is no breaking of the PQ symmetry. If, instead, the soft mass is tachyonic, i.e. $c_2<0$, the PQ symmetry is broken at the scale:
\begin{equation}
f_\zeta=\langle|\phi|\rangle \simeq \frac{M_p}{\mathcal{V}^{p_2-p_3}} \qquad\Rightarrow\qquad \langle\tau_\dP\rangle \simeq \frac{1}{\mathcal{V}^{2 (p_2- p_3)-1}}\,.
\end{equation}
The value of $|\phi|$ sets the decay constant of the open string phase $\zeta$, $f_\zeta\sim \langle |\phi|\rangle$, which should be compared with the decay constant of the closed string axion $\theta_\dP$:
\begin{equation}
f_{\theta_\dP}^2\simeq \frac{\partial^2 K}{\partial\tau_\dP^2}\sim \frac{M_p^2}{\mathcal{V}}\sim M_s^2\,.    
\end{equation}
The powers $p_2$ and $p_3$ can take different values depending on model building details \cite{Blumenhagen:2009gk,Aparicio:2014wxa}. In the absence of sequestering where all soft terms are of order the gravitino mass, they are $p_2=2$ and $p_3=1$. On the other hand, when the visible sector is sequestered, they can be either $p_2=3$ and $p_3=2$, or $p_2=4$ and $p_3=2$. Let us consider each case separately: 
\begin{eqnarray}
(1)\,\, p_2&=&2\quad p_3=1:\qquad f_\zeta \simeq \frac{M_p}{\mathcal{V}}\ll f_{\theta_\dP}\simeq \frac{M_p}{\sqrt{\mathcal{V}}}\qquad m_0\simeq \frac{M_p}{\mathcal{V}}\qquad \langle\tau_\dP\rangle \simeq\frac{1}{\mathcal{V}}\ll 1\,,  \nonumber \\ 
(2)\,\, p_2&=&3\quad p_3=2:\qquad f_\zeta \simeq \frac{M_p}{\mathcal{V}}\ll f_{\theta_\dP}\simeq \frac{M_p}{\sqrt{\mathcal{V}}}\qquad m_0\simeq \frac{M_p}{\mathcal{V}^{3/2}}\qquad \langle\tau_\dP\rangle \simeq\frac{1}{\mathcal{V}}\ll 1\,, \nonumber \\
(3)\,\, p_2&=&4\quad p_3=2:\qquad f_\zeta \simeq \frac{M_p}{\mathcal{V}^2}\ll f_{\theta_\dP}\simeq \frac{M_p}{\sqrt{\mathcal{V}}}\qquad m_0\simeq \frac{M_p}{\mathcal{V}^2}\qquad \langle\tau_\dP\rangle \simeq\frac{1}{\mathcal{V}^3}\ll 1\,. \nonumber
\end{eqnarray}
Note that in all cases $f_\zeta\ll f_{\theta_\dP}$, implying that the eaten axion is indeed $\theta_\dP$, and the value of the del Pezzo modulus is $\langle\tau_\dP\rangle \ll 1$, ensuring that this $4$-cycle is in the singular regime. In the first two cases, reproducing an meV axion with $f_\zeta\sim 5\times 10^9$ GeV, requires $\mathcal{V}\sim 10^8$ which gives $M_s\sim 10^{14}$ GeV, and high scale soft terms around $m_0\sim m_{3/2}\sim 10^{10}$ GeV for case (1), and $m_0\sim 10^6$ GeV for case (2). Note that in both cases the model does not suffer from any CMP since the mass of the volume mode is $m_{\tau_b}\sim M_p/\mathcal{V}^{3/2}\sim 10^6$ GeV. Case (3) instead can reproduce an meV axion for $\mathcal{V}\sim 10^4$, which correlates with $M_s\sim 10^{16}$ GeV, $m_0\sim 10^{10}$ GeV and $m_{\tau_b}\sim 10^{12}$ GeV. 

It is worth pointing out that in all cases the decay constant of the open string axion is suppressed with respect to the string scale, allowing for the possibility to realise a standard post-inflationary scenario with $f_\zeta\ll H_{\rm inf}$. Due to the absence of any CMP, this model seems more promising to realise an meV axion. The main challenge is however the control over the EFT around the singularity, in particular the form of perturbative corrections to the K\"ahler potential, as well as potential sources of desequestering due to moduli redefinitions \cite{Conlon:2010ji}.

\section{meV axions as cold dark matter}
\label{sec:meV_axions_as_cold_dark_matter}


Relic axions can be produced thermally by particle reactions initiated by the primordial plasma or by a number of non-thermal processes. 
Today, the former would have momenta of the order of the temperature of the Universe, $T\sim 2.3\times 10^{-4}$ eV, 
smaller than a meV. However, they would have become non-relativistic late enough in the history of the Universe to behave as ``hot" dark matter or dark radiation for most cosmological observables. The next generations of precision cosmological surveys will sharpen our measurements of $\Lambda$CDM parameters and could reveal a number of surprises in observables like the power spectrum of fluctuations or $N_{\rm eff}$, cf. Sec.~\ref{sec:dark_radiation} for a complete discussion. 

\begin{figure}
    \centering    \includegraphics[width=0.45\linewidth]{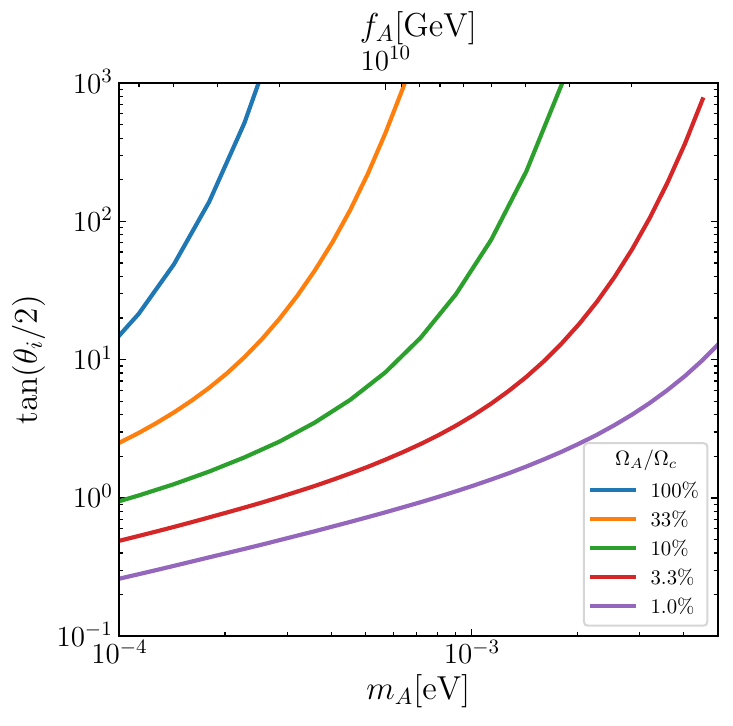}
\includegraphics[width=0.45\linewidth]{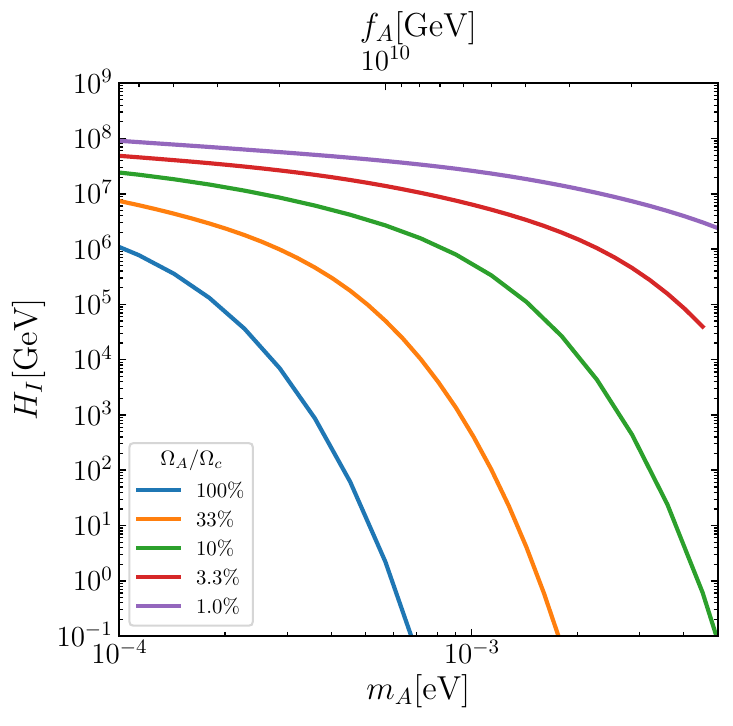}
    \caption{\emph{Left.} Value of the initial misalignment angle required to obtain different fractions of axion cold dark matter. \emph{Right.} Upper bound on the CDM abundance of meV axions in the LMA scenario as a function of the inflationary rate $H_I$. Figure reproduced from Ref.~\cite{IAXO:2019mpb}.}
    \label{fig:upper}
\end{figure}

In the following, we will discuss non-thermally produced QCD axion cold dark matter. We focus on vanilla cosmology, assuming radiation-domination (RD) started at temperatures $T_{\rm RH}\gg $ GeV.  Depending on the timing of the PQ symmetry breaking, either before or after inflation, the dominant production mechanisms are the misalignment mechanism or axion production from the decay of topological defects. The characteristic momentum of the ensued axions is tied to the Hubble scale, which is suppressed by factors of $T/M_{\rm Planck}$ with respect to the temperature, making them perfect candidates for CDM. 
In the pre-inflationary scenario, the ``standard" misalignment mechanism for \textit{untuned} initial conditions far from $\theta_0\simeq (2n+1)\pi$, with $n\in\mathbb{Z}$, predicts the observed CDM yield to be on the per-mil level~\cite{Ballesteros:2016xej}, 
\begin{equation}
    \frac{\Omega_a h^2}{0.12} \simeq 2.7\times 10^{-3}\theta_i^2 \left(\frac{\rm meV}{m_a}\right)^{1.17}.
\end{equation}
The yield can be significantly increased by a few effects.  In the \textit{large misalignment angle} (LMA) scenario \cite{Arvanitaki:2019rax}, the initial angle $\theta_i$ is very close to the top of the potential, $\theta\approx\pi$, delaying the onset of axion oscillations, see Fig.~\ref{fig:upper} (left). Assuming the $-\chi_T\cos\theta$ potential motivated by the dilute instanton gas approximation (DIGA) and following the standard calculation, see e.g.~\cite{Borsanyi:2016ksw}, one needs $\theta_0-\pi\sim 10^{-32}$ for $m_a = 6$ meV axions to be dark matter. 

Such a finely tuned result is very easy to invalidate after taking into account fluctuations. Adiabatic perturbations might not spoil the prediction, but isocurvature fluctuations are potentially enhanced and very constrained~\cite{Kobayashi:2013nva}. 
Since the latter are controlled by the inflationary expansion rate $H_I$, imposing a lower bound on $H_I$ gives us an upper bound on the CDM abundance as a function of axion mass as shown in Fig. \ref{fig:upper}. 

In the \textit{kinetic misalignment} (KM) scenario \cite{Co:2019jts}, the axion field at GeV temperatures still has some residual velocity from its non-trivial history during the early Universe. 
The onset of oscillations, which happens at a time $t_1$ given by $m_a(t_1)=H(t_1)$ in the standard misalignment case, is delayed until the kinetic energy drops below the potential energy and the field is ``trapped" by the QCD potential. Axions with meV masses require a very significant velocity and delay of the trapping with respect to the standard $\dot \theta_1=0$ case to account for a sizable fraction of the CDM abundance, 
\begin{equation}
\Omega_a h^2 \simeq 
0.12\left(\frac{\rm meV}{m_a}\right)^{1.176} \frac{\dot\theta_1/H_1}{545}. 
\end{equation}
Adiabatic (and potentially isocurvature) fluctuations become $\mathcal{O}(1)$ for the large velocities required in the meV region due to non-linearities~\cite{Eroncel:2022vjg}. This affects the relic density calculation, but not as radically as in the LMA case~\cite{Fasiello:2025ngh}. In the extreme cases of the misalignment mechanism relevant for the meV axion scenario discussed here, the nonlinear dynamics can give rise to interesting phenomena with direct implications for experiments, such as the formation of axion miniclusters, which are typically considered a distinctive feature of post-inflationary axions, cf. Refs.~\cite{Eroncel:2022efc, KM:2026}. 

In the post-inflationary scenario, the evolution of the resulting cosmic string network releases axions as string loops collapse, oscillate and intersect. The energy radiated by the network, quantified by the \textit{string tension} $\mu$, is enhanced by the large gradient energy near the axion strings, which is proportional to $\ell\equiv \log d_H/d_c$, the logarithm of the ratio of the typical inter-string distance $\mathord{\sim}H^{-1}$ and the inverse string core size $\mathord{\sim}f_a^{-1}$ which at around the QCD scale is $f_a/H\approx 10^{30}$, or $\ell\approx 70$. 

The axion radiation from strings is measured as the instantaneous emission spectrum $\mathcal{F}$, which in the scaling regime is expected to follow a power-law $\mathcal{F}\sim k^{-q}$, with spectral index $q$ between IR and UV cut-offs limited by causality (parameterized by a multiple of the Hubble scale, $x_0H$) and the string core size ($\mathord{\sim}f_a^{-1}$), respectively. 
The absolute normalisation is proportional to the string length per Hubble patch $\xi$.  The resulting density of axions has been the subject of a longstanding debate, which has been revived in the last decade, with the hope of settling it with state-of-the-art numerical simulations~\cite{Gorghetto:2018myk, Gorghetto:2020qws, Hindmarsh:2021zkt, Buschmann:2019icd, Buschmann:2021sdq,Saikawa:2024bta, Kim:2024wku, Benabou:2024msj,Correia:2024cpk,Correia:2025nns}. 

Most groups simulate the axion field as the angular degree of freedom of a complex scalar field $\phi$ with potential $V_{\rm PQ} \sim \lambda\left(|\phi|^2-v^2\right)^2$. Current high-performance computing (HPC) clusters have limited the hierarchy of scales to a $\mathcal{O}(1)$ fraction of the grid size, typically $N^3 = (16k)^3$ or less grid sites for regular grids and up to effectively $N^3 = (262k)^3$ using Adaptive Mesh Refinement (AMR) \cite{Benabou:2024msj}. A clever model with two complex scalars with charges $q_{1,2}$ under a vector field is able to simulate strings with much higher ``effective" tensions, $\mu\sim (q_1^2 + q_2^2)$, using the same regular grids~\cite{Klaer:2017ond, Klaer:2017qhr,Klaer:2019fxc}.
The results can be summarized as follows. 
The string length parameter $\xi$ shows an attractor-like behavior whose fixed point increases with $\ell$~\cite{Fleury:2015aca,Gorghetto:2018myk,Vaquero:2018tib}. Simulations with fixed $\ell$ confirm the dependence and allow us to interpret the evolution as a tracking solution where $\xi$ approaches a growing moving target $\xi(\ell)$~\cite{Klaer:2019fxc}. 
This behavior has been established up to $\ell\sim 9$ but is expected to plateau at some level, because in the $\ell\to\infty$ limit, global strings are expected to evolve as Nambu-Goto (NG) strings with renormalised tension, which are insensitive to the value of $\ell$. However, the plateau could be very large. Early NG simulations find $\xi\sim 13$ \cite{Bennett:1989yp} and recent ones $\xi\sim 4$ \cite{Daverio:2015nva,Klaer:2017qhr}. Extrapolating the simulations of Ref. \cite{Saikawa:2024bta} leads to $\xi\sim 7(3)$.  The lower cut-off parameter, $x_0$, has been found to depend on this value \cite{Saikawa:2024bta}, which can  make a difference if large $\xi$ values are assumed at the physical values of $\ell\sim 70$. 

The most relevant parameter for the final abundance is $q$, which controls whether many $k\sim H$ ($q>1$) or few $k\sim f_a$ ($q<1$) axions are emitted by strings.  
Some theoretical expectations can be drawn based on simulations of Nambu-Goto strings coupled to a Kalb-Ramond field $B_{\mu\nu}$, which plays the role of the axion via the duality $\nabla\theta_{\mu} \sim \nabla_{[\nu} B_{\rho\sigma]}$ in four spacetime dimensions \cite{Vilenkin:1986ku, Davis:1988rw}. Long strings radiate low energy axions and loops lead to $q\sim 4/3$ except in some cases like the circular loop, $q\approx1$. 
Thus, one expects $q\gtrsim 1$ and a significant enhancement of the axion abundance. 
On the other hand, some authors argue that $q=1$ is the correct expectation for global strings \cite{Buschmann:2021sdq,Benabou:2024msj,Correia:2025nns}. 
Network simulations show either $q\simeq1$~\cite{Buschmann:2021sdq,Benabou:2024msj} or $q<1$ with a linearly increasing trend with $\ell$ approaching $q\sim 1$~\cite{Gorghetto:2018myk,Gorghetto:2020qws,Saikawa:2024bta,Kim:2024wku}. The results appear to be compatible at the moment, given that the first AMR simulation \cite{Buschmann:2021sdq} did not use initial conditions in the attractor but slightly underdense, which increases $q$ \cite{Saikawa:2024bta}. 
The same applies to the updated results from an even larger AMR simulation of the same group~\cite{Benabou:2024msj}. The attractor simulations extrapolate well linearly to $q>1$ at $\ell\sim 70$ but a saturating solution $q\to 1$ cannot be excluded, particularly given the large discretisation effects affecting the spectrum at large $\ell$ \cite{Saikawa:2024bta}. 

A recent study considered the uncertainties in the simulations to extrapolate the axion number radiated by networks taking into account different extrapolation models \cite{Saikawa:2024bta}. If $q>1$ at $\ell\sim 70$ is assumed, the axion yield is so large that $\langle |\dot\theta|^2\rangle $ is significant at $t_1$, in analogy to the kinetic misalignment in the pre-inflation scenario. Cosmic strings and walls tend to be subdominant at this stage. The delay in the oscillations covers a period where nonlinearities actually decrease the axion number, moderating the enhancement gained by the IR emission spectrum~\cite{Gorghetto:2020qws}. If $q=1$ is assumed, these effects are not large.
The combined theoretical uncertainty in the post-inflationary 
axion CDM scenario within a standard RD cosmology leads to the mass range \cite{Saikawa:2024bta},
\begin{equation}
   50\ \mu \mathrm{eV} \lesssim m_a \lesssim 0.5\ \mathrm{meV}, 
\end{equation}
which encompasses the results of Refs.~\cite{Benabou:2024msj,Gorghetto:2020qws, Kim:2024wku}, lying near the lower and upper ends of this range, respectively\footnote{For a detailed overview and an in-depth discussion of the predictions and methodologies adopted by the different groups, we refer to Sec. 4.2 of Ref.~\cite{Arza:2026rsl}}. 

\begin{figure}
    \centering
    \includegraphics[width=0.5\linewidth]{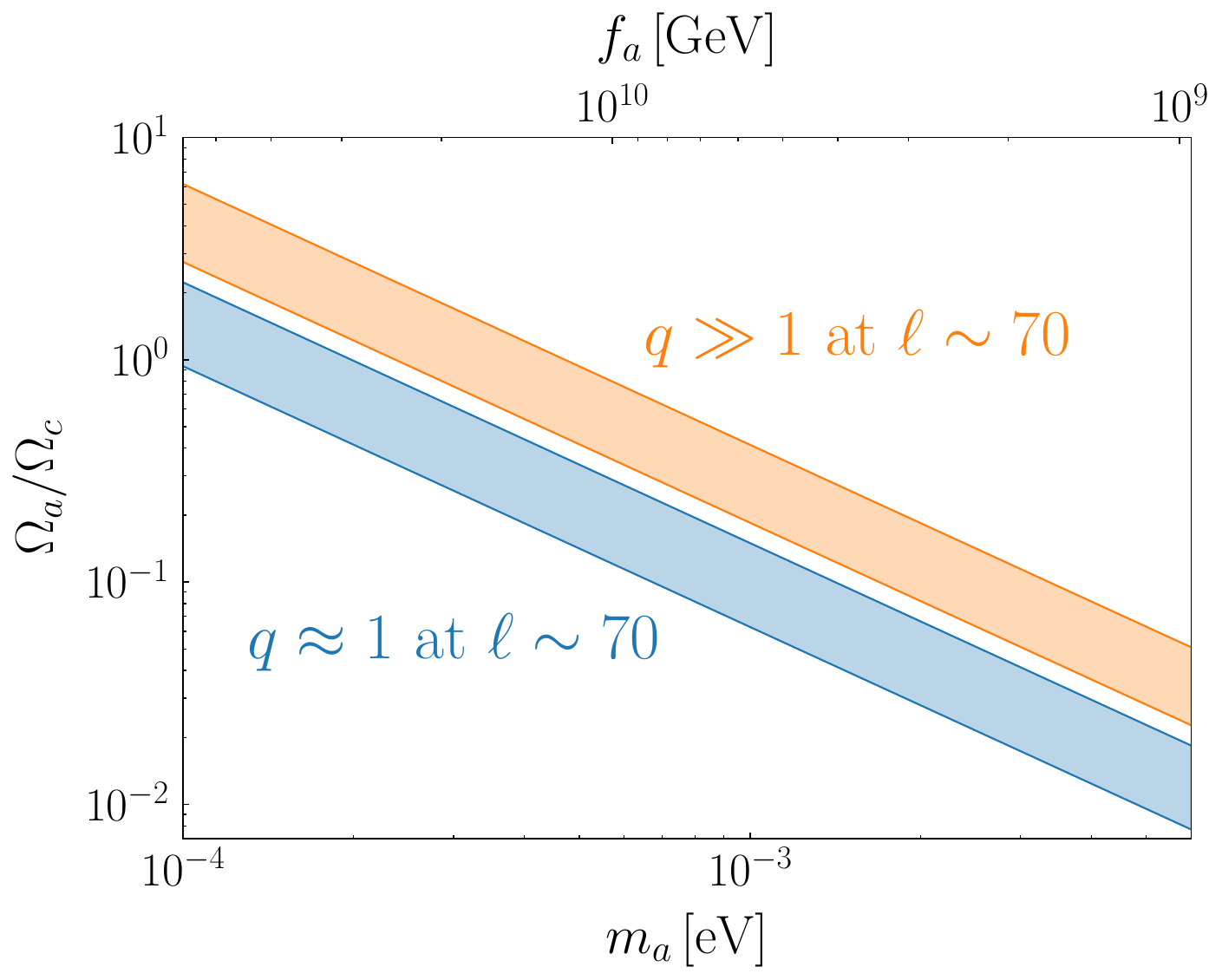}
    \caption{Abundance of $N_{\rm DW}=1$ post-inflationary axions from strings as fraction of the observed CDM. Different colors correspond to different assumptions about the spectral index, $q \approx 1$ (blue) and $q \gg 1$ (orange) evaluated at the physical value of $\ell\sim 70$, and the latter includes the effect of nonlinearities discussed in Ref.~\cite{Gorghetto:2020qws}.
    The colored regions also show the systematic uncertainties associated with the extrapolation of the simulation results estimated in Ref.~\cite{Saikawa:2024bta}.}
    \label{fig:NDW1Posti}
\end{figure}

The axion cold DM abundance from strings scales approximately as $f_a^{1.17}$, neglecting nonlinear effects. Using this scaling to extrapolate to the meV-mass regime suggests that post-inflationary 
axions contribute at the level of $1-50\%$ of the CDM, cf. Fig. \ref{fig:NDW1Posti}. 

The above estimate of the axion CDM abundance in the post-inflationary scenario can be applied to the models with $N_{\rm DW} = 1$, where $N_{\rm DW}$, dubbed as a ``domain wall number", is an integer representing the number of degenerate minima of the axion potential.
For $N_{\rm DW}>1$, the picture is very different, because in this case the axion field has $N_{\rm DW}$ $CP$-conserving, degenerate vacua, all of which get statistically equally populated in the Universe. After the QCD phase transition, the Universe would be a collection of domains separated by domain walls (DWs), sheet-like topological defects associated with the spontaneous breaking of a discrete $\mathbb{Z}_{N_{\rm DW}}$ subgroup of the $U(1)_{\rm PQ}$ symmetry. 
The contribution of these walls can eventually dominate the energy density of the Universe, which is observationally excluded~\cite{Zeldovich:1974uw,Sikivie:1982qv}, but small modifications of the standard scenario change this conclusion dramatically. As examples, we mention the effect of small PQ-breaking terms and models with more than one axion. 
The PQ-breaking terms, which may be completely natural from quantum gravity if PQ is a purely global symmetry, will lift these vacua by favouring the least energetic configuration~\cite{Kallosh:1995hi}. This drives the collapse of the DWs and restores a Universe compatible with observations~\cite{Sikivie:1982qv,Gelmini:1988sf,Larsson:1996sp}. But these ``biases" imply that the axion will not exactly cancel the $CP$ violation of the SM in the vacuum and they are thus very strongly constrained by the upper limits on the neutron EDM \cite{Chang:1998tb,Hiramatsu:2012sc,Kawasaki:2014sqa,Ringwald:2015dsf}. Limiting bias terms to moderate $CP$ violation requires making them very small or very fine-tuned to zero in the true vacuum. If they are small, then the pressure difference between the domains is small, and DWs release their energy much later than the epoch of the QCD phase transition, leading to a huge boost in CDM such that it surpasses the observed CDM abundance. Therefore, if not excluded, meV axions tend to account for $100\%$ of the CDM in this scenario. A small but non-vanishing $CP$ violation induced by the bias terms for the collapse of DWs also leads to observable effects in the experimental searches for axion-mediated long-range forces~\cite{Arvanitaki:2014dfa}, cf. Ref.~\cite{Giannotti:2017hny}. Furthermore, the late-time collapse of DWs could lead to the production of gravitational waves~\cite{Hiramatsu:2012sc,Saikawa:2017hiv} and the formation of primordial black holes~\cite{Ferrer:2018uiu}, which could be a smoking gun for this scenario. 
Alternatively, if there are two PQ $U(1)$ symmetries, and consequently two Nambu-Goldstone bosons, the string network will contain two types of strings and the domain walls may be eventually locked into cosmologically not-so-harmful string-bundles ~\cite{Lee:2024toz,Lee:2025zpn}, which might survive until today.

Despite several potentially interesting phenomenological consequences, the $N_{\rm DW}>1$ 
scenario has not been studied numerically as intensively as the $N_{\rm DW}=1$ case, see Refs.~\cite{Hiramatsu:2012sc,Kawasaki:2014sqa,Gorghetto:2022ikz}. More detailed studies are needed and much welcomed in view of the relevance and potential impact of the meV frontier detailed in this paper. 

In summary, meV axions can account for a sizeable fraction, if not all, of the observed cold DM density and have very distinctive observational consequences. The misalignment mechanism requires a finely tuned initial condition with potential isocurvature fluctuations discoverable in the cosmic microwave background and other features~\cite{Arvanitaki:2019rax}. The kinetic misalignment with large velocity implies also the presence of small and characteristic axion miniclusters~\cite{Eroncel:2022efc,KM:2026}. Even in the most constrained $N_{\rm DW}=1$ post-inflationary scenario, meV axions contribute easily a few percent of the CDM abundance (and up to a 50\%) and also imply axion miniclusters, although their size is typically larger and somehow uncertain~\cite{Pierobon:2023ozb}. Finally, in the $N_{\rm DW}>1$ post-inflationary case, axions can naturally constitute all of the CDM, requiring new $CP$-violating physics with observable consequences and predicting primordial gravitational waves, black holes, and different (larger) axion miniclusters.

\section{meV axions as dark radiation}
\label{sec:dark_radiation}
\bigskip

Axions interact with SM particles via anomalous gauge couplings and derivative couplings to fermions, giving rise to cosmological signatures that extend beyond their role as DM. In the primordial thermal bath, which consists of a relativistic plasma of SM particles in thermal equilibrium at temperature $T$, axions can be efficiently produced through scatterings and  decays~\cite{Chang:1993gm,Masso:2002np,Graf:2010tv,Brust:2013ova,Salvio:2013iaa,DEramo:2014urw,Baumann:2016wac,Ferreira:2018vjj,DEramo:2018vss,Arias-Aragon:2020qtn,Arias-Aragon:2020shv,Ferreira:2020bpb,Green:2021hjh,DEramo:2021psx,DEramo:2021lgb,DEramo:2021usm,DEramo:2022nvb,Caloni:2022uya,Notari:2022ffe,Bianchini:2023ubu,DEramo:2023nzt,Bouzoud:2024bom,Caloni:2024olo,DEramo:2024jhn,Badziak:2024qjg,Badziak:2025mkt}. If these interactions are sufficiently frequent, inverse processes drive axions into thermal equilibrium with the bath. If the couplings are instead too weak to ensure thermalization, axions can still be generated in appreciable abundances, potentially leaving observable cosmological imprints. In both scenarios, cosmic expansion eventually leads to a phase in which axions propagate freely along geodesics of the expanding Universe, described by the Friedmann--Lemaître--Robertson--Walker (FLRW) metric. In the former case, free propagation begins after \emph{freeze-out}, while in the latter axions are produced through \emph{freeze-in}.

The existence of a primordial thermal bath is firmly established at temperatures below the MeV scale, as evidenced by the successful prediction of light element abundances through Big Bang Nucleosynthesis (BBN). Extending this well-tested picture of the early Universe to higher temperatures is reasonable, although it must eventually break down during inflationary reheating. We only require that SM particles form a thermal bath during the epochs relevant for axion production, implying a reheat temperature just above the weak scale.

Once these two conditions are satisfied, namely the presence of an axion coupled to SM fields and a primordial thermal bath with temperatures reaching the weak scale, the subsequent cosmological evolution is fully determined. The resulting axion population is therefore \emph{inevitable} and must be computed with care. In practice, the implications of this framework are explored by first calculating the axion abundance produced in the early Universe and then translating it into observable signatures.

\subsection{Thermal production of axions}
\label{sec:Thermal_production_of_axions}

Thermal axions are produced at early times when the bath temperature is much larger than their mass, so their average kinetic energy greatly exceeds their mass. They arise in the ultrarelativistic regime, and we refer to this population as \emph{hot axions}. As the Universe expands, axions follow FLRW geodesics and their momenta redshift, causing energies to decrease inversely with the scale factor. The thermal bath cools simultaneously, with its temperature decreasing according to entropy conservation. Up to order-one corrections from changes in the effective number of degrees of freedom, the axion ``temperature'' tracks that of the plasma. The quotation marks emphasize that this does not imply thermal equilibrium. Even after decoupling, or if equilibrium is never reached, the axion momentum distribution remains close to an equilibrium form, reflecting its thermal origin. {{For axions in the meV mass range, which remain relativistic at recombination, momentum and energy are effectively equivalent. Consequently, the average axion energy remains, up to order-one factors, comparable to the bath temperature, allowing it to be reliably tracked throughout cosmic evolution.}} 

As long as they remain relativistic, axions contribute to the radiation energy density of the Universe. This contribution is constrained at two key epochs: during BBN, which probes the radiation content at temperatures around $\mathrm{MeV}$, and at the time of CMB formation, when the temperature is approximately $0.3\,\mathrm{eV}$. In both cases, axions with masses in the $\mathrm{meV}$ range remain relativistic and act as an additional radiation component. This extra contribution, commonly referred to as \emph{dark radiation}, is parameterized in terms of an effective number of additional neutrino species,
\begin{equation}
\Delta N_{\rm eff} \equiv \frac{8}{7} \left(\frac{11}{4}\right)^{4/3} \frac{\rho_a}{\rho_\gamma} \,,
\end{equation}
where $\rho_a$ and $\rho_\gamma$ denote the energy densities of relativistic axions and photons, respectively. The numerical prefactor accounts for the difference in quantum statistics between fermions and bosons, as well as for the temperature mismatch between the cosmic neutrino background and the CMB.

Once the axion couplings are specified, the problem reduces to determining the abundance of axions produced and their energy spectrum. This requires choosing a computational strategy. The central ingredient is the production rate, which for a given channel quantifies the number of axions produced per unit time and volume. When needed, the rate can be expressed differentially in momentum, a refinement useful below. In this section, we outline three common methods to evaluate the thermal axion contribution to $\Delta N_{\rm eff}$. A comprehensive overview is provided in Ref.~\cite{DEramo:2023nzt}, where the methods are presented in increasing order of complexity and compared to assess their reliability. Fully general treatments are computationally demanding, so approximate methods are preferred whenever applicable.

\begin{itemize}

\item \textbf{Instantaneous Decoupling.} The simplest method to estimate the axion abundance is the \emph{instantaneous decoupling} approximation. It relies on two key assumptions: (i) axions thermalize with the plasma at early times; (ii) decoupling is sudden, occurring at a single temperature $T_D$ for all momenta. The decoupling temperature is estimated by equating the production rate $\Gamma$ to the Hubble rate $H$, $\Gamma(T_D) \simeq H(T_D)$. Consequently, the contribution to dark radiation can be expressed as $\Delta N_{\rm eff} \simeq 13.69 \, g_{\star s}(T_D)^{-4/3}$, where $g_{\star s}(T)$ denotes the effective number of degrees of freedom contributing to the entropy density of the primordial thermal bath.

\item \textbf{Tracking the number density.} The previous method is oversimplified for several reasons. It fails if axions never reach thermal equilibrium, missing freeze-in production, and the estimate $\Gamma(T_D)\simeq H(T_D)$ only captures the correct order of magnitude. Since $\Delta N_{\rm eff}$ is sensitive to $T_D$, particularly below the weak scale where the effective number of degrees of freedom changes rapidly, this approach can produce errors exceeding current experimental uncertainties. A more refined, still approximate, method follows the axion \emph{number density} $n_a$ via the Boltzmann equation
\begin{equation}
\frac{dn_a}{dt}+3Hn_a=\gamma_a\left(1-\frac{n_a}{n_a^{\rm eq}}\right)  \ ,
\end{equation}
where $\gamma_a$ is the axion production rate per unit volume and $n_a^{\rm eq}$ the equilibrium density. This equation assumes Maxwell--Boltzmann statistics for all particles and kinetic equilibrium throughout production (see, e.g., the derivation in App.~C of Ref.~\cite{DEramo:2023nzt}). While the first assumption is generally valid due to small occupation numbers in the early Universe, the second can have a significant impact. The equation is conveniently solved using the comoving density $Y_a\equiv n_a/s$, which approaches a constant $Y_a^\infty$ once production ceases, yielding $\Delta N_{\rm eff}\simeq 74.85\,(Y_a^\infty)^{4/3}$.

\item \textbf{Full phase-space analysis.} The limitation of the approach described above can be overcome by performing a full phase-space analysis. Specifically, one can follow the time evolution of the \emph{axion distribution} in momentum space, $f_a(k, t)$, by solving the integro-differential Boltzmann equation
\begin{equation}
\omega \frac{df_a(k, t)}{dt} = C_a[f_a(k, t)] \ .
\end{equation}
This approach allows one to track the evolution of each momentum mode individually. The collision operator $C_a$ encodes all relevant production processes and consistently incorporates the quantum statistics of all participating particles. The Boltzmann equation can be coupled to the evolution equation for the thermal bath energy density, thereby accounting for energy exchange between the visible sector and dark radiation. Together with the Friedmann equation, these relations form a closed system that can be solved to follow the axion evolution in momentum space without assuming instantaneous or momentum-independent decoupling~\cite{DEramo:2023nzt}. This framework naturally captures spectral distortions arising from non-instantaneous decoupling, which are missed by simpler methods. Compared to commonly used approximations, the resulting differences in the predicted $\Delta N_{\rm eff}$ can exceed the sensitivity of future CMB observations, emphasizing the need for this level of precision.

\end{itemize}

\subsection{Dark radiation as a cosmological consequence}
\label{sec:Dark_radiation_as_a_cosmological_consequence}

In the following, we discuss the predicted contribution to dark radiation from axion couplings to gluons and to SM fermions, respectively.

\begin{figure}[t]
\begin{center}
\includegraphics[width = 0.48\textwidth]{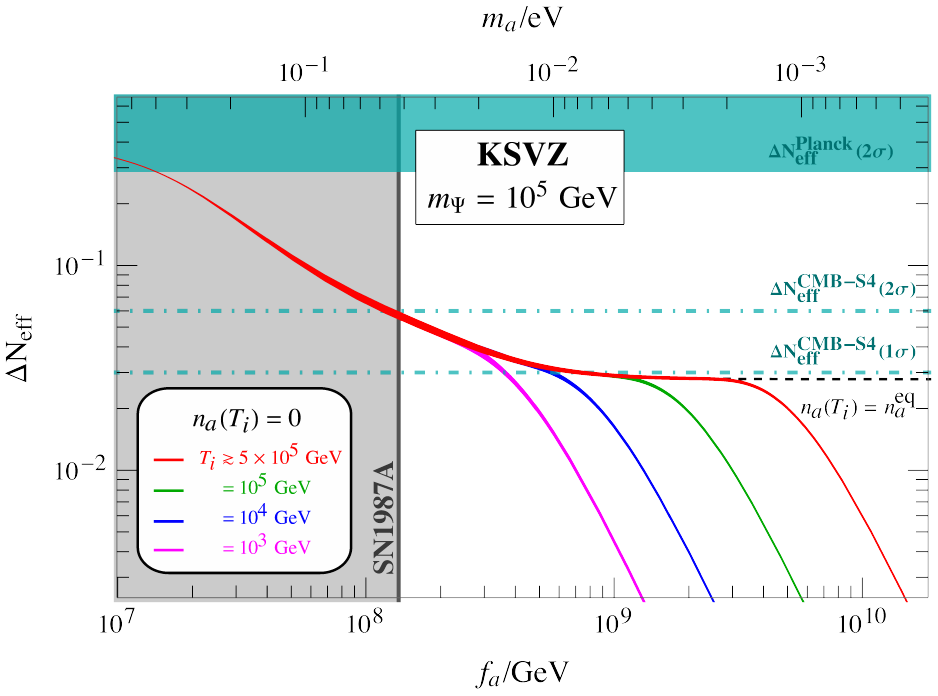} \qquad 
\includegraphics[width = 0.42\textwidth]{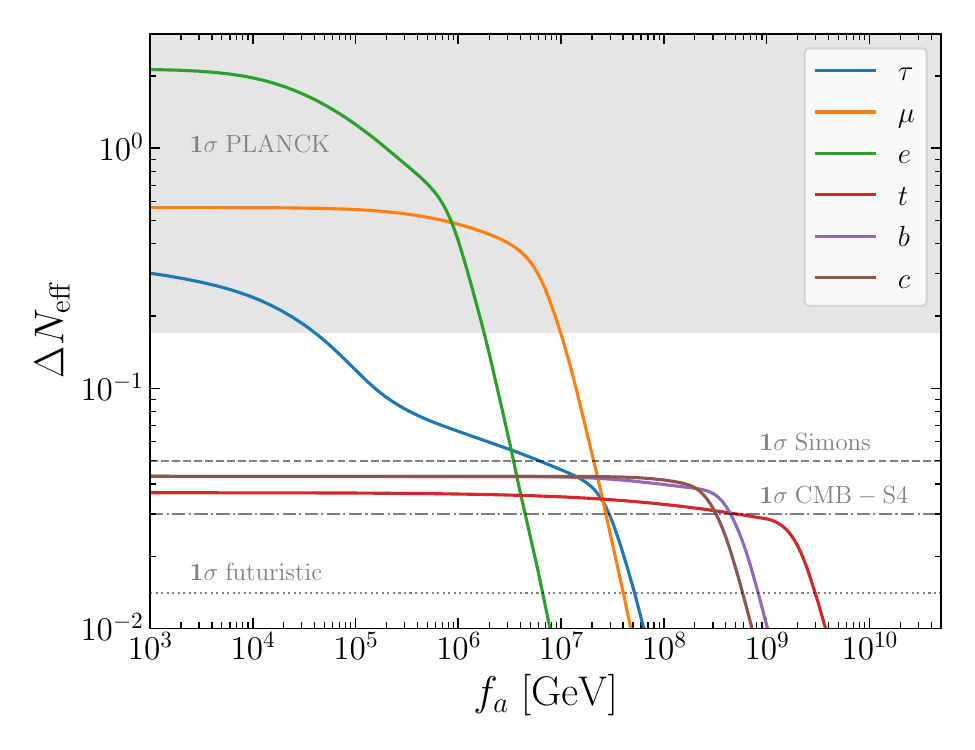}
\tiny\caption{Predicted $\Delta N_{\rm eff}$ from thermal axion production as a function of $f_a$, compared with current and future CMB sensitivities. \emph{Left.} KSVZ axion framework, with the corresponding $m_a$ shown on the upper horizontal axis~\cite{DEramo:2021lgb}. \emph{Right.} Axion couplings to SM fermions with unit Wilson coefficient, based on the full phase-space analysis of Ref.~\cite{DEramo:2024jhn}.}
\label{fig:darkradiation}
\end{center}
\end{figure}

The left panel of Fig.~\ref{fig:darkradiation} shows the predicted contribution to dark radiation from thermally produced axions, expressed as $\Delta N_{\rm eff}$, as a function of the axion decay constant $f_a$. The results are obtained within the KSVZ axion framework, where interactions with gluons dominate axion production; for this reason, the upper horizontal axis also indicates the corresponding axion mass $m_a$, which is in one-to-one correspondence with $f_a$. The axion abundance is computed by solving the Boltzmann equation for the number density, assuming a vanishing initial population. This conservative choice ensures that any axions produced during reheating or via other nonthermal mechanisms would only increase the final abundance, so the curves represent the minimal thermal contribution. Different curves correspond to different initial temperatures for the Boltzmann evolution; once this temperature exceeds the relevant fermion mass threshold, the resulting axion abundance becomes insensitive to this choice, reflecting the IR-dominated nature of production above this mass scale. The shaded region shows the current bounds set by Planck~\cite{Planck:2018vyg}, which lie entirely within the parameter space already excluded by supernova bounds. Although these bounds are subject to astrophysical uncertainties, they underscore the value of complementary cosmological probes. In this context, measurements by the Simons Observatory and future CMB surveys~\cite{SimonsObservatory:2018koc,CMB-S4:2022ght} will extend sensitivity to currently allowed regions, enabling the exploration of axion masses down to the meV scale.

The right panel of Fig.~\ref{fig:darkradiation} shows predictions for $\Delta N_{\rm eff}$ as a function of the axion decay constant $f_a$, now arising from couplings to fermions. A derivative, flavor-conserving interaction is assumed with a Wilson coefficient set to unity, though other choices can be accommodated by rescaling $f_a$. In this case, the axion mass is unrelated to $f_a$, so the results are valid as long as $m_a$ remains below the temperature at the last scattering surface, in particular down to the meV scale. The results are taken from Ref.~\cite{DEramo:2024jhn}, where only couplings to charged leptons and heavy quarks are considered. Since charged leptons carry no color, their production rates can be reliably computed across the QCD confinement scale. For heavy quarks, only the top quark is sufficiently heavier than the GeV scale for robust predictions, while the bottom quark case is borderline and the charm quark case requires caution; these subtleties and their associated uncertainties are discussed in detail in Ref.~\cite{DEramo:2024jhn}. A key novelty of this study is the inclusion of a proper phase-space treatment of axion production. As advocated in Ref.~\cite{DEramo:2023nzt}, a rigorous momentum-space framework is essential for accurate predictions of dark radiation in future data. Solving the full Boltzmann equation for the phase-space distribution goes beyond the instantaneous decoupling approximation and the assumption of kinetic equilibrium, while consistently incorporating quantum statistical effects. Non-instantaneous decoupling across different momentum modes leads to spectral distortions that affect $\Delta N_{\rm eff}$. Compared with conventional approximations, these corrections can exceed the projected sensitivity of upcoming CMB experiments, motivating this detailed analysis. Overall, the results update earlier approximate studies and quantify the ability of future CMB surveys to probe axion couplings to SM fermions; in particular, for heavier fermions, cosmological probes are expected to be more sensitive than terrestrial experiments.

\subsection{Thermal axions: what's next?}
\label{sec:Thermal_axions_whats_next}

Thermal axion production is an inevitable consequence of the early Universe once couplings to SM fields exist and the primordial plasma reaches at least electroweak-scale temperatures. Fig.~\ref{fig:darkradiation} quantifies this by showing the predicted $\Delta N_{\rm eff}$ as a function of the axion decay constant $f_a$. Further theoretical progress is desirable to sharpen these predictions and extend their validity. The results in the left panel of Fig.~\ref{fig:darkradiation} are obtained within a framework combining different theoretical treatments of the axion production rate via the coupling to gluons: leading order in perturbative QCD above the confinement scale, leading order in chiral perturbation theory in the confined phase, and a smooth interpolation in the intermediate regime. The production rate itself is not directly observable and enters only as an input to the Boltzmann equation computing $\Delta N_{\rm eff}$; accordingly, sizeable modifications of the rate do not necessarily translate into equally large changes in the final observable (see App.~D of Ref.~\cite{DEramo:2021lgb}). Nevertheless, overcoming the current limitations in both QCD phases and treating the intermediate regime more quantitatively remain important goals. In particular, the convergence of the perturbative expansion at finite temperature may be less efficient than at zero temperature, both above~\cite{Bouzoud:2024bom,Notari:2022ffe} and below~\cite{Notari:2022ffe,DiLuzio:2021vjd,DiLuzio:2022gsc} confinement. Fully momentum-resolved predictions currently exist for axion couplings to SM fermions~\cite{DEramo:2024jhn,Badziak:2024qjg,Badziak:2025mkt} or pions~\cite{Notari:2022ffe,Bianchini:2023ubu}, but extending such analyses to gauge bosons, particularly gluons, remains a crucial challenge. The right panel of Fig.~\ref{fig:darkradiation}, obtained from a momentum-space treatment, highlights additional limitations: production via charm quarks occurs too close to the confinement scale to be fully reliable, while for light quarks a careful matching to a hadronic description is required in a delicate regime. While these challenges await resolution to achieve fully robust and precise predictions, axion dark radiation remains a key observable for testing early-Universe physics and constraining fundamental particle interactions in the era of precision cosmology.

\section{meV axions astrophysics}
\label{sec:meV_axions_astrophysics}

\bigskip

Light axions with feeble couplings can be plentifully produced in stellar interiors. Various direct and indirect searches have been devised to test this emission. As no evidence has emerged so far, astrophysical data result in some of the most stringent existing bounds on axions~\cite{Caputo:2024oqc,Carenza:2024ehj}. Axions with masses $10^{-3}-10^{-2}\,\rm eV$ are predicted in top-down model building approaches~(Sec.~\ref{sec:meV_axions_model_building}), can constitute dark matter in post-inflationary scenarios~(Sec.~\ref{sec:meV_axions_as_cold_dark_matter}), and could also show up as thermal radiation at future CMB surveys~(Sec.~\ref{sec:dark_radiation}). Coincidentally, such axions lie just at the edge of current stellar sensitivity.

The most suitable astrophysical probes depend on the axion coupling structure. In the KSVZ model---and in the DFSZ-I model for $\sin^2\beta\lesssim0.2$---the dominant constraints come from axion interactions with nucleons. Therefore, the most powerful ``axion factories'' are dense nuclear environments such as cooling neutron stars (NSs)~\cite{Iwamoto:1984ir} and the protoneutron stars (PNSs) formed in core-collapse supernovae (SNe) (for early work see e.g.~\cite{Raffelt:1987yt, Ellis:1987pk, Turner:1987by, Mayle:1987as, Mayle:1989yx, Brinkmann:1988vi, Burrows:1988ah, Burrows:1990pk, Janka:1995ir, Keil:1996ju, Hanhart:2000ae}). These two systems probe similar physics, as the ratio of axion to neutrino luminosities is nearly temperature-independent, implying comparable relative effects of exotic cooling~\cite{Fiorillo:2025zzx}. Observations of isolated NSs require that axion emission do not excessively accelerate cooling, which yields a bound $m_a\lesssim16\,\mathrm{meV}$ for the KSVZ axion~\cite{Buschmann:2021juv}.

The classic SN constraints are also based on a cooling argument~\cite{Raffelt:1987yt, Ellis:1987pk, Turner:1987by, Mayle:1987as, Mayle:1989yx, Brinkmann:1988vi, Burrows:1988ah, Burrows:1990pk, Janka:1995ir, Keil:1996ju, Hanhart:2000ae}. The PNS cools over $\sim5$-$10$ s mainly by neutrino emission, as confirmed by the SN~1987A neutrino burst. Excessive axion losses would shorten the burst by increasing the cooling rate, implying the condition $L_a\lesssim L_\nu$ at around 1~second post-bounce, when axion emission is peaking~\cite{Raffelt:2006cw}. Numerical simulations including axion cooling reproduce this qualitative effect~\cite{Burrows:1988ah, Burrows:1990pk}, although they did not include PNS convection, which significantly impacts the neutrino burst~\cite{Fiorillo:2023frv}. Nevertheless, it is still a commonly adopted order-of-magnitude requirement that $L_a\lesssim L_\nu$ to avoid impacting the SN~1987A observations.

For large couplings, axions are not emitted volumetrically from the entire PNS, but rather from a decoupling surface inside of which they are in thermal equilibrium~\cite{Burrows:1990pk, Caputo:2022rca}. The axion emission in this regime is diffusive, acting like an effective non-standard channel of thermal conduction~\cite{Fiorillo:2025yzf}; as usual, the thermal conductivity is proportional to the axion mean free path, so that at large couplings the effiency of non-standard PNS cooling becomes negligible.

\subsection{Axion production in dense nuclear media}
\label{sec:Axion_production_in_dense_nuclear_media}

Estimating the axion luminosity $L_a$ remains challenging because of the poorly known properties of dense nuclear matter in SN cores; an incomplete list of works addressing this challenge includes Refs.~\cite{Raffelt:1987yt, Ellis:1987pk, Turner:1987by, Mayle:1987as, Brinkmann:1988vi,Mayle:1989yx,Raffelt:1991pw,Turner:1991ax, Raffelt:1993ix, Janka:1995ir,Keil:1996ju, Raffelt:1998pa, Sedrakian:2000kc, vanDalen:2003zw, Lykasov:2008yz,Chang:2018rso,Carenza:2019pxu,Carenza:2020cis, Choi:2021ign,Ho:2022oaw, Lella:2022uwi, Lella:2023bfb, Fiorillo:2025gnd}. Even the concept of well-defined nucleon quasiparticles becomes questionable: nuclear matter at these densities is only a marginal Fermi liquid. The typical nucleon–nucleon interaction rate~\cite{Fiorillo:2025gnd} is $\Gamma\sim n_n v_{\rm rel}\sigma_{nn} F_{\rm deg}$,
with $n_n\simeq10^{38}\,\mathrm{cm^{-3}}$ the nuclear density, $v_{\rm rel}\sim0.3c$ the relative velocity, $\sigma_{nn}\sim40\,\mathrm{mb}$ the cross section at the relevant energies, and $F_{\rm deg}\sim0.6$ a mild degeneracy factor. This gives $\Gamma\sim30$–$40\,\mathrm{MeV}$, comparable to the nucleon kinetic energy $K\sim50$–$100\,\mathrm{MeV}$. Consequently, individual nucleon excitations are short-lived, and treatments of axion production based on single-nucleon processes carry intrinsic uncertainties of order unity. While such approaches can yield reliable order-of-magnitude estimates of $L_a$, achieving precision better than a factor of $\sim2$ is likely impossible within this framework.

Even assuming a single-particle picture, a detailed treatment of axion emission is far from trivial. The dominant emission process is nucleon-nucleon bremsstrahlung $NN\to NNa$, but a first-principle treatment requires a modeling of the nucleon-nucleon potential, and furthermore the potential itself is not guaranteed to be perturbative at the relevant center-of-mass energies.
Light axions are sourced by the spin-isospin fluctuations of the nuclear medium, so the information on their emission is primarily encoded in the rate of spin fluctuations of a nucleon $\Gamma_\sigma$~\cite{Raffelt:2006cw,Fiorillo:2025gnd}. If the coupling is predominantly to neutrons, then axion production can be triggered only by a fluctuation in the total spin of the pair of colliding neutrons $nn\to nna$; in turn, total spin fluctuations are predominantly caused by the tensor interaction dominated by one-pion-exchange~\cite{Raffelt:1987yt, Ellis:1987pk, Turner:1987by, Mayle:1987as, Mayle:1989yx, Brinkmann:1988vi}, although at such large center-of-mass energies the amplitude for this exchange is substantially reduced, as phenomenologically modeled with heavier meson exchanges~\cite{ericson1989axion,Carenza:2019pxu}. However, for $np\to npa$ scattering, if the axion couples with different strengths to the nucleons (e.g. the KSVZ axion couples predominantly to protons), then it suffices that only the proton spin fluctuates. In this case, non-tensor interaction channels can contribute~\cite{Fiorillo:2025gnd}, which are dominated by contact interactions and therefore not well represented by one-pion-exchange. It goes without saying that the interaction amplitude for all these channels might itself suffer strong renormalization in the dense nuclear medium that are not fully controlled (e.g. Refs.~\cite{Schwenk:2003pj,Schwenk:2003bc,Shternin:2018dcn}). 

In this uncertain setting, a very flexible way forward is to consider an energy-independent $\Gamma_\sigma$ as a parametric measure of uncertainty~\cite{Raffelt:2006cw}; in this approximation, the axion emission spectrum per unit stellar mass $M$ and axion energy $E_a$ is simply
\begin{equation}\label{eq:emission_rate_bremsstrahlung}
    \frac{d\dot{N}_a}{dE_a dM}=\frac{g_{ap}^2 Y_p}{8\pi^2m_p^3}\,
    \frac{E_a}{e^{E_a/T}+1}\,\frac{\Gamma_\sigma}{1+(\Gamma_\sigma/2E_a)^2},
\end{equation}
where $T$ is the medium temperature, $Y_p$ is the proton mass fraction, and $m_p$ is the proton mass. The properties of the internal core can vary within quite a wide range among different SNe~\cite{Fiorillo:2025gnd}, depending on the mass of the central PNS. Overall, a good measure of the range that different SNe can attain is in the total number of axions emitted; this can vary in the interval $N_a\sim 10^{56-57}\, g_9^2$, with $g_9=g_{ap}/10^{-9}$, with heavier PNSs tending to the upper edge of this interval. The average energy of these axions is surprisingly stable $E_{\rm av}\sim 60-80\,\mathrm{MeV}$, much larger than the neutrino energy since these axions are emitted volumetrically over the entire core.

Axions may also be produced by pionic processes $\pi^- p\to n a$~\cite{Turner:1991ax,Raffelt:1993ix,Keil:1996ju}, a possibility recently revived in Ref.~\cite{Fore:2019wib,Carenza:2020cis,Fischer:2021jfm}. A simple measure of how much this channel could contribute is the ratio between the number of axions emitted from pionic processes and from bremsstrahlung, which for a light PNS is roughly $\mathcal{R}\sim 33 Y_\pi$, where $Y_\pi$ is the pion fraction in the core. According to the estimates in Ref.~\cite{Fiorillo:2025gnd}, $Y_\pi\sim 0.03$ would lead to a correction $\mathcal{R}\sim 1$; the effects on the cooling rate of the PNS would be even larger, since pion-induced axion emission peaks above the pion mass, larger by a factor $2-3$ than the average energy of axions from bremsstrahlung. However, the authors of Ref.~\cite{Fore:2019wib} underestimated the upward energy shift of pions caused by their refraction in the dense nuclear medium; their later work~\cite{Fore:2023gwv} leads to a range $\Delta E_\pi\sim 100-150\,\mathrm{MeV}$. For comparison, parametric textbook estimates would lead to $\Delta E_\pi\sim 40-50\,\mathrm{MeV}$~\cite{ericson1988pions}. Hence, the fraction $Y_\pi$ would be suppressed by an additional factor $e^{-\Delta E_\pi/T}\sim 10^{-3}-10^{-1}$, depending on how large $\Delta E_\pi$ ultimately is, and assuming a typical temperature $T\sim 30\,\mathrm{MeV}$. Hence, thermal pions remain an intriguing possibility for axion production, but uncertain by orders of magnitude.

Overall, excluding the pions, the cooling constraints from SN~1987A based  on emission of axions with only  bremsstrahlung shows a surprising stability among the many works in the literature~\cite{Raffelt:2006cw,Chang:2018rso,Carenza:2019pxu,Carenza:2020cis,Lella:2023bfb,Fiorillo:2025gnd}, with the most recent estimate setting it at ${m_a\lesssim 12\,\mathrm{meV}}$~\cite{Fiorillo:2025gnd}; a reasonable uncertainty on this constraint may span perhaps a factor 3 in mass, corresponding to roughly an order-of-magnitude uncertainty in the axion emission.

\subsection{Axion conversion in astrophysical magnetic fields}
\label{sec:Axion_conversion_in_astrophysical_magnetic_fields}

An intriguing perspective to probe axions emitted from SNe is via their subsequent conversions to high-energy gamma-rays in astrophysical magnetic fields. This has long been considered in the literature in the context of conversions in the Galactic magnetic field~\cite{Brockway:1996yr, Grifols:1996id, Payez:2014xsa,Calore:2020tjw,Calore:2021hhn, Hoof:2022xbe}, which could produce a 100-MeV-gamma-ray burst in correspondence of a SN. For SN~1987A, this would have been visible by the Solar Maximum Mission (SMM)~\cite{1980SoPh...65...15F} (see also Refs.~\cite{Chupp:1989kx,Oberauer:1993yr}). However, SMM did not observe any such burst, leading to stringent bounds on axion-like particles coupling to photons or to photons and nucleons simultaneously. However, conversions in the Galactic magnetic field become inefficient for relatively light axions $m_a\gtrsim 10^{-6}\,\mathrm{eV}$. At such masses, the QCD axion has very feeble couplings, which cannot be reached by this method.

Recently, Ref.~\cite{Manzari:2024jns} has proposed the intriguing idea that the magnetic field of the progenitor star of the SN might be a much more efficient axion-photon converter. The distances involved in the conversion are in this case much shorter, of the order of tens to hundreds of solar radii, compared with the kiloparsec scale of the Galactic magnetic field. This means that axions have much less space to convert into photons, so their overall conversion probability is smaller. However, it also implies that axions with much larger masses can convert efficiently. 

For SN~1987A, the relatively weak sensitivity to the 100-MeV signal does not allow one to obtain any novel constraint, especially because of the largely unknown situation with regards to the surface magnetic field of its progenitor, Sanduleak $-69\,202$. The latter was a blue supergiant (BSG), a class of supergiants whose measured magnetic fields is in the ballpark of $10-100\,\mathrm{G}$; Ref.~\cite{Manzari:2024jns} obtained, with an optimistic value of $100\,\mathrm{G}$, constraints on the ALP-photon coupling stronger than the level of the CAST exclusion. However, after removing the contribution of pionic emission as discussed above, as well as correcting for the adopted data sample from the SMM (since an artificially rescaled version from Ref.~\cite{Hoof:2022xbe} had been used) and accounting for the uncertainty in the progenitor radius, the overall constraints~\cite{Fiorillo:2025gnd} turn out to be comparable with the CAST exclusion. One should stress, though, that without a direct observation of the magnetic field of Sanduleak $-69\, 202$, one cannot truly rely on these regions as excluded.

The next galactic SN might be observable with telescopes much more sensitive than the SMM satellite in the 100-MeV energy range. Our instrument of choice at present is the Fermi Large Area Telescope (LAT). With the typical effective area of Fermi-LAT, Ref.~\cite{Manzari:2024jns} shows that a galactic SN at 10~kpc from Earth might lead to a detection of a gamma-ray burst from the QCD axion. On the other hand, this detection prospect includes a surface magnetic field of 1~kG, as well as the pionic contribution to axion production. As argued in Ref.~\cite{Fiorillo:2025gnd}, such fields are actually not observed in BSGs, which seem to have fields at most as large as $\sim 100\,\mathrm{G}$. When accounting for typically lower magnetic fields, Ref.~\cite{Fiorillo:2025gnd} finds that a more realistic horizon of detection from a BSG progenitor is around $1\,\mathrm{kpc}$. Moreover, BSG progenitors are quite rare, about $1\%$ of core-collapse SNe~\cite{Graur+2017,Ma+2025}; for the much more common RSG progenitors, the typical coherent fields are significantly lower, around $1-10\,\mathrm{G}$, although it is speculated that they may host turbulent small-scale fields reaching up to $100\,\mathrm{G}$. In this case, the prospects for detection of RSGs might be quite close to those of BSGs at large axion masses~\cite{Manzari:2024jns}. 

On the other hand, as recently advocated in Ref.~\cite{Candon:2025sdm}, the best candidate for axion-photon conversion in a supernova progenitor would in fact be a Type Ibc SN, associated with more compact progenitors ($R\sim 0.1-1\,R_\odot$)~\cite{Tauris:2015xra} and correspondingly larger observed magnetic fields, ranging from hundreds of G~\cite{2014ApJ...781...73D,2016MNRAS.458.3381H,2020MNRAS.499L.116H,2023MNRAS.524L..21J} up to even $43\,\mathrm{kG}$ in an extreme case~\cite{Shenar:2023zoa}. Their generally larger magnetic field and incidence rate (about $30\%$ of all core-collapse SNe~\cite{Kleiser:2011ms,Pessi:2025wht,Graur+2017,Ma+2025}) make them an ideal target for gamma-ray burst detection from QCD axion conversion, reaching up even to SNe at a distance of $10\,\mathrm{kpc}$, exploding in the Galactic Center~\cite{Candon:2025sdm}.

Regardless of the nature of the next galactic SN, the 100-MeV-gamma-ray burst induced by axion conversion will only last a few seconds, the typical timescale over which the SN core cools down. Hence, Fermi-LAT, which at any moment only sees about a fifth of the sky, might easily miss it altogether. For this reason, the Berkeley group~\cite{Manzari:2024jns} has advocated for a constellation of satellites that might achieve Fermi-LAT sensitivity with a $4\pi$ field of view, ensuring coverage of the sky during the next galactic SN.

A final perspective for the future, mostly motivated by the recent case of GW170817~\cite{LIGOScientific:2017vwq,LIGOScientific:2017ync,LIGOScientific:2017zic,LIGOScientific:2018hze}, is the production of axions in the merger of two neutron stars (NSMs), which produces a hot remnant very similar to the PNS formed within a core-collapse SN. Depending on the masses of the colliding neutron stars, this remnant is often expected to be unstable and collapse into a black hole over shorter timescales than in a SN, ranging from tens of milliseconds to a few seconds. More importantly, these events are much rarer than SNe, and therefore expected at extragalactic distances of tens of Mpc. In principle, they can be nearly as efficient as SNe in emitting axions, limited mainly by the shorter lifetime of the remnant. However, due to the larger distances, they are generally much less competitive when it comes to the observation of the gamma-ray burst from axion-photon conversion in the Galactic magnetic field~\cite{Fiorillo:2022piv}. Conversion within the magnetic field of the remnant, which can reach up to even $10^{14}\,\mathrm{G}$, might be significantly more efficient at large masses~\cite{Manzari:2024jns}. Despite such large magnetic fields, when realistic conditions for the merger remnant are considered, in particular by virtue of its shorter lifetime than SNe, its time-integrated axion emission and the larger expected distances conspire to make NSMs a much less sensitive probe of axion-photon conversion~\cite{Lecce:2025dbz}. In fact, the situation is even worse, since the merger produces a sizable amount of dynamical ejecta which impede axion-photon conversion for a large fraction of the emitted axions~\cite{Fiorillo:2025gnd}, so that even for a future NSM one does not expect to probe any novel parameter space by this strategy.

\begin{figure*}
    \includegraphics[width=\textwidth]{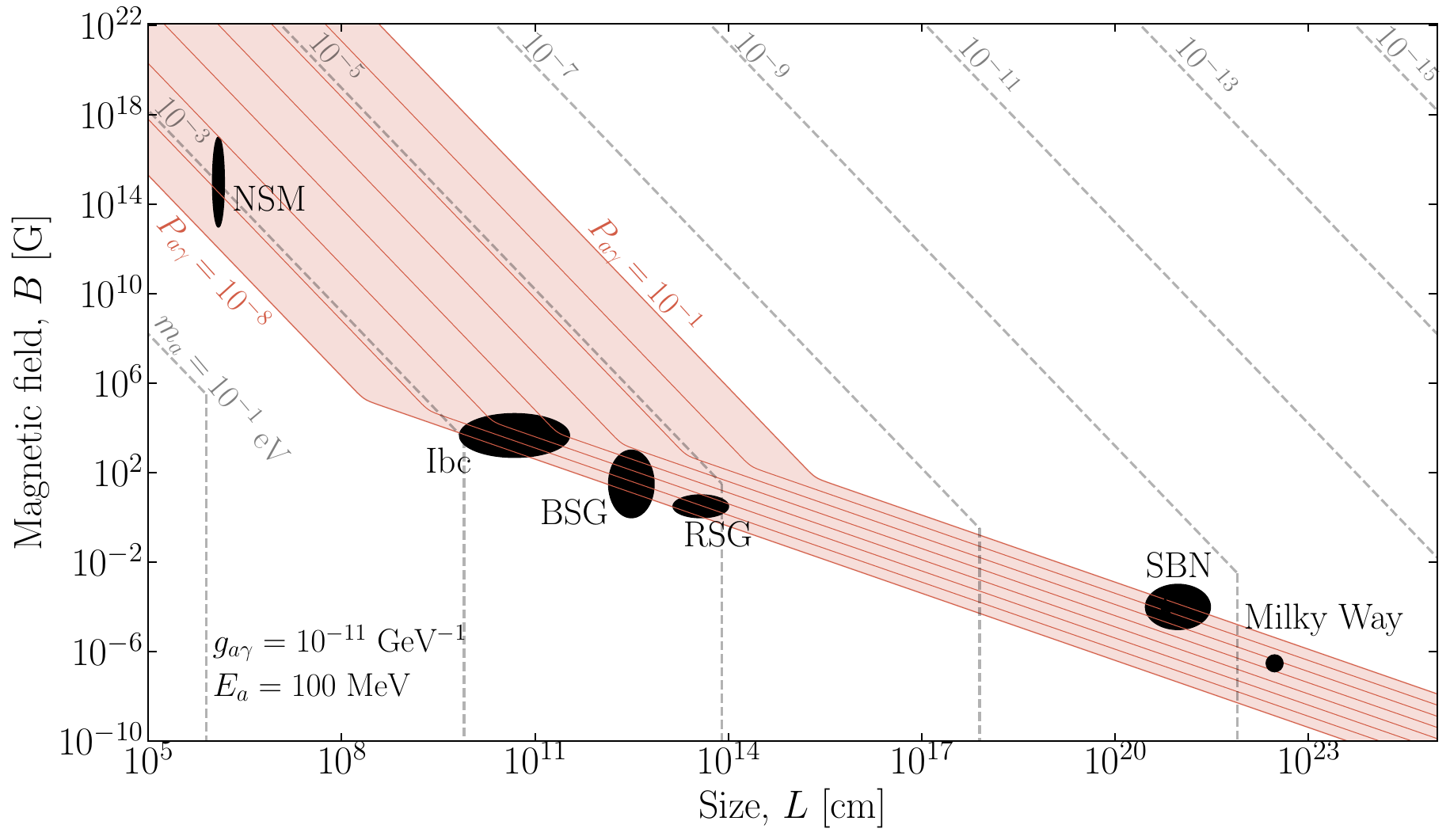}
    \caption{Hillas plot of the astrophysical sources as axion-photon converters. The colored lines identify the probability of axion-photon conversion for massless axions, while the gray lines identify the axion masses at which coherence in the axion-photon conversion is lost. Adapted from Ref.~\cite{Fiorillo:2025gnd}.}\label{fig:hillas}
\end{figure*}

Figure~\ref{fig:hillas} summarizes succinctly the panorama of axion-photon conversion in astrophysical sources, using the format of the Hillas plot, widely known in cosmic-ray physics. The size and strength of the magnetic field is what determines both the maximum energy to which charged particles can be accelerated, and the axion-photon conversion probability. We see here that large sources with feeble magnetic fields, such as starburst nuclei~\cite{Lecce:2025vjc} and the Milky Way, can efficiently interconvert axions and photons, due to their large coherence length. However, for the same reason, it takes a very small axion mass to disrupt the coherence and dramatically suppress the conversion efficiency. Compact sources, like the supernovae progenitors, have significantly lower conversion probabilities, but can maintain coherence up to significantly larger masses. Finally, neutron star mergers, with their very intense fields, are located in a region where the conversion probability is suppressed---visible from the change in slope of the red lines in Fig.~\ref{fig:hillas}---so they are overall \textit{less} efficient than stellar environments at converting axions to photons, even though their compact radii allows for an efficient conversion of slightly heavier axions.


\subsection{Supernova axions in large-volume Cherenkov detectors}
\label{sec:Supernova_axions_in_large-volume_Cherenkov_detectors}

\bigskip


Large water Cherenkov detectors, such as Hyper-Kamiokande (HK)~\cite{Abe:2011ts, Hyper-Kamiokande:2018ofw}, are primarily designed to observe neutrino fluxes from astrophysical sources, most notably core-collapse supernovae (SNe). In these environments, neutrinos dominate the cooling of the newly formed proto--neutron star. However, the extreme densities and temperatures reached during core collapse also make SNe efficient laboratories for the production of other light particles with feeble interactions with matter, and thus able to contribute to the cooling process.

Among such hypothetical particles, the QCD axion focus of this review is particularly well motivated. If axions couple to nucleons, they can be copiously produced in SN cores and escape the dense environment, carrying away energy. In the case of a sufficiently nearby SN, the resulting axion flux may be detectable on Earth through interactions in large-volume detectors.

Water Cherenkov experiments offer a natural detection strategy for such axion fluxes. In addition to neutrino-induced events, axions can be absorbed in water via inelastic scattering on nucleons, producing energetic secondary particles that initiate Cherenkov cascades. The goal of this section is to review the theoretical framework and detection prospects for axions in water Cherenkov detectors, with particular emphasis on axion--nucleon interactions and on spectral features that allow axion-induced signals to be distinguished from the neutrino background. The discussion closely follows the effective-field-theory treatment developed in Ref.~\cite{Cavan-Piton:2025nsj}.

\vspace{0.2cm}

Axion detection in water Cherenkov experiments has also been discussed in the context of axion couplings that arise independently of the derivative axion--nucleon interaction. In particular, the axion--gluon coupling generically induces an effective interaction between the axion field and the nucleon electric dipole moment (EDM). This so-called nucleon EDM portal represents a model-independent feature of the QCD axion and leads to axion--photon--nucleon interactions even in scenarios where the standard axion couplings to photons and fermions are suppressed.

The phenomenology of SN axions produced through the nucleon EDM portal, as well as their detection via axion-induced photon signals in large water Cherenkov detectors, has been investigated in detail in Ref.~\cite{Lucente:2022vuo}, which provides a benchmark study of this detection strategy. In that framework, axion emission is dominated by Compton-like processes involving photons in the SN medium, while detection proceeds primarily through the inverse process $a+p\to p+\gamma$. The resulting Cherenkov signatures were shown to exhibit distinctive spectral features at energies well above the bulk of the neutrino signal.

The present discussion adopts a complementary perspective, following the approach developed in Ref.~\cite{Cavan-Piton:2025nsj}. Rather than focusing on the EDM-induced axion--photon--nucleon interaction, attention is restricted to the most general derivative axion--nucleon couplings encoded in the low-energy chiral effective theory. This framework enables a unified treatment of axion emission and absorption driven by axion--nucleon interactions, and highlights the role of hadronic final states---notably neutral pions---as efficient initiators of Cherenkov cascades in water.

\vspace{0.2cm}

From the perspective of the broader experimental landscape reviewed in this work, searches for SN axions in large-volume Cherenkov detectors occupy a distinct and complementary role. Helioscopes target axions thermally produced in the solar interior, haloscopes search for axions comprising the local dark matter halo, and light-shining-through-walls experiments rely on controlled laboratory sources---all three primarily probing axion--photon interactions. By contrast, Cherenkov detectors probe axions produced in the extreme environments of core-collapse supernovae and are directly sensitive to axion--nucleon couplings through absorption processes. As such, they provide access to regions of parameter space that are largely orthogonal to those explored by photon-based experiments.

At the same time, the relevant axion mass range is set by the typical energies of axions thermally produced in SN cores, which naturally extends into the meV regime~\cite{Raffelt:1990yz,Caputo:2024oqc}. In this sense, SN axion searches at Cherenkov detectors complement ongoing efforts to extend experimental sensitivity to meV-scale axions using helioscopes, haloscopes, and condensed-matter systems. Although the reliance on rare astrophysical transients limits the achievable event rate, the use of existing large-scale detectors and the direct sensitivity to hadronic couplings make Cherenkov-based searches a valuable component of the experimental program for meV axions.

\subsubsection{Detection Framework and Theoretical Setup}

The detection framework considered here consists of three main components: axion production in a core-collapse SN, propagation of the axion flux to Earth, and axion absorption in a water Cherenkov detector.

Axions are assumed to be produced in the SN core through their couplings to nucleons. The resulting flux depends on the axion--nucleon couplings, the core temperature, and the nuclear equation of state. Once produced, axions stream freely out of the SN and propagate essentially without attenuation to Earth.

At the detector, axions can be absorbed through processes of the form
\begin{equation}
  a + N \to N + \gamma, \qquad
  a + N \to N + \pi^0,
\end{equation}
where $N$ denotes a nucleon. The photon or neutral pion in the final state initiates an electromagnetic cascade in water, producing Cherenkov light. These processes are analogous, at the level of detector response, to neutrino-induced interactions, but lead to distinct spectral signatures.

Both axion emission and absorption are computed within a unified effective field theory framework, namely chiral perturbation theory (ChPT) extended to include the axion field~\cite{Georgi:1986df}. This ensures a consistent treatment of hadronic physics at the source and in the detector.

At energies well below the QCD confinement scale $\Lambda_{\rm QCD} \approx 1~\mathrm{GeV}$, the relevant degrees of freedom are hadrons rather than quarks and gluons. The interactions of the light pseudoscalar meson octet are governed by the approximate chiral symmetries of QCD, which form the basis of ChPT.

The QCD axion carries the same quantum numbers as neutral pseudoscalar mesons such as the $\pi^0$ and the $\eta$. As a result, it can be incorporated naturally into ChPT as an additional pseudoscalar field~\cite{Georgi:1986df}.
In this framework, the interactions of the axion with mesons and baryons are fixed by chiral symmetry and by the underlying axion--quark couplings. At the quark level, the axion interaction takes the schematic form
\begin{equation}
  \mathcal{L}_{aqq}
  = \frac{\partial_\mu a}{f_a}
  \left(
    \bar q \, \gamma^\mu_L k_L \, q_L
    + \bar q \, \gamma^\mu_R k_R \, q_R
  \right),
\end{equation}
where $f_a$ is the axion decay constant and $k_{L,R}$ are coupling matrices acting on the light-quark triplet $q=(u,d,s)^T$.

Matching onto the hadronic theory yields, among other terms, derivative axion--nucleon interactions of the form (e.g. \cite{Chang:1993gm})
\begin{equation}
  \mathcal{L}_{aNN}
  =
  \frac{\partial_\mu a}{2 f_a}
  \sum_{N=p,n}
  C_{aNN}\,
  \bar N \gamma^\mu \gamma_5 N~, \nonumber
\end{equation}
where the coefficients $C_{aNN}$ are calculable functions of the underlying axion--quark couplings.

The effective axion--nucleon couplings $C_{aNN}$ can be specified in several ways. 
Representative scenarios include:

\begin{enumerate}
\item \textbf{Benchmark UV-complete models.}  
In the KSVZ model~\cite{Kim:1979if,Shifman:1979if}, the axion--nucleon   couplings are determined by running effects alone, giving
\begin{equation}
  C_{ann}|_{\rm KSVZ} = 0.012~, \qquad C_{app}|_{\rm KSVZ} = -0.452~.
\end{equation}
In the DFSZ model~\cite{Dine:1981rt,Zhitnitsky:1980tq}, the couplings 
depend on the ratio $\tan\beta$ of the two Higgs vacuum expectation values,
\begin{align}
  C_{ann}|_{\rm DFSZ} &= -0.123 + 0.406\,\sin^2\beta~, \\
  C_{app}|_{\rm DFSZ} &= -0.169 - 0.430\,\sin^2\beta~,
\end{align}
with $\tan\beta \in [0.25, 170]$. The couplings' maximum magnitudes are $|C_{app}|\simeq 0.60$ and $|C_{ann}| \simeq 0.28$.

\item \textbf{Agnostic effective scenario.}  
In this approach, $C_{app}$ and $C_{ann}$ are treated as independent parameters subject only to observational constraints. The most relevant constraints arise from:
\begin{itemize}
  \item neutron-star cooling~\cite{Buschmann:2021juv}, which requires \\
  $C_{ann}\,m_N/f_a \lesssim 1.15 \times 10^{-9}$ and 
  $C_{app}\,m_N/f_a$ $\lesssim$ $1.09 \times 10^{-9}$;
  \item supernova cooling, implemented through the Raffelt bound~\cite{Raffelt:1990yz,Burrows:1988ah}, requiring the axion luminosity to satisfy $L_a \lesssim L_\nu$, as inferred from SN1987A~\cite{Bionta:1987qt,IMB:1988suc,Kamiokande-II:1987idp,Hirata:1988ad,Alekseev:1987ej,Alekseev:1988gp}.
\end{itemize}
In each case, the stronger of the two bounds is applied. For the benchmark SN parameters used in Fig.~\ref{fig:summary} (see below), the SN bound is more constraining and yields $|C_{app}| < 2.1$ and $|C_{ann}| < 2.0$~\cite{Cavan-Piton:2025nsj,3121183}.
\end{enumerate}

\subsubsection{Axion Detection in Water Cherenkov Detectors}

In a core-collapse SN, the emission of axions may accompany the neutrino burst. Both species can be detected on Earth through interactions with nucleons in a water Cherenkov detector, leading to processes of the general form
\begin{equation}
  \alpha + N \to N' + X~,
\end{equation}
where $\alpha=\nu,a$ and $X$ is a ``Cherenkov parton'', a particle, such as $e^\pm$, $\gamma$, or $\pi^0$, capable of initiating a cascade leading to Cherenkov light. For axions, the dominant Cherenkov-producing channels are
\begin{equation}
  a + p \to p + \gamma,
  \qquad
  a + N \to N + \pi^0 .
\end{equation}
The associated differential cross sections $\sigma_{\alpha X}$ are computed within the same chiral EFT framework used for axion production.
From them, the differential number spectrum of a Cherenkov-emitting particle $X$ induced by a flux of coolers $\alpha$ is given by
\begin{equation}
  \frac{dN_X^{(\alpha)}}{dE_X}
  =
  \frac{N_t}{4\pi d^2}
  \int dE_\alpha\,
  \frac{d\sigma_{\alpha X}(E_\alpha,E_X)}{dE_X}\,
  \frac{dN_\alpha}{dE_\alpha}~,
\end{equation}
where $N_t$ is the number of target nucleons in the detector and $d$ is the SN distance. The expected event yield scales linearly with detector size and as $1/d^2$ with distance.

\subsubsection{Spectral Features and Detectability}

A defining advantage of large-volume water Cherenkov detectors in the context of SN axions lies in the markedly different spectral imprints left by axion-induced and neutrino-induced interactions. These differences arise both from the underlying particle physics governing axion absorption on nucleons and from the kinematics of the resulting final states that initiate Cherenkov cascades in water.

At a qualitative level, axion absorption through inelastic nucleon scattering,
\begin{equation}
a + N \to N + X \, , \nonumber
\end{equation}
with $X=\gamma,\pi^0$, produces secondary particles with characteristic energies of order $\mathcal{O}(100)\,\mathrm{MeV}$ or higher. By contrast, the dominant neutrino signal in water Cherenkov detectors is generated by inverse beta decay, $\bar{\nu} p \to n e^+$, whose positron spectrum peaks at significantly lower energies, typically $\mathcal{O}(10)\,\mathrm{MeV}$. This kinematic separation is a robust consequence of the much larger axion energies expected from thermal emission in the SN core, combined with the non-negligible hadronic mass scales entering the axion–nucleon absorption processes.

\medskip

\begin{figure}[t]
  \centering
  \includegraphics[width=0.49\columnwidth]{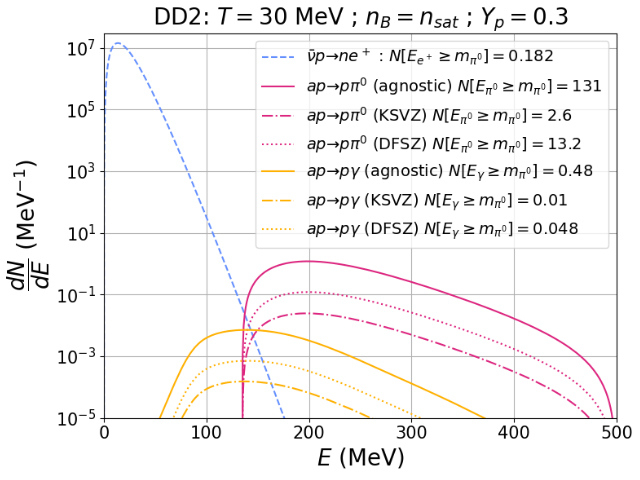}\hfill
  \includegraphics[width=0.49\columnwidth]{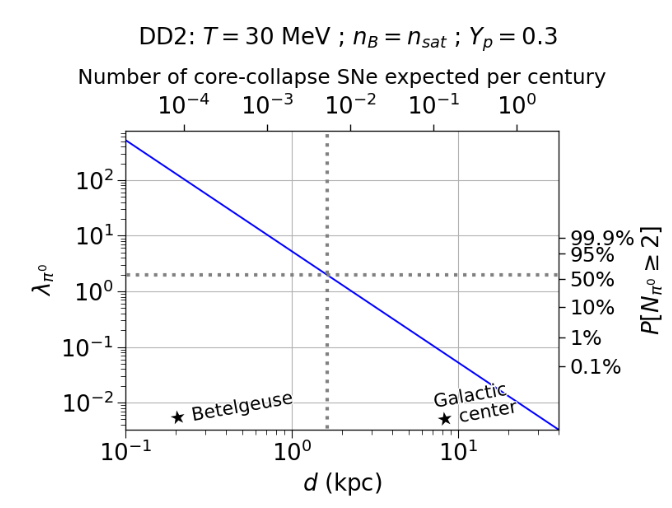}
  \captionsetup{width=\linewidth} 
  \caption{
  \emph{Left.} Comparison between neutrino-induced and axion-induced Cherenkov spectra for a representative core-collapse SN, illustrating the separation in characteristic energies. Different line styles correspond to the KSVZ, DFSZ, and agnostic coupling scenarios (see text). The axion signal is dominated by the $\pi^0$ production channel and peaks well above the neutrino background.
  \emph{Right.} Expected number of axion-induced $\pi^0$ events as a function of the SN distance. The horizontal line indicates a nominal detectability threshold of two events. For details on the microphysics and SN parameters used, see text. Both panels reproduced from Ref.~\cite{Cavan-Piton:2025nsj}.
}  
  \label{fig:summary}
\end{figure}
The results shown in Fig.~\ref{fig:summary} are obtained for a representative set of SN and detector parameters. Specifically, axion emission is computed using the DD2 nuclear equation of state~\cite{Hempel:2009mc}, with core temperature $T = 30~\mathrm{MeV}$, baryon number density $n_B = n_{\rm sat}$, proton fraction $Y_p = 0.3$, and a SN-to-Earth distance of $d = 0.2~\mathrm{kpc}$. The three line styles in the left panel correspond to the KSVZ, DFSZ, and agnostic coupling scenarios described above. For an extensive discussion of the microphysics and SN parameters, 
see Refs.~\cite{Cavan-Piton:2025nsj,3121183}.
The left panel of Fig.~\ref{fig:summary} illustrates this separation by comparing the Cherenkov-parton spectra induced by neutrino absorption with those arising from axion absorption for a representative core-collapse SN model. The neutrino-induced positron spectrum peaks at low energies and rapidly falls off above $\sim 50$--$60~\mathrm{MeV}$. In contrast, the axion-induced spectra populate a much harder energy range. Among the latter, the $\pi^0$ production channel,
\begin{equation}
a + N \to N + \pi^0 \, , \nonumber
\end{equation}
is particularly prominent. Its spectrum peaks at energies of order $100$--$200~\mathrm{MeV}$, well beyond the bulk of the neutrino background. The single-photon channel $a p \to p \gamma$ is also present but is subdominant both in rate and in characteristic energy.

This hierarchy between the $\pi^0$ and $\gamma$ channels has a clear physical origin. In ChPT, the axion couples derivatively to the nucleon axial current, which naturally enhances pion emission relative to radiative processes. As a result, once the kinematic threshold for $\pi^0$ production is crossed, the $\pi^0$ channel becomes the dominant source of axion-induced Cherenkov light. For energies above the pion mass, the axion-induced $\pi^0$ spectrum typically overwhelms both the photon channel and the residual neutrino background.

\medskip

From the experimental point of view, this spectral separation is crucial. Energies above $\sim 140$--$150~\mathrm{MeV}$ lie beyond the regime where neutrino-induced events are abundant, and detector simulations~\cite{Cavan-Piton:2025nsj} indicate that the background in this range is expected to be negligible on the timescale of a SN burst. Moreover, neutral pions' prompt decay into two energetic photons produces distinctive multi-ring topologies that can be efficiently reconstructed. For Hyper-Kamiokande the $\pi^0$ reconstruction efficiency in this energy range can reach the level of $\mathcal{O}(80\%)$, with the majority of events classified as two-ring $\pi^0$-like. These features make the $\pi^0$ channel a particularly clean experimental handle on axion absorption.

\medskip

While the spectral properties determine the intrinsic discriminating power of Cherenkov detectors, the ultimate detectability of a SN axion signal is governed by the expected event rate. For a given axion model and SN realization, the number of axion-induced $\pi^0$ events in an energy range $R$ can be expressed as
\begin{equation}
\lambda(C_{app},C_{ann};R)
=
\int_R dE_{\pi^0}\,
\frac{dN_{\pi^0}^{(a)}}{dE_{\pi^0}} \, ,\nonumber
\end{equation}
and depends sensitively on the axion--nucleon couplings, the core temperature, and the distance of the SN. Cooling constraints from neutron stars and from SN~1987A typically limit $\lambda$ to at most a few hundred events even for very nearby explosions (0.2 kpc).

The dependence on the SN core temperature is particularly pronounced. Increasing the temperature from $30$ to $40~\mathrm{MeV}$ enhances the axion emissivity by nearly an order of magnitude, reflecting its steep scaling with temperature. As a consequence, predicted event rates can vary by almost a factor of ten across this temperature range. This sensitivity motivates adopting conservative benchmark temperatures when assessing detectability prospects.

\medskip

A further, and unavoidable, limitation arises from the geometric dilution of the axion flux. Since the event rate scales as $1/d^2$ with the SN distance $d$, only explosions within a few kiloparsecs can yield an observable axion signal. This behavior is illustrated in the right panel of Fig.~\ref{fig:summary}, which shows the expected number of axion-induced $\pi^0$ events as a function of distance for representative SN models and axion couplings. Requiring a minimal detectability threshold of two events typically restricts the observable range to $d \lesssim 2~\mathrm{kpc}$.

Such nearby core-collapse SNe are rare. While neutrino bursts are expected from the Milky Way at a rate of $\mathcal{O}(1)$ per century, only a small fraction of these events occur within the distance required for axion detection via Cherenkov light. Crude estimates suggest that the rate of SNe capable of producing a detectable axion burst is at the level of $\mathcal{O}(10^{-2})$ per century. This rarity underscores the importance of {\em maximizing the axion signal yield per event}.

\medskip

In this context, the enhancement provided by the $\pi^0$ channel relative to single-photon emission should be regarded as a necessary but not sufficient step toward realistic detection prospects. While the hard $\pi^0$-initiated spectrum provides a clean experimental handle with negligible neutrino background above $\sim 150~\mathrm{MeV}$, the overall sensitivity remains limited by the steep $1/d^2$ suppression of the event rate with supernova distance, and calls for further spectrum-enhancement mechanisms.

In summary, water Cherenkov detectors provide a conceptually clean and experimentally mature avenue for probing axion--nucleon interactions through axion absorption from nearby core-collapse supernovae. While neutrino bursts are detectable from much larger distances, axion detection requires a fortuitously {\em close} event.

Among the available absorption channels, axion-induced $\pi^0$ production offers the most promising signal, owing to its higher rate and higher characteristic energy. The estimates presented here are intentionally conservative, as they include only axion absorption on free protons in water.

A central challenge for the future is therefore to identify and quantify additional sources of signal enhancement beyond this baseline treatment. Natural candidates include axion absorption on neutrons; alternative absorption processes, such as $a + N \to Y + X$, where $Y$ is a baryon other than a nucleon and $X$ produces Cherenkov light (these processes exploit in reverse order the interactions considered in Ref.~\cite{Cavan-Piton:2024ayu}); a dedicated treatment of axion--oxygen scattering, which remains largely unexplored and could substantially increase the effective target mass. Establishing the size and robustness of such effects is essential given the rarity of sufficiently nearby SNe.

In this respect, SN axion searches at Cherenkov detectors should be viewed not as competitors to laboratory-based experiments, but as a complementary probe of meV-scale axions that is uniquely sensitive to axion--nucleon interactions in dense nuclear environments.


\section{meV axions experiments}
\label{sec:meV_axions_experiments}

{Experimentally probing the meV axion frontier is particularly challenging. In this section, we focus in detail on a representative subset of experimental directions, which were discussed as part of the ``meV Mass
Axion Frontier" workshop, including helioscope searches such as CAST ~\cite{CAST:2024eil} and BabyIAXO~\cite{BabyIAXO_Conceptual,IAXO_PhysPot}; and proposed meV haloscope searches such as CADEx~\cite{Aja:2022csb,Aja:2025pul}, and experimental proposals based on antiferromagnets~\cite{Catinari:2024ekq},  highly excited electron cycltotron ~\cite{Fan:2024mhm}, fifth force searches~\cite{Grossman:2025cov}. }

{However, we emphasise that a broad variety of complementary strategies is currently being developed. For example, the QUAX experiment searches for the axion-electron coupling using a ferromagnetic haloscope as well as the axion-photon coupling using a standard resonant cavity for masses around $36\,\mu\mathrm{eV}$~\cite{Barbieri:2016vwg,QUAX:2023gop,QUAX:2024fut}. ALPHA is a plasma haloscope which uses tunable metamaterials to target masses up to $186\,\mu\mathrm{eV}$~\cite{Lawson:2019brd,ALPHA:2022rxj}. There are also several broadband concepts targeting this mass range, including BREAD~\cite{BREAD:2021tpx}, which is a dish-antenna haloscope using a parabolic reflector to focus the resulting signal; and LAMPOST~\cite{Chiles:2021gxk}, which uses a multi-layer dielectric haloscope, read out using a superconducting nanowire single-photon detector, to target the meV-eV mass range. In addition, ARIADNE aims to probe axion-mediated `magnetic' forces in the mass range 0.1-10 meV~\cite{Arvanitaki:2014dfa,ARIADNE:2017tdd}.  A number of recent proposals have further expanded the experimental landscape by exploring novel resonant, interferometric, and condensed-matter-based techniques adapted to meV-scale axions~\cite{Mitridate:2020kly,Berlin:2023ubt,Arvanitaki:2024php}.}





\subsection{Extending axion mass sensitivity in CAST}
\label{sec:Extending_axion_mass_sensitivity_in_CAST}

\bigskip

The CERN Axion Solar Telescope (CAST) was a ground-based experiment located in Geneva (Switzerland) that searched for axions originating from the Sun. To look for axions, CAST pointed a decommissioned LHC prototype dipole magnet toward the Sun, with various X-ray detectors installed at both ends of the magnet. To date, CAST has been the most sensitive existing helioscope, making use of a prototype superconducting LHC dipole magnet that provided a magnetic field of $\rm{9\,T}$. CAST was able to follow the Sun twice a day during sunset and sunrise for a total of $\rm{3\,h}$. At both ends of the $\rm{10\,m}$ long magnet, X-ray detectors \cite{CAST_TPC,CAST_MM, CAST_CCD, CAST_GRIDPix} were mounted to search for photons from Primakoff conversion. The analysis of the data acquired during the first phase of the experiment yielded the most restrictive experimental upper limit on the axion–photon coupling constant for axion masses up to about $\rm{0.02\,eV/c^{2}}$. During the second phase, CAST extended its mass sensitivity by tuning the electron density in the magnetic conversion region. Hence, providing the photons with an effective mass $m_{\gamma}$ \cite{VanBibber_Raffelt}: 
\\
\begin{equation}
\Big( \frac{m_{a}^{2}}{\rm{eV}^{2}} \Big) \ll \Big( \frac{m_{\gamma}^{2}}{\rm{eV}^{2}} \Big) + 2\cdot \Big( \frac{E_{a}/\rm{eV}}{L\cdot \rm{eV}} \Big)
\end{equation}
\\
so that for $m_{a}$ values in the neighborhood of the chosen $m_{\gamma}$ the maximum sensitivity is restored.
\\
During the operation of $^{4}\mathrm{He}$ as a buffer gas in the magnet bores CAST established an upper limit on the axion--photon coupling of $g_{a\gamma} < 1.47 \times 10^{-10}\,\mathrm{GeV^{-1}}$ at 95\% CL. for axion masses $0.02 \lesssim m_{a} \lesssim 0.39~\mathrm{eV}$ \cite{CAST_He4_new} and after the upgrade to operate with $^{3}\mathrm{He}$ CAST established an upper limit on the axion--photon coupling of $g_{a\gamma} < 2.3 \times 10^{-10}\,\mathrm{GeV^{-1}}$ at 95\% CL. for axion masses $0.39\lesssim m_{a} \lesssim 0.64 ~\mathrm{eV}$ \cite{CAST_He3} and $g_{a\gamma} < 3.3 \times 10^{-10}\,\mathrm{GeV^{-1}}$ at 95\% CL. for axion masses $0.64\lesssim m_{a} \lesssim 1.19 ~\mathrm{eV}$ \cite{CAST_He3_new}.

\subsubsection{Effective photon masses at CAST}
\label{sec:Effective_photon_masses_at_CAST}

\begin{figure}[!h]
\begin{center}
\includegraphics[width = 1.\textwidth]{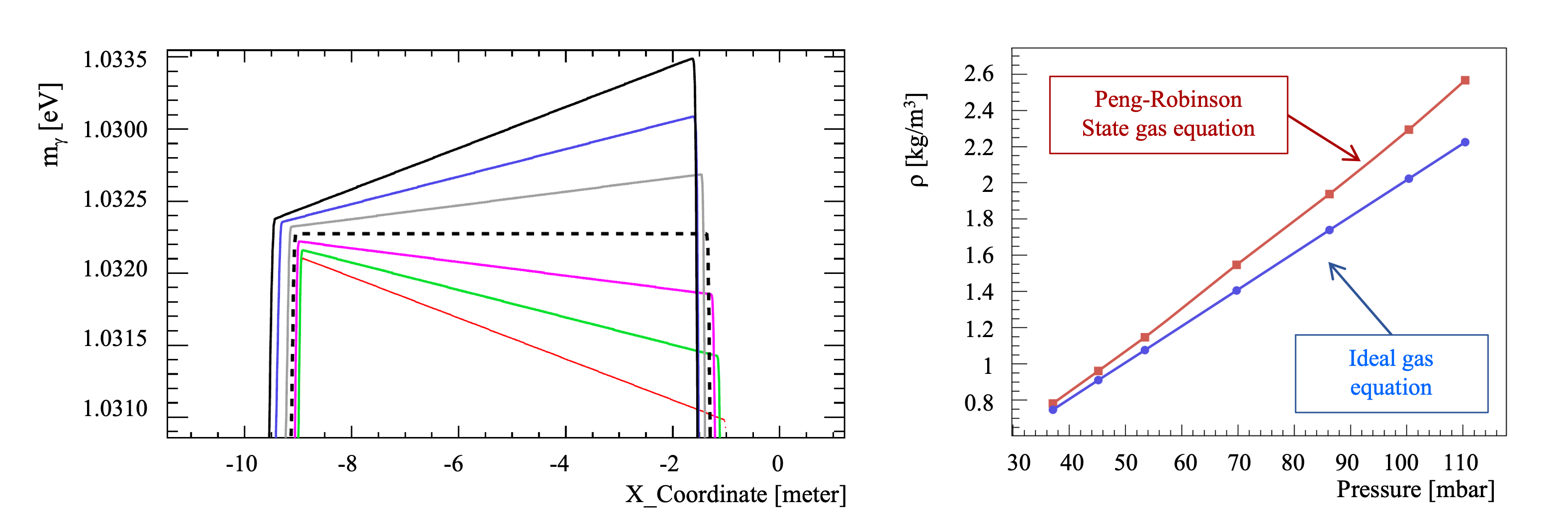}
\tiny\caption{\emph{Left.} Effective photon mass in the CAST cold bore as a function of the magnet tilt during solar tracking, shown together with the corresponding gas density in the bore. \emph{Right.} Evolution of the cold-bore gas density at rest, calculated using the Ideal Gas equation of state (blue) and the Peng–Robinson equation of state (red), both consistent with CAST measurements. Adapted from \cite{JRuz_IEEE_mass}. \label{fig:coldbore_mass}}
\end{center} 
\end{figure}

The electron density $(n_{e})$ of the gas filling the cold bore drives the physics of the axion Primakoff conversion at CAST. The effective mass of the photon arising can be obtained as:
\\
\begin{equation}
m_{\gamma}^{2}=\omega_{p}^{2}=4\pi n_{e}\frac{e^{2}}{m_{e}}=4\pi n_{e}r_{e}
\end{equation}
\\
being $\omega_{p}$ the plasma frequency and $r_{e}$ the electron radius. At CAST, Helium-3 and helium-4 were chosen due  the experimental constraints such as temperature of magnet operation.  
\\
CAST data taking stepped over different densities of gas to allow the search of different axion masses. Achieving the target gas density presented a demanding challenge, requiring not only computational fluid dynamics simulations of the full system, including effects such as hydrostatics, convection, and buoyancy, but also a highly precise gas-injection system capable of controlling the density to an accuracy of a few tenths of parts per million. A detailed examination of the experimental data and simulations revealed clear deviations from the ideal gas law. At extremely low temperatures, gas molecules move much more slowly, making intermolecular forces no longer negligible. Under these conditions, equations of state such as the Peng–Robinson model provide a more accurate description by accounting not only for molecular interactions and compressibility but also for the non-spherical nature of real molecules (see right panel of Fig. \ref{fig:coldbore_mass}). 
\\
\\
At experiments like CAST, the Primakoff conversion of axions depends on the density of electrons present in the system as well as parameters such as the intensity of the magnetic field, and the length of the path that axions travel through the media, 
\\
\begin{equation}
P_{a\rightarrow \gamma} = \frac{1}{4}g_{a\gamma}^{2}\exp(-\Gamma L)\Big\vert \int_{L} \exp(\Gamma l/2 +iql)B_{\perp}(l)dl\Big\vert^{2}.
\end{equation}
\\
Here, we would like to bring the attention to the reader that 
 convection currents of the gas between the X-ray windows and the inner regions of the cold bore translated to a magnetic field profiles dependent on the electron density and tilt of the experiment (see left panel of Fig. \ref{fig:coldbore_mass}).
 
Although the superfluid that cools down CAST magnet assures stable temperatures in the inner regions of the cold bore, the temperature of the X-ray windows and the density of the gas present in the system influence the impact of the convection currents to the effective magnet length. For instance, if we let the temperature of the X-ray windows to drift together with the cold bore gas while increasing the density of the system, the effective length of the magnet follows a linear dependence with the pressure achieved in the system.
\\
\begin{figure}[!t]
\begin{center}
\includegraphics[width = 0.65\textwidth]{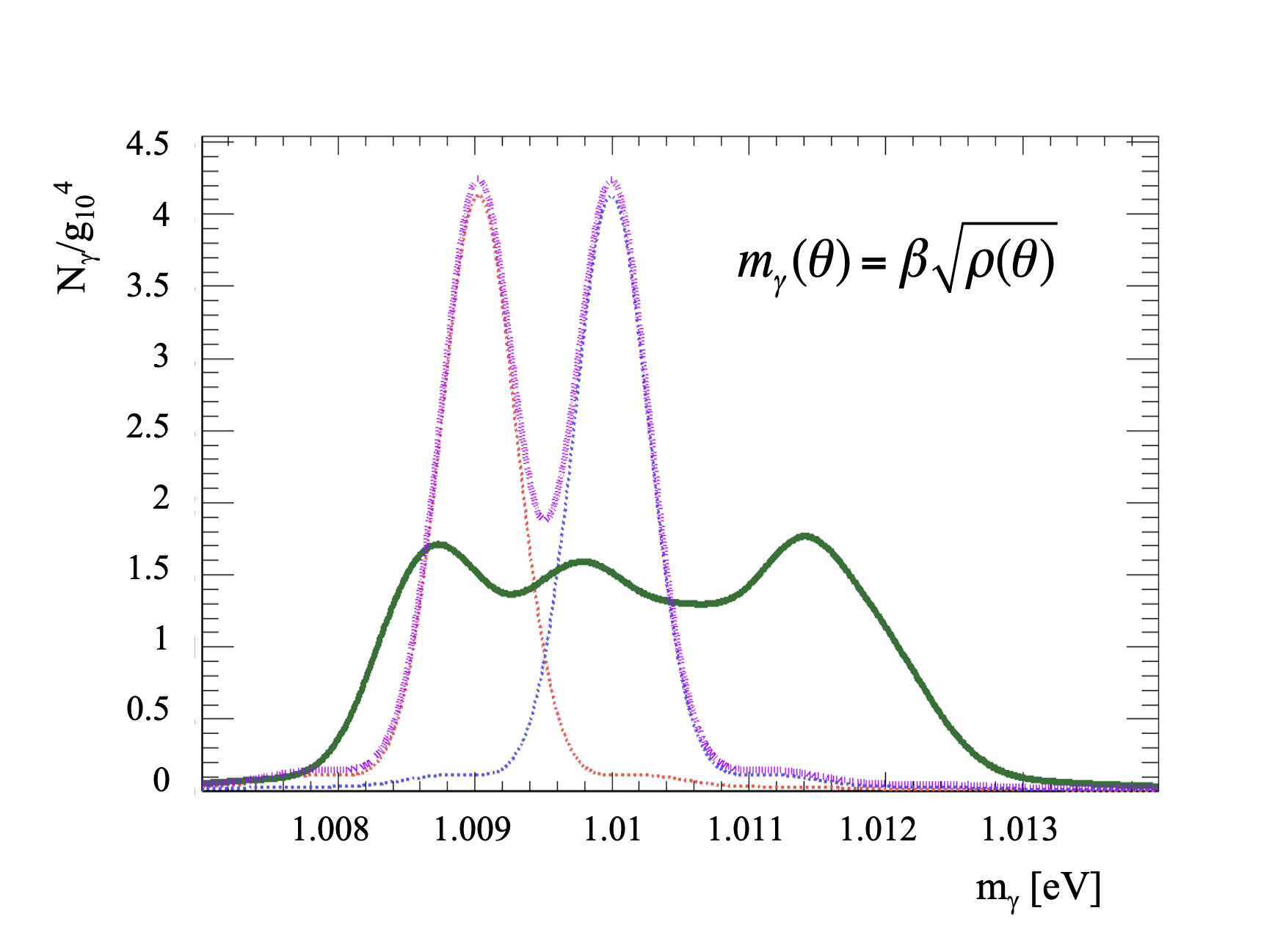}
\tiny\caption{Theoretical recovery of the coherence condition at selected axion masses (dashed blue and red curves) and their combined contribution (dashed purple), compared with the effective response of the experiment (green curve) once all experimental boundary conditions are included. Adapted from \cite{JRuz_IEEE_mass}. \label{fig:effective_mass_tracking}}
\end{center} 
\end{figure}
\\
Moreover, as previously discussed, gravitational forces can introduce a density gradient along the gas column in the magnet region. In its second-phase analysis, CAST accounted for this effect by reproducing the various data-taking conditions using computational fluid dynamics, with the measured temperatures and pressures along the coldbore serving as boundary conditions. 

\begin{figure} [!t]
\begin{center}
\includegraphics[width = 0.55\textwidth]{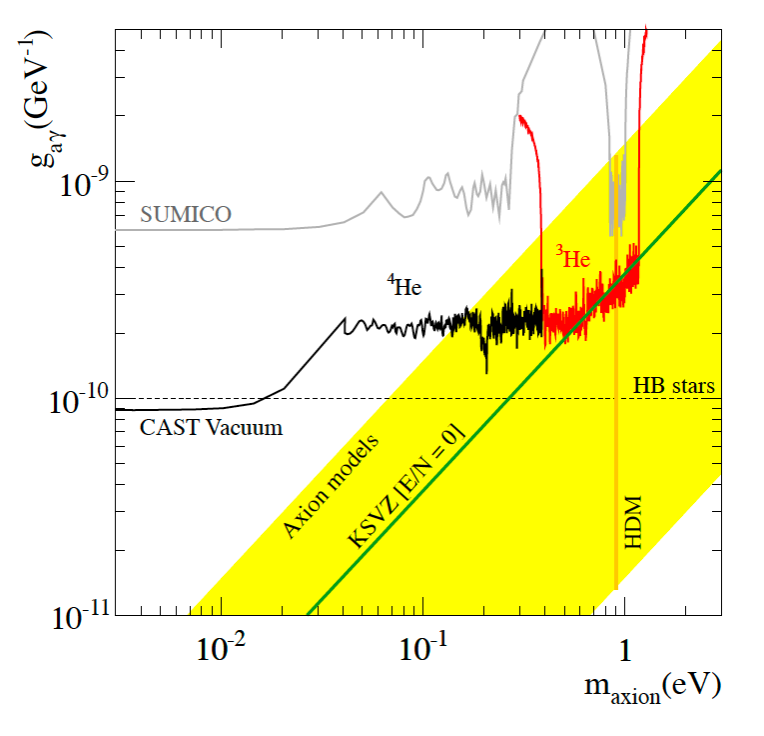}
\caption{\label{exclusion}CAST exclusion limits on the axion–photon coupling at 95\% CL using all data from Phase I and Phase II \cite{CAST_He3_new}. The CAST sensitivity is shown alongside the Horizontal Branch (HB) star bound \cite{Raffelt:1996wa}. The yellow band denotes the range of predicted QCD axion models, and the green curve represents the KSVZ model with E/N = 0.  \label{fig:th_CAST_CBmass}}
\end{center} 
\end{figure}

This procedure yields an effective photon mass that depends not only on the system pressure but also on the polar angle at which that pressure is measured. Using this effective density, together with CAST’s exposure to the Sun CAST was able to determine precisely the effective photon mass during every single solar tracking   (see Fig.~\ref{fig:th_CAST_CBmass} for an example).
\\
\\

\subsection{Axion Mass-Scanning Strategy of BabyIAXO}
\label{sec:Axion_Mass-Scanning_Strategy_of_BabyIAXO}

\bigskip

Following the operational strategy established by CAST, the BabyIAXO physics program is organized into two data-taking phases, each corresponding to approximately 1.5 years of exposure with an effective 12 hours of solar tracking per day~\cite{BabyIAXO_Conceptual,IAXO_PhysPot}. During the first phase, the magnetic field region will be operated under vacuum, providing sensitivity to the axion–photon coupling $g_{a\gamma}$ for axion masses up to several tens of meV, where axion–photon conversion remains coherent. At higher masses the coherence condition is lost and a low-$Z$ buffer gas will be introduced to extend the mass sensitivity to larger values. Building upon CAST experience, BabyIAXO will operate the magnet bores near room temperature to maintain the validity of the ideal-gas approximation and minimize convection effects that could otherwise arise.  Helium-4 is the chosen gas, and axion masses up to $m_a=0.25\,\rm{eV}$ are planned to be scanned, requiring at least 73 discrete gas-density nodes. 

\begin{figure} [!h]
    \centering
    \includegraphics[width=0.8\linewidth]{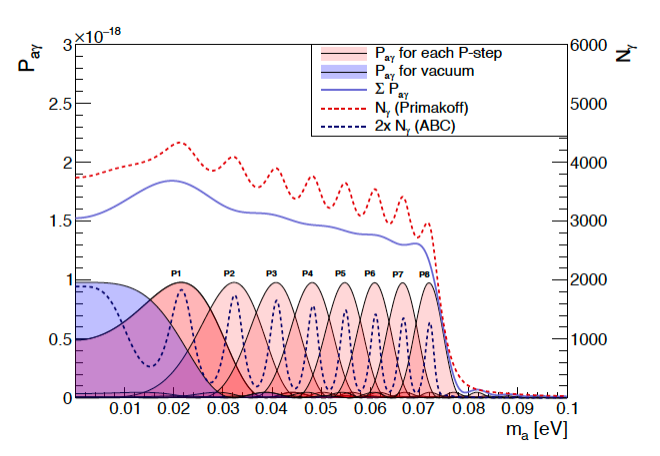}
    \caption{   Axion--photon conversion probability for 
$g_{a\gamma} = 10^{-10}\,\mathrm{GeV^{-1}}$ in a 2~T magnetic field of 10~m length for a photon energy of 4.2~keV. The probability is shown for the vacuum configuration (blue shading) and for the first eight 
gas-density settings (red shading) that enable a continuous axion-mass scan. The figure also displays the probability integrated at 4.2~keV, as well as the corresponding expected photon yield after convolution 
with the axion flux (including both Primakoff and ABC components). The flux is integrated over a 70~cm diameter magnet and an exposure time of 540~hours per setting, with the ABC flux evaluated assuming $g_{ae} = 10^{-12}$. Figure adapted from~\cite{BabyIAXO_raytracing}. 
    \label{fig:BIAXO_steps}}
\end{figure}

Continuous coverage in axion mass is obtained by choosing the gas-density steps such that adjacent resonances overlap at their full width at half maximum (FWHM). Figure \ref{fig:BIAXO_steps} illustrates the planned mass-scan structure, where the relative modulation amplitudes of the expected photon counts $N_\gamma$ from Primakoff and ABC axion components indicate that the Primakoff flux is particularly sensitive to the exact resonance spacing.

The exposure time assigned to each density setting is chosen to scale as $t_{\rm exp} \propto m_a^{-4}$ in order to reproduce the expected KSVZ-like mass dependence of the sensitivity curve. However, for the lowest-mass density settings this requirement is relaxed, as it would exceed the total 1.5~yr allocated to the gas phase. The exposure of the first five density settings is reduced and redistributed uniformly among them. This strategy is chosen because for the lowest-density settings the resonance width spans a comparatively large mass range, leading to non-negligible contributions across several mass nodes. 



\subsection{Post-discovery strategies for axion mass determination}
\label{sec:Post-discovery_strategies_for_axion_mass_determination}

\bigskip

Helioscope searches such as BabyIAXO offer an additional method for determining the axion mass. When an axion signal is detected during the gaseous phase, the coherence condition creates a direct correlation between the effective photon mass in the medium and the incident axion mass, enabling a precise inference of the latter. In addition, the spectral distribution of the detected axions also provides information relevant to characterizing the particle. Both the Primakoff and axion–electron spectra change depending on whether the axion is detected under resonant or non-resonant conditions.

\begin{figure} [!t]
    \centering
    \includegraphics[width=0.95\linewidth]{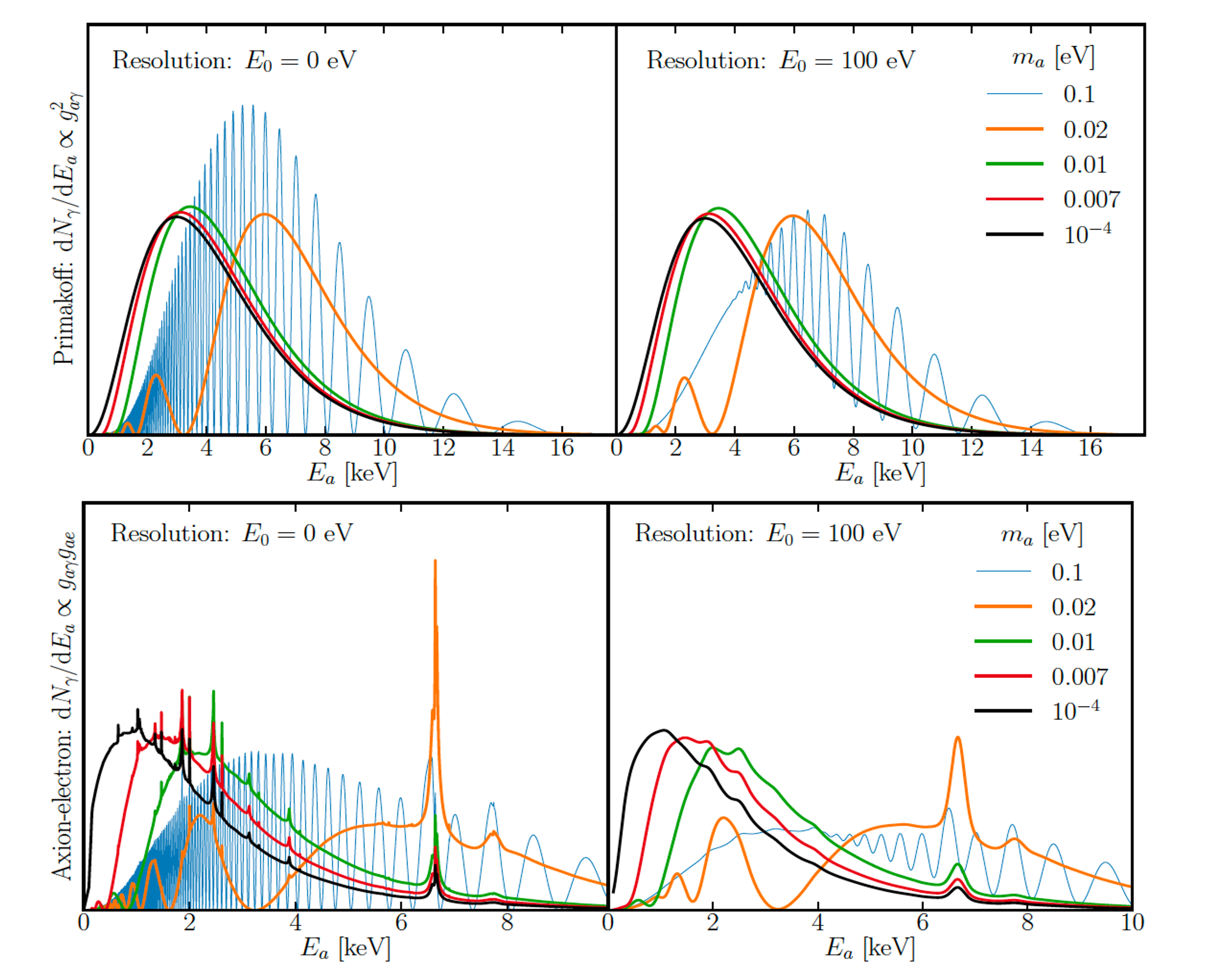}
    \caption{Differential X-ray spectra as a function of energy from solar axion conversion inside a 20\,m long, 2.5\,T magnet. Spectra are shown for several axion masses $m_{a}$ for both the solar Primakoff (\emph{upper panels}) and axion--electron (\emph{lower panels}) flux components. 
The left panels present the raw spectra, while the right panels show the spectra after convolution with a Gaussian energy resolution of width $E_{0} = 100\,\mathrm{eV}$. All spectra are normalized to unity for comparison; the inset shows the total integrated number of events $N_{\gamma}$ as a function of the five masses, assuming $g_{a\gamma} = 10^{-11}\,\mathrm{GeV^{-1}}$. Figure obtained from~\cite{Biljana_PRD}.
    \label{fig:PotDiscovery_scenario}}
\end{figure}

Figure \ref{fig:PotDiscovery_scenario} shows the expected spectra for axions with masses above 0.02 eV when the IAXO magnet is operated under vacuum. Axions out of resonance, but that lie close to satisfying the coherence condition produce spectra with a distinct shape, including a shift of the average detected energy toward higher values. As the axion mass moves further away from the resonance condition defined by the experimental setup, the spectrum develops a more pronounced ripple structure. These spectral features provide key information for determining the axion mass and allow for further tuning of the experiment.

Additional strategies can be implemented to determine the mass of a discovered axion in the sub-0.02 eV regime. One viable approach involves reducing the effective length of the magnetic field within the helioscope. Shortening the magnetic region relaxes the coherence requirement and modifies the interference pattern of axion–photon conversion, making the signal more sensitive to small deviations between the axion mass and the effective photon mass. By performing measurements at multiple, deliberately varied magnetic lengths, the experiment can extract the characteristic modulation in conversion probability that encodes the axion mass. This method provides a complementary mass-determination channel in parameter regions where the standard gaseous phase operation is less effective.

\subsection{Pushing the haloscope searches towards the meV range: The CADEx experiment}
\label{sec:Pushing_the_haloscope_searches_towards_the_meV_range_The_CADEx_experiment}

\bigskip

Haloscope experiments aim to detect DM axions from the halo of the Milky Way, through their conversion into photons in a strong magnetic field. The probability of axion-photon conversion is small, given the small expected value of the axion-photon coupling $g_{a\gamma}$~\cite{DiLuzio:2020wdo}. This probability is typically enhanced in haloscope experiments using resonant cavities: when the axion mass matches the resonant frequency of the cavity $f = m_a c^2/(2\pi\hbar)$, the conversion probability is enhanced by the quality factor $Q$ of the cavity. By tuning the resonant frequency of the cavity, haloscopes can probe a range of axion masses. {With typical (unloaded) quality factors of $Q _0 \sim 10^5$, a range of haloscope searches have begun probing the QCD axion parameter space in the range $m_a \in [2, 50]\,\mu\mathrm{eV}$, including ADMX~\cite{ADMX:2018gho,ADMX:2019uok,ADMX:2021nhd}, 
CAPP~\cite{Kim:2022hmg,CAPP:2024dtx}, 
HAYSTAC~\cite{HAYSTAC:2018rwy,HAYSTAC:2020kwv,HAYSTAC:2023cam,HAYSTAC:2024jch} and QUAX~\cite{QUAX:2023gop,QUAX:2024fut}. A number of searches have also provided constraints below the CAST limit~\cite{CAST:2024eil}, up to masses of $\sim 0.1\,\mathrm{meV}$, including GigaBREAD~\cite{GigaBREAD:2025lzq}, GrAHal~\cite{Grenet:2021vbb},  MADMAX~\cite{Garcia:2024xzc}, ORGAN~\cite{McAllister:2017lkb,McAllister:2022ibe,Quiskamp:2022pks,Quiskamp:2024oet}, RADES~\cite{CAST:2020rlf,Ahyoune:2024klt}, TASEH~\cite{TASEH:2022vvu}, UF/RBF~\cite{DePanfilis,Wuensch:1989sa,Hagmann:1996qd}.}

However, pushing haloscope searches to higher masses presents a number of challenges. The resonant frequency of a cavity is typically set by its physical size $L$, such that it matches the photon wavelength $L \sim \lambda = c/f$. As the target axion mass (and therefore the desired resonant frequency) is increased, the size of the cavity must be decreased. This means that the sensitive volume of the cavity, $V\sim L^3\sim\lambda^3$ drops rapidly with increasing axion mass. In addition, the cavity power in the haloscope is typically picked up by an antenna and amplified. This coherent amplification adds an irreducible noise contribution, meaning that the noise level cannot be suppressed below the so-called Standard Quantum Limit, SQL, (at least using standard amplification techniques). This irreducible noise contribution becomes more significant with increasing frequency, further limiting the sensitivity of high-mass axion searches using coherent readouts. 

The Canfranc Axion Detection Experiment (CADEx) is a proposed search for axions in the mass range $330-460\,\mu\mathrm{eV}$ (corresponding to W-band frequencies, in the range $80-110\,\mathrm{GHz}$)~\cite{Aja:2022csb,Aja:2025pul}. CADEx is a resonant-cavity-based search which aims to address the challenges described above in a number of ways. First, the reduction in sensitive volume at high frequencies will be mitigated through the coherent combination of seven flat rectangular cavities. The smallest dimension of each cavity, which sets the resonant frequency, is approximately $a \sim1.7\,\mathrm{mm}$, while the other two dimensions are much larger ($40-60\,a$), thereby maximizing the sensitive volume of each cavity while fitting in a 10 T superconducting magnet bore. In the CADEx experiment, the frequency sweep is performed by tuning the thin dimension of the cavity using a sliding-wall mechanism \cite{Aja:2022csb}, subsequently demonstrated at frequencies around 26 GHz~\cite{McAllister:2023ipr,Quiskamp:2023ehr}. The output from each of the seven cavities will be extracted using horn antennas and coherently combined and routed to the detecting system using a quasi-optical free-space setup, ensuring phase coherence among cavities. 

Second, the detection system is based on an array of superconducting Kinetic Inductance Detectors (KIDs)~\cite{day2003broadband}. KIDs are superconducting resonators whose resonance properties, \textit{i.e.} kinetic inductance, change in response to absorbed radiation. They are not limited by the SQL and their sensitivity surpasses that achieved by traditionally employed heterodyne receivers in this frequency range. An array of multiple KIDs based on Ti/Al bilayers will be used to sense the output signal for the haloscope~\cite{rollano2025dark}. On-chip polarization discrimination will be implemented to perform correlation analysis between the polarized axion-photon conversion signal and the unpolarized background.

The CADEx experiment, planned to be installed at the Laboratorio Subterráneo de Canfranc (LSC), is currently in the design and technological demonstration stage. Typical parameters for the baseline setup of CADEx include a cavity quality factor of $Q_0 = 10^5$, a total cavity volume of $V = 0.08\,\mathrm{L}$ and a noise equivalent power (NEP) of $3 \times 10^{-20}\,\mathrm{W}\sqrt{\mathrm{Hz}}$ for the KIDs sensors. With these, CADEx expects to be sensitive to axions in the QCD band, probing couplings of $g_{a\gamma} \lesssim 1.5 \times 10^{-13}\,\mathrm{GeV}^{-1}$ for an axion mass of $370\,\mu\mathrm{eV}$ over a 3 month exposure. Owing to the broadband nature of KIDs, CADEx serves both as a technological pathfinder for high-frequency axion searches and as a complementary technological platform to proposed broadband haloscope experiments targeting high axion masses, such as BREAD~\cite{BREAD:2021tpx}, BRASS~\cite{Bajjali:2023uis}, DALI~\cite{DeMiguel:2020rpn}.




\subsection{meV Axion Quasiparticles}
\label{sec:meV_Axion_Quasiparticles}

\bigskip

\subsubsection{Discovery of meV axion quasiparticles}

\begin{figure}
    \centering
    \includegraphics[width=0.99\linewidth]{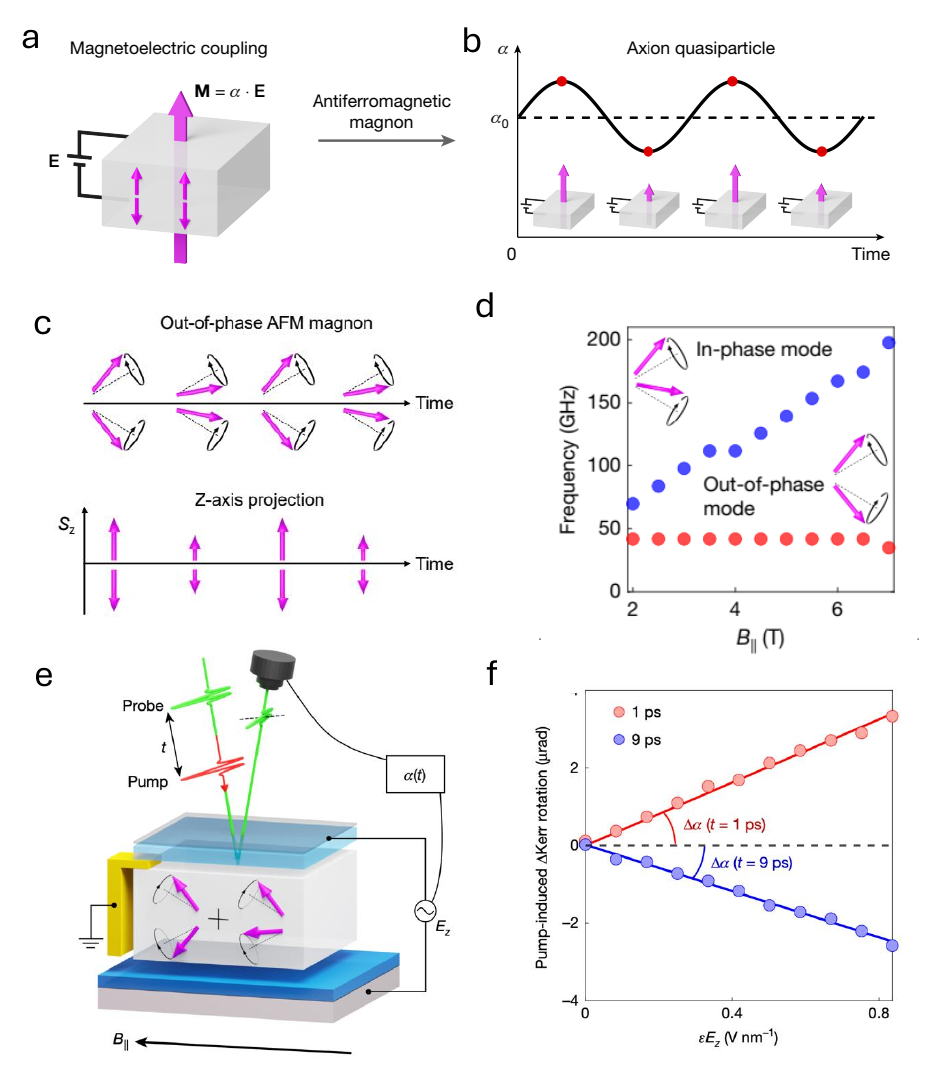}
    \caption{Axion quasiparticle in MnBi$_2$Te$_4$. \textbf{a,b,} In condensed matter, the $\theta$ angle is directly proportional to the trace of the magnetoelectric coupling, $\Theta=\pi\frac{2h}{3e^2}\sum_{i=x,y,z}\alpha_{ii}$). Thus, a time varioation in $\alpha$ induced by an antiferromagnetic magnon corresponds to the axion quasiparticle. \textbf{c,} Time evolution of out-of-phase magnon and $\hat{z}$ projection of the out-of-phase magnon, $S_z$, which resembles the antiferromagnetic amplitude mode. The axion quasiparticle is the coherent oscillation of the magnetoelectric coupling. \textbf{d,}  Experimentally measured frequencies for the in-phase magnon and out-of-phase magnon modes as a function of the in-plane magnetic field $B_{\|}$ of 6-layer MnBi$_2$Te$_4$. \textbf{e,} Schematic experimental setup to probe the time-dependent magnetoelectric coupling. The pump laser launches coherent magnons; The probe laser, combined with the dual gate $E_z$, measures $\frac{d \textrm{Kerr rotation}}{d E_z}$; By varying the delay time $t$, we can measure pump-induced $\Delta\alpha(t)$ with femtosecond resolution. \textbf{f,} Measured pump-induced Kerr rotation $\Delta \textrm{Kerr}(t)$ as a function of $E_z$ and $t$. (Adapted from Ref.~\cite{Qiu:2025bbi})}
    \label{fig:aq_physics}
\end{figure}

\begin{figure}
    \centering
    \includegraphics[width=0.75\linewidth]{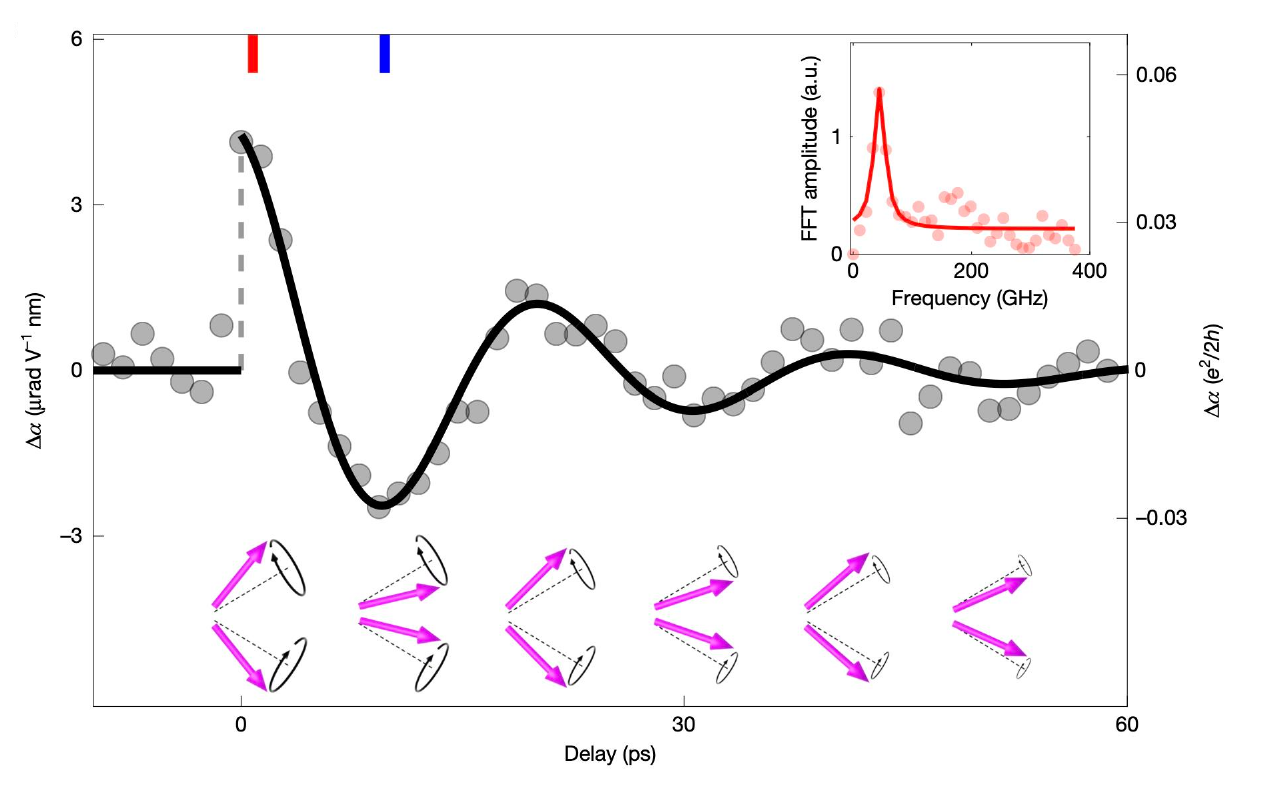}
    \caption{Experimental measurements of the theta angle oscillation for the axion quasiparticle in MnBi$_2$Te$_4$. Measured magnetoelectric coupling $\alpha$ oscillation as a function of time. The theta angle (axion quasiparticle) is directly proportional to the magnetoelectric coupling by $\Theta=\frac{2\pi h}{e^2}\alpha$. Arrows indicate spins of the tilted magnon such that the projection leads to an oscillation in $S_z$ (see Fig.~\ref{fig:aq_physics}) Inset: fourier transform of the signal, indicating the axion quasiparticle frequency is 44 GHz, i.e. $m_\Theta=0.18\text{ meV}$. (adapted from Ref.~\cite{Qiu:2025bbi}) }
    \label{fig:aq_discovery}
\end{figure}

Wilczek first proposed in 1987~\cite{Wilczek:1987mv} the possibility that condensed matter systems may exhibit axion-like physics, in particular realising the axion-photon interaction term with some effective degree of freedom, i.e. axion quasiparticles (AQs). The AQ action is:
\begin{equation}
    S = \int d^4 x\left\{ \frac{f_\Theta^2}{2}\left[(\partial_t\delta\Theta)^2-(v_i^2\partial_i\delta\Theta)^2-m_\Theta^2\delta\Theta^2\right]+\frac{\alpha}{\pi}(\delta\Theta+\Theta_0)\vec{E}\cdot\vec{B}\right\}\, ,
\end{equation}
where $\Theta_0$ is the constant background field (fixed by the material), $\delta\Theta$ is the dynamical AQ, and $m_\Theta$, $f_\Theta$, $v_i$ are the AQ mass, decay constant, and speed respectively.

A significant development in this field occurred in 2009 when Li et al.~\cite{Li:2009tca} proposed that topological magnetic insulators were suitable candidate materials. In such materials, the AQ arises from fluctuations in antiferromagnetic order. The original proposal supposed that this requires fluctuations directly in the order parameter, $S_z$, the ``longitudinal magnon'', which only occurs when the unit cell breaks $P$ and $T$ symmetries. Such a scenario is proposed to be realised in iron-doped Bismuth Selenide, (Bi$_{1-x}$Fe$_x$)$_2$Se$_3$~\cite{Li:2009tca}, or in Manganese Bismuth Telluride 2-2-5, Mn$_2$Bi$_2$Te$_5$~\cite{zhang2020large}. These materials, and the AQ phase in them, proved difficult to realise in the laboratory, in part because excitation of longitudinal magnons is difficult to achieve. 

There is also a class of so-called ``axion insulator'' materials that are more well studied and realised in the lab (e.g. Refs.~\cite{qi2008topological, essin2009magnetoelectric, Nenno2020axion, sekine2021axion}). Axion insulators are characterised by the presence of a constant electromagnetic $\Theta_0$ term with $\Theta_0\neq 0,\pi$, but where symmetries forbid dynamics for $\delta\Theta$. The axion insulator Manganese Bismuth Telluride 1-2-4, MnBi$_2$Te$_4$, is well studied and is known to possess antiferromagnetism below its N\'eel temperature~\cite{otrokov2019prediction}. Recently, Qiu et al.~\cite{Qiu:2025bbi} demonstrated that the application of an in-plane magnetic field to even-layer MnBi$_2$Te$_4$ has the effect to tilt the spins in the model, leading to an effective change in the order parameter projected out of the plane caused by a transverse magnon, which is much easier to excite. Furthermore, Qiu et al. were able to demonstrate that in such a setting the out-of-phase magnon couples to the $\Theta$ (Chern-Simons) term of electromagnetism, inducing rotation in polarisation of an incident laser beam onto the sample, i.e. birefringence. By measuring the frequency of this rotation with  ultrafast pump-probe optics, Qiu et al. demonstrate the existence of the AQ in this material, and measure its mass (frequency) to be $m_\Theta=0.18\text{ meV}$ (44 GHz). 

An illustration of the physics behind AQs and their recent discovery is shown in Fig.~\ref{fig:aq_physics}, while Fig.~\ref{fig:aq_discovery} shows the headline result of the discovery of the AQ and measurement of its mass. 

The basic physics that places the AQ in the meV range is the role of the antiferromagnetic anisotropy field, $H_A$, in setting the mass of the transverse magnons (see e.g. Refs.~\cite{Hofmann:1998pp,Schutte-Engel:2021bqm}), with typical values of $\mu_B H_A\sim \mathcal{O}(\text{meV})$. Thus we expect that should other AQ materials be discovered that exploit antiferromagnetism then they too will have AQs in the meV range.~\footnote{The natural meV range of AFMR implies that an AFMR QUAX-like haloscope following Refs.~\cite{Kakhidze:1990in,Barbieri:2016vwg} may be a candidate experiment for follow up to measure the axion-electron coupling at meV.}

AQ-photon mixing is thought to be strong enough that in the presence of a background magnetic field the AQ hyridizes with the electric field, forming states known as \emph{polaritons}. The polariton dispersion relation has two branches, $\omega_\pm$. In a material with refractive index $n$ and AQ velocity $v\approx 0$, the dispersion relation is:
\begin{equation}
    \omega_\pm(k)^2 = \frac{1}{2}\left[\omega_{\rm LO}^2+\frac{k^2}{n^2}\right]\pm \frac{1}{2}\sqrt{\left(\omega_{\rm LO}^2-\frac{k^2}{n^2}\right)^2-4b^2\frac{k^2}{n^2}}\, ,
\end{equation}
where  $\omega_{\rm LO}$ is the ``longitudinal optical'' frequency:
\begin{equation}
    \omega_{\rm LO}^2 = m_\Theta^2+b^2\, , \quad b^2 = \frac{\alpha}{2\pi n^2}\frac{B_0^2}{f_\Theta^2}\, .
\end{equation}

The techniques employed in Ref.~\cite{Qiu:2025bbi}, which excite the AQ off resonance with an optical probe, have not measured the AQ-polariton. The presence of the polariton can be inferred from the gap in the dispersion relation between $m_\Theta$ and $\omega_{\rm LO}$~\cite{Li:2009tca}. By performing transmission measurements in this range of frequencies, it would be possible to measure $f_\Theta$, as well as the losses in the material (discussed further below)~\cite{Schutte-Engel:2021bqm}. Density functional theory calculations performed in Ref.~\cite{Qiu:2025bbi} predict $f_\Theta=82 \text{ eV}$, close to cruder estimates made in Ref.~\cite{Schutte-Engel:2021bqm}: this prediction remains to be confirmed experimentally.

\subsubsection{Axion quasiparticles for meV axion dark matter}

The AQ-polariton can be exploited for axion DM detection, as first proposed in Ref.~\cite{Marsh:2018dlj}, where it is referred to as ``TOORAD'' (TOpolOgical Resonant Axion Detection). The ``$+$'' polariton branch with dispersion relation $\omega_+$ is the same as that of a massive particle with mass $\omega_{\rm LO}$. Axion DM sources an electric field in the material as usual, but now the electric field hybridizes with the AQ, and the polariton modes are driven. Resonance is achieved when $\omega_{\rm LO}\approx m_a$ (see also Ref.~\cite{Chigusa:2021mci}). 

It can be shown that the AQ-polariton  driven by axion DM can be modelled using an effective refractive index~\cite{Schutte-Engel:2021bqm}:
\begin{equation}
    n_\Theta = n^2\left(1-\frac{b^2}{\omega^2-m_\Theta^2}\right)\, ,
\end{equation}
Resonance occurs when $n_\Theta$ approaches zero. Having $n_\Theta<1$ causes internal reflections and a piece of AQ material starts to behave as a resonant cavity~\cite{Schutte-Engel:2021bqm}. Techniques developed to study dielectric haloscopes~\cite{Millar:2016cjp} can then be employed to compute the boost factor, $\beta$, to the axion DM induced power emitted at frequency $\omega_{\rm LO}$ from the material surface. In previous works, the correct physics giving rise to the AQ (in plane $B$-field to tilt the transverse magnon) was not known, but it turns out that this difference has only a minor effect on the range of $B$-induced tuning predicted for the AQ-polariton. 

Ref.~\cite{Schutte-Engel:2021bqm} estimated the relevant contributions to the material losses, $\Gamma$, and the dominant losses are thought to be due to magnetic impurities, $\Gamma_m$, and finite resistivity, $\Gamma_\rho$ (i.e. conductive losses). Ref.~\cite{Qiu:2025bbi} computed the complex refractive index in MnBi$_2$Te$_4$ using density functional theory, which gives $n=6.4$, $\Gamma_\rho/\omega=0.2\times 10^{-3}$, again largely confirming cruder estimates~\cite{Schutte-Engel:2021bqm}. Magnetic impurity losses can be estimated to be $\Gamma_m/\omega=0.7\times 10^{-3}$ following Ref.~\cite{PhysRevLett.111.017204}. 

In the absence of losses, the AQ-polariton resonance allows for axion DM detection at high frequency in a totally volume-independent way, which is already a significant feature. The presence of losses, however, leads to a maximum in the boost factor $\beta$ set by when the total optical path (accounting for finesse) reaches the skin depth. For the estimated parameters of MnBi$_2$Te$_4$ the maximum thickness is of order 0.4~mm, which is already extremely thick in material terms: a realistic method to reach such thickness would be to deposit layers sandwiched with dielectric~\cite{Qiu:2025bbi}.

\begin{figure}
    \centering
    \includegraphics[width=0.75\linewidth]{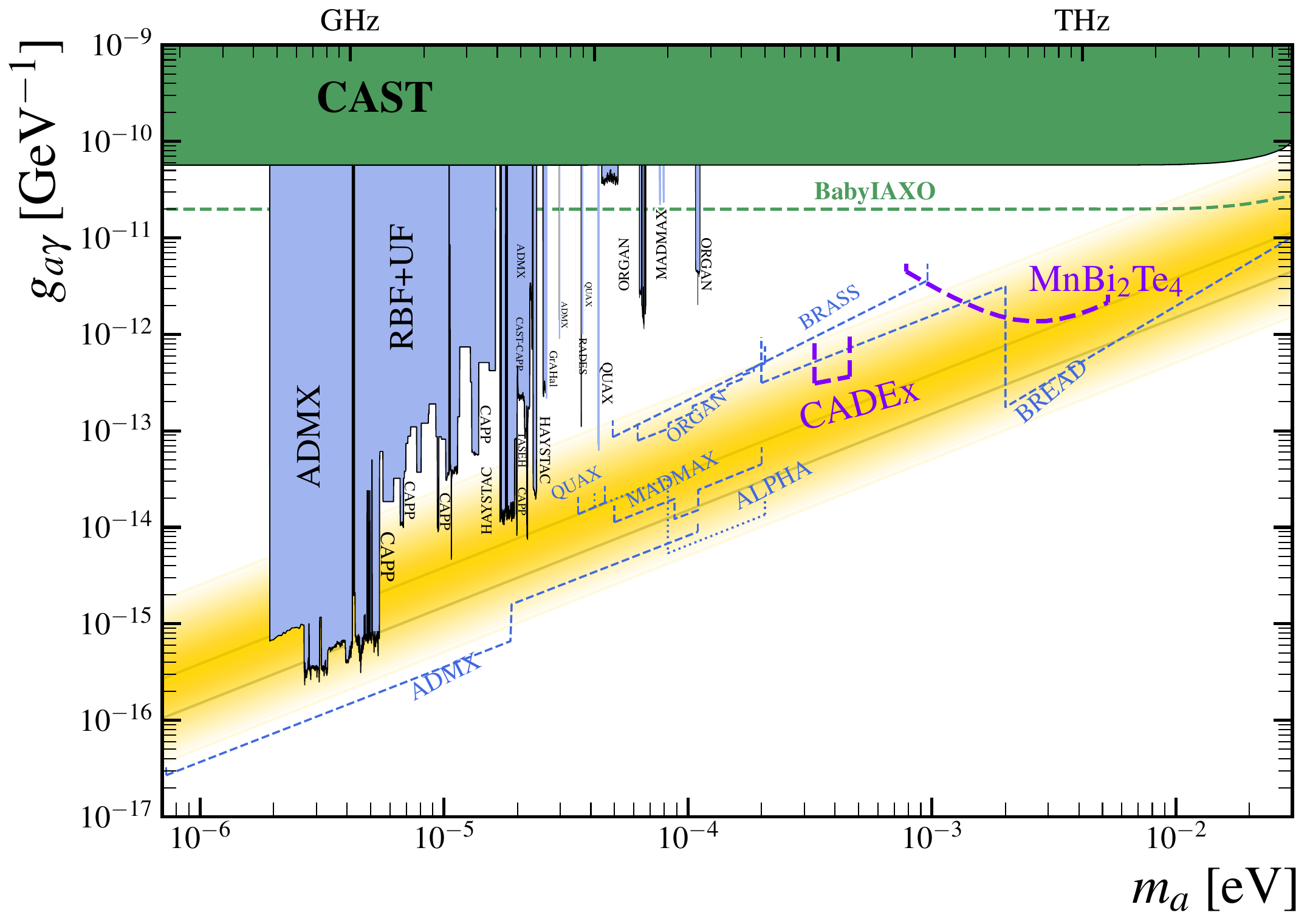}
    \caption{Existing (solid) and projected (dashed) constraints on the axion photon coupling, $g_{a\gamma}$. Highlighted in purple are the proposed DM haloscopes discussed in this paper: CADEx (Sec.~\ref{sec:Pushing_the_haloscope_searches_towards_the_meV_range_The_CADEx_experiment}) and the AQ material MnBi$_2$Te$_4$ (Sec.~\ref{sec:meV_Axion_Quasiparticles}). The yellow band indicates QCD axion models. CADEx is sensitive to the QCD axion for $m_a\approx 0.4\text{ meV}$. The AQ material is sensitive to the QCD axion for $m_a\approx 1-7\text{ meV}$. Figure produced using \textsc{AxionLimits}~\cite{AxionLimits} (where references for the other shown limits can be found).}
    \label{fig:AxionPhoton_projections}
\end{figure}

A significant feature of the AQ-polariton approach to axion DM detection is in resonance tuning: the resonant frequency is controlled by the externally applied $B$-field. In the set-up of Ref.~\cite{Qiu:2025bbi} the bending of spins caused by the in-plane magnetic field $B_{||}$ is precisely what is being exploited to generate the AQ in the first place, and the AQ was shown to be present and stable in the range 2 to 7 T. The same in-plane $B$-field will cause axion DM to drive the AQ-polariton. Thus, in Ref.~\cite{Qiu:2025bbi} the range of resonant frequencies was chosen by varying $B_{||}$ in the range 1 to 10 T, which corresponds to:
\begin{equation}
    0.7\text{ meV}\lesssim m_a \lesssim 7\text{ meV}\, \quad (\text{scanning range of MnBi}_2\text{Te}_4\text{ for }B=1-10\text{ T})\, .
\end{equation}
Refs.~\cite{Qiu:2025bbi,Schutte-Engel:2021bqm} showed that (under some assumptions about future detectors in this range) it will be possible to search for QCD axion DM above 1 meV inside its model band~\cite{DiLuzio:2020wdo} using these techniques. Fig.~\ref{fig:AxionPhoton_projections} shows the projected sensitivity of an AQ-based haloscope using MnBi$_2$Te$_4$, with measured and calculated parameters and experimental assumptions as outlined in Ref.~\cite{Qiu:2025bbi}. 

\section{Discussion and conclusions}
\label{sec:conclusion}

The meV mass range has emerged as a focal point in axion physics, where advances across 
theory, cosmology, astrophysics, and experiment converge with remarkable coherence. 
In this review, we have surveyed the current status of meV axion research, highlighting both the theoretical motivations that single out this mass window and the diverse experimental strategies poised to probe it in the coming years.

From the theoretical perspective, field-theoretic approaches to the PQ quality problem generically favour decay constants around $f_a \sim 5\times 10^9$ GeV, corresponding to axion masses in the meV range. This regime can also be the outcome of large classes of well-motivated ultraviolet completions of the axion framework. In fact, the low-energy limit of string compactifications on CY manifolds can realise an meV-scale QCD axion either as a closed string mode arising from the dimensional reduction of higher dimensional forms, or as the phase of an open string mode charged under an anomalous $U(1)$ on branes. The two different model building options can lead to a pre- as well as to a post-inflationary scenario. Experimental programmes targeting the meV axion regime can therefore be exploited as probes of the structure of the string landscape. Moreover, since axions come with saxion partners tied to the same moduli dynamics, sensitivity to meV axions can also inform us about saxion masses and couplings.

The cosmological phenomenology of meV axions is rich and multifaceted. While the 
standard misalignment mechanism with $\mathcal{O}(1)$ initial angles yields a 
subdominant dark matter contribution at these masses, alternative production 
scenarios---including large-misalignment tuning, kinetic misalignment, and 
post-inflationary dynamics with topological defects---can accommodate meV axions 
as viable cold dark matter candidates. The precise predictions for the latter 
scenario remain subject to ongoing theoretical investigation, particularly 
regarding the spectral properties of axion emission from cosmic strings. 
Independently of their role as dark matter, thermal production of axions in 
the early Universe represents an inevitable cosmological consequence that 
will be probed by next-generation CMB experiments. The projected sensitivity to $\Delta N_\mathrm{eff}$ will test thermal axion populations 
precisely in the meV mass range, offering a powerful and complementary 
discovery channel.

Astrophysical observations currently provide the most stringent constraints 
on meV axions. {Observations of} neutron-star cooling and the SN~1987A neutrino burst independently limit the axion mass to $m_a \lesssim 10$--$20$~meV for 
KSVZ-type couplings, with residual uncertainties stemming from the 
challenging physics of dense nuclear matter. These bounds place the meV 
frontier tantalisingly close to current sensitivity thresholds. Looking 
forward, the next Galactic core-collapse supernova offers a unique 
opportunity: axion-to-photon conversion in the magnetic field of the 
progenitor star could produce a detectable gamma-ray burst, providing 
a direct probe of QCD axion couplings at meV masses. The detection 
prospects depend sensitively on the progenitor type, with compact 
Type~Ibc supernovae
(representing 30\% of the total core-collapse SN rate)
emerging as particularly promising targets,   due to 
their stronger magnetic fields
than both red and blue supergiants.
 Ensuring 
full-sky gamma-ray coverage during the next Galactic supernova 
constitutes a compelling science case for future satellite missions. Additionally, large-volume Cherenkov detectors provide 
a complementary probe of axion-nucleon couplings through 
absorption of supernova axions, with the $\pi^0$ production 
channel yielding distinctive spectral signatures above the 
neutrino background.

On the experimental front, the landscape has been transformed by 
a convergence of mature and emerging technologies. The helioscope 
programme, building on the pioneering results of CAST, is advancing 
through BabyIAXO toward the full IAXO experiment, which will 
systematically scan the meV mass range with sensitivity reaching 
the QCD axion band. Buffer-gas techniques enable coherent 
axion-photon conversion at elevated masses, while improved 
X-ray optics and low-background detectors enhance the signal-to-noise 
ratio. High-frequency haloscope concepts, exemplified by CADEx, 
are pushing resonant-cavity searches toward the 100~$\mu$eV--meV 
regime, complementing the helioscope approach with sensitivity to 
the local dark matter density rather than the solar axion flux.

Finally, a particularly exciting recent development is the experimental 
discovery of axion quasiparticles in topological antiferromagnets. 
The observation of an axion quasiparticle with mass 
$m_\Theta \simeq 0.18$~meV in MnBi$_2$Te$_4$ not only validates 
the underlying axion electrodynamics in a controlled laboratory 
setting but also opens new avenues for dark matter detection. 
The axion-polariton hybridisation in these materials creates 
volume-independent resonant enhancement that can be tuned 
across the meV range by varying an external magnetic field, 
offering a qualitatively new approach to high-frequency axion 
searches.

Several open questions and challenges remain. On the theoretical 
side, a complete understanding of axion production from cosmic 
string networks---and its implications for post-inflationary 
cosmology---awaits further progress in numerical simulations 
and analytical modelling. The physics of axion emission from 
dense nuclear matter, while qualitatively understood, carries 
irreducible uncertainties at the {factor-of-few}  level that 
propagate into astrophysical bounds. Experimentally, the 
primary challenge lies in extending sensitivity to the QCD 
axion coupling strength across the full meV mass range, a 
goal that will require sustained investment in detector 
technology and infrastructure.

Despite these challenges, the overall picture is one of 
remarkable convergence and opportunity. The meV axion 
hypothesis is now subject to a coordinated assault from 
multiple independent directions: helioscopes probing 
solar axion production, haloscopes and quasiparticle 
detectors searching for dark matter signals, precision 
cosmology constraining thermal relics, and astrophysical 
observations setting limits from stellar and supernova 
environments. This multiplicity of approaches offers not 
only redundancy but also the prospect of cross-validation 
in the event of a positive signal, enabling robust 
characterisation of axion properties including mass, 
couplings, and cosmological abundance.

The coming decade will be decisive for meV axion physics. 
The commissioning of BabyIAXO and the continued development 
of IAXO will bring helioscope sensitivity into the QCD axion 
band. Future CMB experiments will probe thermal axion contributions to dark 
radiation with unprecedented precision. The maturation of 
axion quasiparticle techniques may enable tunable resonant 
searches across the meV range. And the next Galactic supernova, 
whenever it occurs, will provide a singular opportunity to 
detect or constrain axion emission from the most extreme 
astrophysical environments. Together, these developments 
position the meV mass range as one of the most compelling 
and experimentally accessible frontiers in the search for 
physics beyond the Standard Model.

\section*{Acknowledgments}
This article is based upon work from COST Action COSMIC WISPers (CA21106).
We warmly thank Federico Mescia for the kind hospitality at the Laboratori Nazionali di Frascati during the workshop ``The meV Mass Axion Frontier: Challenges and Opportunities''  (\url{https://agenda.infn.it/event/48801/}) and for providing us with good food and wine that helped to keep our spirits up.

Maurizio Giannotti acknowledges support from the Spanish Agencia Estatal de Investigaci\'on under grant PID2019-108122GB-C31, funded by MCIN/AEI/10.13039/501100011033, and from the ``European Union NextGenerationEU/PRTR'' (Planes complementarios, Programa de Astrof\'isica y F\'isica de Altas Energ\'ias). 
Additionally, Maurizio Giannotti acknowledges funding from the European Union's Horizon 2020 research and innovation programme under the European Research Council (ERC) grant agreement ERC-2017-AdG788781 (IAXO+).

The work of Mathieu Kaltschmidt, Maurizio Giannotti and Javier Redondo is supported by the grants PGC2022-126078NB-C21 and PID2024-160228NB-I00 funded by MCIN/\-AEI/\-10.13039\-501100011033 and ERDF - A way of making Europe, and DGA-FSE 2023-E21-23R by the Government of Arag\'{o}n, Spain, and the European Union NextGenerationEU Recovery and Resilience Program on Astrof\'{i}sica y F\'{i}sica de Altas Energ\'{i}as CEFCA-CAPA-ITAINNOVA. Additionally, Mathieu Kaltschmidt is supported by the Government of Aragón, Spain, with a PhD fellowship as specified in DGA-ORDEN-CUS/702/2022. 

The work of Ken'ichi Saikawa is supported by JSPS KAKENHI Grant Number JP24K07015.

Edoardo Vitagliano and Alessandro Lella are supported by the Italian MUR through the FIS 2 project FIS-2023-01577 (DD n. 23314 10-12-2024, CUP C53C24001460001).

Giuseppe Lucente (SLAC) acknowledges support from the U.S. Department of Energy under contract number DE-AC02-76SF00515.

David J. E. Marsh is supported by an Ernest Rutherford Fellowship (Grant No. ST/T004037/1) and a consolidator grant (Grant No. ST/X000753/1) from the Science and Technologies Facilities Council, United Kingdom.

The work of Alessandro Mirizzi is partially supported by the research grant number 2022E2J4RK "PANTHEON: Perspectives in Astroparticle and
Neutrino THEory with Old and New messengers" under the program PRIN 2022 (Mission 4, Component 1,
CUP I53D23001110006) funded by the Italian Ministero dell'Universit\`a e della Ricerca (MUR) and by the European Union – Next Generation EU. 

The work of Francesco D'Eramo and Edoardo Vitagliano is supported in part by the Italian MUR Departments of Excellence grant 2023-2027 ``Quantum Frontiers''.


The work of Luca Di Luzio and Federico Mescia  is partially supported by the European Union - Next Generation EU and by the Italian Ministry of University and Research (MUR) via the PRIN 2022 project n. 2022K4B58X - AxionOrigins

The work of Francesco D'Eramo, Alessandro Lella, Federico Mescia, Alessandro Mirizzi and Edoardo Vitagliano is supported by Istituto Nazionale di Fisica Nucleare (INFN) through the Theoretical Astroparticle Physics (TAsP) project.

Alicia Gomez thanks the Spanish Agencia Estatal de Investigaci\'on (AEI, MICIU) for their support under the Project  PID2022-137779OB-C41 financed by MCIN /AEI/10.13039 /501100011033/FEDER, EU and from “Tecnologías avanzadas para la exploración del Universo y sus componentes” (PR47/21 TAU-CM) project funded by Comunidad de Madrid and “NextGenerationEU”/PRTR.

The work of Diego Guadagnoli has received funding from the French ANR, under contract ANR-23-CE31-0018 (`InvISYble'), that is gratefully acknowledged.

Bradley J.~Kavanagh thanks the Spanish Agencia Estatal de Investigaci\'on (AEI, MICIU) for their support under the Project \textsc{DMpheno2lab} (PID2022-139494NB-I00) financed by MCIN /AEI /10.13039/501100011033 / FEDER, EU.

Elisa Todarello has received funding from the European Union’s Horizon 2020 research and innovation programme under the Marie Skłodowska-Curie grant agreement No. 101204903, and from the  STFC Consolidated Grant [Grant No. ST/T000732/1]. 

Nicole Righi is supported by the ERC NOTIMEFORCOSMO, 101126304.

The work in Su-Yang Xu group was supported by CATS, an Energy Frontier Research Center (EFRC) funded by the US Department of Energy (DOE) Office of Science, through the Ames National Laboratory under contract DE-AC0207CH11358, the Office of Naval Research (ONR) grant the N000142512285, and the National Science Foundation (NSF) Career Grant No. DMR-2143177, the Air Force Office of Scientific Research (AFOSR) grant FA9550-23-1-0040.

Damiano F. G. Fiorillo was supported by the Alexander von Humboldt Foundation (Germany) for most of the completion of the project.


\appendix

\section{List of acronyms}
\label{app:acronyms}

\begin{center}
\begin{longtable}{@{}ll@{}}
\caption{List of acronyms used in this work.} \label{tab:acronyms} \\
\toprule
\textbf{Acronym} & \textbf{Meaning} \\
\midrule
\endfirsthead

\toprule
\textbf{Acronym} & \textbf{Meaning} \\
\midrule
\endhead

\bottomrule
\endfoot

ABC  & Atomic, Bremsstrahlung and Compton processes \\
ALP  & Axion-Like Particle \\
AMR  & Adaptive Mesh Refinement \\
AQ   & Axion Quasiparticle \\
BBN  & Big-Bang Nucleosynthesis \\
BSG  & Blue Supergiant \\
CADEx & Canfranc Axion Detection Experiment \\
CAST & CERN Axion Solar Telescope \\
CDM  & Cold Dark Matter \\
ChPT & Chiral Perturbation Theory \\
CL   & Confidence Level \\
CMB  & Cosmic Microwave Background \\
CP   & Charge-Parity (symmetry/violation) \\
DFSZ & Dine--Fischler--Srednicki--Zhitnitsky  (benchmark axion model)\\
DIGA & Dilute Instanton Gas Approximation \\
DM   & Dark Matter \\
DW   & Domain Wall \\
EDM  & Electric Dipole Moment \\
EFT  & Effective Field Theory \\
Fermi-LAT & Fermi Large Area Telescope \\
FLRW & Friedmann-Lema\^{i}tre-Robertson Walker (metric) \\
FRST & Fine, Regular, and Star (triangulations in string theory) \\
FWHM & Full Width at Half Maximum \\
GLSM & Gauged Linear Sigma Model \\
HB   & Horizontal Branch (star) \\
HK   & Hyper-Kamiokande \\
HPC  & High-Performance Computing \\
IAXO & International Axion Observatory \\
IIB  & Type IIB (string theory) \\
KIDs & Kinetic Inductance Detectors \\
KM   & Kinetic Misalignment \\
KSVZ & Kim--Shifman--Vainshtein--Zakharov (benchmark axion model)\\
LHC  & Large Hadron Collider \\
LMA  & Large Misalignment Angle \\
LSC  & Laboratorio Subterr\'aneo de Canfranc \\
nEDM & neutron Electric Dipole Moment \\
NEP  & Noise Equivalent Power \\
NG   & Nambu--Goto \\
NS   & Neutron Star \\
NSM  & Neutron Star Merger \\
PNS  & Proto-Neutron Star \\
PQ   & Peccei--Quinn \\
QCD  & Quantum Chromodynamics \\
RD   & Radiation Domination \\
RSG  & Red Supergiant \\
SM   & Standard Model \\
SMM  & Solar Maximum Mission \\
SN   & Supernova \\
SQL  & Standard Quantum Limit \\
SUSY & Supersymmetry \\
UV   & Ultraviolet \\
VEV  & Vacuum Expectation Value \\
WD   & White Dwarf \\
\end{longtable}
\end{center}

\bibliography{bibbo}

@article{Raffelt:2011ft,
    author = "Raffelt, Georg G. and Redondo, Javier and Viaux Maira, Nicolas",
    title = "{The meV mass frontier of axion physics}",
    eprint = "1110.6397",
    archivePrefix = "arXiv",
    primaryClass = "hep-ph",
    reportNumber = "MPP-2011-11",
    doi = "10.1103/PhysRevD.84.103008",
    journal = "Phys. Rev. D",
    volume = "84",
    pages = "103008",
    year = "2011"
}

@article{Arkani-Hamed:2003xts,
    author = "Arkani-Hamed, Nima and Cheng, Hsin-Chia and Creminelli, Paolo and Randall, Lisa",
    title = "{Extra natural inflation}",
    eprint = "hep-th/0301218",
    archivePrefix = "arXiv",
    reportNumber = "HUTP-03-A006",
    doi = "10.1103/PhysRevLett.90.221302",
    journal = "Phys. Rev. Lett.",
    volume = "90",
    pages = "221302",
    year = "2003"
}

@article{Aja:2022csb,
    author = "Aja, Beatriz and others",
    title = "{The Canfranc Axion Detection Experiment (CADEx): search for axions at 90 GHz with Kinetic Inductance Detectors}",
    eprint = "2206.02980",
    archivePrefix = "arXiv",
    primaryClass = "hep-ex",
    doi = "10.1088/1475-7516/2022/11/044",
    journal = "JCAP",
    volume = "11",
    pages = "044",
    year = "2022"
}

@article{Aja:2025pul,
    author = "Aja, Beatriz and others",
    title = "{The CADEx Experiment: A new haloscope axion search in the 330-460 micro-eV mass range at the Canfranc Underground Laboratory (LSC)}",
    doi = "10.22323/1.474.0039",
    journal = "PoS",
    volume = "COSMICWISPers2024",
    pages = "039",
    year = "2025"
}

@article{Catinari:2024ekq,
    author = "Catinari, Pier Giuseppe and Esposito, Angelo and Pavaskar, Shashin",
    title = "{Hunting axion dark matter with antiferromagnets: A case study with nickel oxide}",
    eprint = "2411.11971",
    archivePrefix = "arXiv",
    primaryClass = "hep-ph",
    doi = "10.1103/k6pg-bkwh",
    journal = "Phys. Rev. D",
    volume = "112",
    number = "3",
    pages = "035007",
    year = "2025"
}

@article{Fan:2024mhm,
    author = "Fan, Xing and Gabrielse, Gerald and Graham, Peter W. and Ramani, Harikrishnan and Wong, Samuel S. Y. and Xiao, Yawen",
    title = "{Highly excited electron cyclotron for QCD axion and dark-photon detection}",
    eprint = "2410.05549",
    archivePrefix = "arXiv",
    primaryClass = "hep-ph",
    reportNumber = "FERMILAB-PUB-24-0893-SQMS-V",
    doi = "10.1103/PhysRevD.111.075022",
    journal = "Phys. Rev. D",
    volume = "111",
    number = "7",
    pages = "075022",
    year = "2025"
}

@article{Grossman:2025cov,
    author = "Grossman, Yuval and Yu, Bingrong and Zhou, Siyu",
    title = "{Axion forces in axion backgrounds}",
    eprint = "2504.00104",
    archivePrefix = "arXiv",
    primaryClass = "hep-ph",
    doi = "10.1007/JHEP01(2026)145",
    journal = "JHEP",
    volume = "01",
    pages = "145",
    year = "2026"
}

@article{ADMX:2018gho,
    author = "Du, N. and others",
    collaboration = "ADMX",
    title = "{A Search for Invisible Axion Dark Matter with the Axion Dark Matter Experiment}",
    eprint = "1804.05750",
    archivePrefix = "arXiv",
    primaryClass = "hep-ex",
    reportNumber = "FERMILAB-PUB-18-101-AD-AE",
    doi = "10.1103/PhysRevLett.120.151301",
    journal = "Phys. Rev. Lett.",
    volume = "120",
    number = "15",
    pages = "151301",
    year = "2018"
}

@article{ADMX:2019uok,
    author = "Braine, T. and others",
    collaboration = "ADMX",
    title = "{Extended Search for the Invisible Axion with the Axion Dark Matter Experiment}",
    eprint = "1910.08638",
    archivePrefix = "arXiv",
    primaryClass = "hep-ex",
    reportNumber = "FERMILAB-PUB-19-569-AD-AE-PPD",
    doi = "10.1103/PhysRevLett.124.101303",
    journal = "Phys. Rev. Lett.",
    volume = "124",
    number = "10",
    pages = "101303",
    year = "2020"
}

@article{ADMX:2021nhd,
    author = "Bartram, C. and others",
    collaboration = "ADMX",
    title = "{Search for Invisible Axion Dark Matter in the 3.3\textendash{}4.2\,\,\ensuremath{\mu}eV Mass Range}",
    eprint = "2110.06096",
    archivePrefix = "arXiv",
    primaryClass = "hep-ex",
    reportNumber = "FERMILAB-PUB-21-774-DI-PPD-SQMS",
    doi = "10.1103/PhysRevLett.127.261803",
    journal = "Phys. Rev. Lett.",
    volume = "127",
    number = "26",
    pages = "261803",
    year = "2021"
}

@article{HAYSTAC:2018rwy,
    author = "Zhong, L. and others",
    collaboration = "HAYSTAC",
    title = "{Results from phase 1 of the HAYSTAC microwave cavity axion experiment}",
    eprint = "1803.03690",
    archivePrefix = "arXiv",
    primaryClass = "hep-ex",
    doi = "10.1103/PhysRevD.97.092001",
    journal = "Phys. Rev. D",
    volume = "97",
    number = "9",
    pages = "092001",
    year = "2018"
}

@article{HAYSTAC:2020kwv,
    author = "Backes, K. M. and others",
    collaboration = "HAYSTAC",
    title = "{A quantum-enhanced search for dark matter axions}",
    eprint = "2008.01853",
    archivePrefix = "arXiv",
    primaryClass = "quant-ph",
    doi = "10.1038/s41586-021-03226-7",
    journal = "Nature",
    volume = "590",
    number = "7845",
    pages = "238--242",
    year = "2021"
}

@article{HAYSTAC:2023cam,
    author = "Jewell, M. J. and others",
    collaboration = "HAYSTAC",
    title = "{New Results from HAYSTAC's Phase II Operation with a Squeezed State Receiver}",
    eprint = "2301.09721",
    archivePrefix = "arXiv",
    primaryClass = "hep-ex",
    month = "1",
    year = "2023"
}

@article{HAYSTAC:2024jch,
    author = "Bai, Xiran and others",
    collaboration = "HAYSTAC",
    title = "{Dark Matter Axion Search with HAYSTAC Phase II}",
    eprint = "2409.08998",
    archivePrefix = "arXiv",
    primaryClass = "hep-ex",
    month = "9",
    year = "2024"
}

@article{Kim:2022hmg,
    author = "Kim, Jinsu and others",
    title = "{Near-Quantum-Noise Axion Dark Matter Search at CAPP around 9.5 $\mu$eV}",
    eprint = "2207.13597",
    archivePrefix = "arXiv",
    primaryClass = "hep-ex",
    month = "7",
    year = "2022"
}

@article{CAPP:2024dtx,
    author = "Ahn, Saebyeok and others",
    collaboration = "CAPP",
    title = "{Extensive search for axion dark matter over 1 GHz with CAPP's Main Axion eXperiment}",
    eprint = "2402.12892",
    archivePrefix = "arXiv",
    primaryClass = "hep-ex",
    month = "2",
    year = "2024"
}

@article{McAllister:2017lkb,
    author = "McAllister, Ben T. and Flower, Graeme and Ivanov, Eugene N. and Goryachev, Maxim and Bourhill, Jeremy and Tobar, Michael E.",
    title = "{The ORGAN Experiment: An axion haloscope above 15 GHz}",
    eprint = "1706.00209",
    archivePrefix = "arXiv",
    primaryClass = "physics.ins-det",
    doi = "10.1016/j.dark.2017.09.010",
    journal = "Phys. Dark Univ.",
    volume = "18",
    pages = "67--72",
    year = "2017"
}

@article{McAllister:2022ibe,
    author = "McAllister, Ben T. and Quiskamp, Aaron and O'Hare, Ciaran and Altin, Paul and Ivanov, Eugene and Goryachev, Maxim and Tobar, Michael",
    title = "{Limits on Dark Photons, Scalars, and Axion-Electromagnetodynamics with The ORGAN Experiment}",
    eprint = "2212.01971",
    archivePrefix = "arXiv",
    primaryClass = "hep-ph",
    month = "12",
    year = "2022"
}

@article{Quiskamp:2024oet,
    author = "Quiskamp, Aaron P. and Flower, Graeme and Samuels, Steven and McAllister, Ben T. and Altin, Paul and Ivanov, Eugene N. and Goryachev, Maxim and Tobar, Michael E.",
    title = "{Near-quantum limited axion dark matter search with the ORGAN experiment around 26 $\mu$eV}",
    eprint = "2407.18586",
    archivePrefix = "arXiv",
    primaryClass = "hep-ex",
    month = "7",
    year = "2024"
}

@article{Quiskamp:2022pks,
    author = "Quiskamp, Aaron P. and McAllister, Ben T. and Altin, Paul and Ivanov, Eugene N. and Goryachev, Maxim and Tobar, Michael E.",
    title = "{Direct search for dark matter axions excluding ALP cogenesis in the 63- to 67-\ensuremath{\mu}eV range with the ORGAN experiment}",
    eprint = "2203.12152",
    archivePrefix = "arXiv",
    primaryClass = "hep-ex",
    doi = "10.1126/sciadv.abq3765",
    journal = "Sci. Adv.",
    volume = "8",
    number = "27",
    pages = "abq3765",
    year = "2022"
}

@article{CAST:2020rlf,
    author = "Melc\'on, A. \'Alvarez and others",
    collaboration = "CAST",
    title = "{First results of the CAST-RADES haloscope search for axions at 34.67 $\mu$eV}",
    eprint = "2104.13798",
    archivePrefix = "arXiv",
    primaryClass = "hep-ex",
    reportNumber = "CERN-EP-2021-070",
    doi = "10.1007/JHEP10(2021)075",
    journal = "JHEP",
    volume = "21",
    pages = "075",
    year = "2020"
}

@article{Ahyoune:2024klt,
    author = "Ahyoune, S. and others",
    title = "{RADES axion search results with a High-Temperature Superconducting cavity in an 11.7 T magnet}",
    eprint = "2403.07790",
    archivePrefix = "arXiv",
    primaryClass = "hep-ex",
    reportNumber = "CERN-EP-2024-076, MPP-2024-55",
    month = "3",
    year = "2024"
}

@article{TASEH:2022vvu,
    author = "Chang, Hsin and others",
    collaboration = "TASEH",
    title = "{First Results from the Taiwan Axion Search Experiment with a Haloscope at 19.6\,\,\ensuremath{\mu}eV}",
    eprint = "2205.05574",
    archivePrefix = "arXiv",
    primaryClass = "hep-ex",
    doi = "10.1103/PhysRevLett.129.111802",
    journal = "Phys. Rev. Lett.",
    volume = "129",
    number = "11",
    pages = "111802",
    year = "2022"
}

@article{Grenet:2021vbb,
    author = "Grenet, Thierry and Ballou, Rafik and Basto, Quentin and Martineau, Killian and Perrier, Pierre and Pugnat, Pierre and Quevillon, J\'er\'emie and Roch, Nicolas and Smith, Christopher",
    title = "{The Grenoble Axion Haloscope platform (GrAHal): development plan and first results}",
    eprint = "2110.14406",
    archivePrefix = "arXiv",
    primaryClass = "hep-ex",
    month = "10",
    year = "2021"
}

@article{Hagmann:1996qd,
    author = "Hagmann, C. and others",
    editor = "Cline, D. B.",
    title = "{First results from a second generation galactic axion experiment}",
    eprint = "astro-ph/9607022",
    archivePrefix = "arXiv",
    reportNumber = "UCRL-JC-124162",
    doi = "10.1016/S0920-5632(96)00516-6",
    journal = "Nucl. Phys. B Proc. Suppl.",
    volume = "51",
    pages = "209--212",
    year = "1996"
}

@article{Wuensch:1989sa,
    author = "Wuensch, Walter and De Panfilis-Wuensch, S. and Semertzidis, Y. K. and Rogers, J. T. and Melissinos, A. C. and Halama, H. J. and Moskowitz, B. E. and Prodell, A. G. and Fowler, W. B. and Nezrick, F. A.",
    title = "{Results of a Laboratory Search for Cosmic Axions and Other Weakly Coupled Light Particles}",
    reportNumber = "FERMILAB-PUB-89-185-E, BNL-43010",
    doi = "10.1103/PhysRevD.40.3153",
    journal = "Phys. Rev. D",
    volume = "40",
    pages = "3153",
    year = "1989"
}

@article{DePanfilis,
  title = {Limits on the abundance and coupling of cosmic axions at 4.5$<{m}_{a}<$5.0 \ensuremath{\mu}eV},
  author = {DePanfilis, S. and Melissinos, A. C. and Moskowitz, B. E. and Rogers, J. T. and Semertzidis, Y. K. and Wuensch, W. U. and Halama, H. J. and Prodell, A. G. and Fowler, W. B. and Nezrick, F. A.},
  journal = {Phys. Rev. Lett.},
  volume = {59},
  issue = {7},
  pages = {839--842},
  numpages = {0},
  year = {1987},
  month = {Aug},
  publisher = {American Physical Society},
  doi = {10.1103/PhysRevLett.59.839},
  url = {https://link.aps.org/doi/10.1103/PhysRevLett.59.839}
}

@article{Garcia:2024xzc,
    author = "Garcia, B. Ary dos Santos and others",
    title = "{First search for axion dark matter with a Madmax prototype}",
    eprint = "2409.11777",
    archivePrefix = "arXiv",
    primaryClass = "hep-ex",
    month = "9",
    year = "2024"
}

@article{GigaBREAD:2025lzq,
    author = "Hoshino, Gabe and others",
    collaboration = "GigaBREAD",
    title = "{First Axionlike Particle Results from a Broadband Search for Wavelike Dark Matter in the 44 to 52{\,}{\,}{\ensuremath{\mu}}eV Range with a Coaxial Dish Antenna}",
    eprint = "2501.17119",
    archivePrefix = "arXiv",
    primaryClass = "hep-ex",
    reportNumber = "FERMILAB-PUB-25-0076-PPD",
    doi = "10.1103/PhysRevLett.134.171002",
    journal = "Phys. Rev. Lett.",
    volume = "134",
    number = "17",
    pages = "171002",
    year = "2025"
}

@article{Berlin:2023ubt,
    author = "Berlin, Asher and Millar, Alexander J. and Trickle, Tanner and Zhou, Kevin",
    title = "{Physical signatures of fermion-coupled axion dark matter}",
    eprint = "2312.11601",
    archivePrefix = "arXiv",
    primaryClass = "hep-ph",
    reportNumber = "FERMILAB-PUB-23-779-SQMS-T, SLAC-PUB-17758",
    doi = "10.1007/JHEP05(2024)314",
    journal = "JHEP",
    volume = "05",
    pages = "314",
    year = "2024"
}

@article{Mitridate:2020kly,
    author = "Mitridate, Andrea and Trickle, Tanner and Zhang, Zhengkang and Zurek, Kathryn M.",
    title = "{Detectability of Axion Dark Matter with Phonon Polaritons and Magnons}",
    eprint = "2005.10256",
    archivePrefix = "arXiv",
    primaryClass = "hep-ph",
    doi = "10.1103/PhysRevD.102.095005",
    journal = "Phys. Rev. D",
    volume = "102",
    number = "9",
    pages = "095005",
    year = "2020"
}

@article{Arvanitaki:2024php,
    author = "Arvanitaki, Asimina and Engel, Jonathan and Geraci, Andrew A. and Madden, Amalia and Hepburn, Alexander and Van Tilburg, Ken",
    title = "{Detecting the QCD Axion via the Ferroaxionic Force with Piezoelectric Materials}",
    eprint = "2411.10516",
    archivePrefix = "arXiv",
    primaryClass = "hep-ph",
    doi = "10.1103/6xw1-1715",
    journal = "Phys. Rev. Lett.",
    volume = "136",
    number = "8",
    pages = "081803",
    year = "2026"
}

@article{Arvanitaki:2014dfa,
    author = "Arvanitaki, Asimina and Geraci, Andrew A.",
    title = "{Resonantly Detecting Axion-Mediated Forces with Nuclear Magnetic Resonance}",
    eprint = "1403.1290",
    archivePrefix = "arXiv",
    primaryClass = "hep-ph",
    doi = "10.1103/PhysRevLett.113.161801",
    journal = "Phys. Rev. Lett.",
    volume = "113",
    number = "16",
    pages = "161801",
    year = "2014"
}

@article{ARIADNE:2017tdd,
    author = "Geraci, A. A. and others",
    editor = "Carosi, Gianpaolo and Rybka, Gray and van Bibber, Karl",
    collaboration = "ARIADNE",
    title = "{Progress on the ARIADNE axion experiment}",
    eprint = "1710.05413",
    archivePrefix = "arXiv",
    primaryClass = "astro-ph.IM",
    doi = "10.1007/978-3-319-92726-8_18",
    journal = "Springer Proc. Phys.",
    volume = "211",
    pages = "151--161",
    year = "2018"
}

@article{ALPHA:2022rxj,
    author = "Millar, Alexander J. and others",
    collaboration = "ALPHA",
    title = "{Searching for dark matter with plasma haloscopes}",
    eprint = "2210.00017",
    archivePrefix = "arXiv",
    primaryClass = "hep-ph",
    reportNumber = "FERMILAB-PUB-22-739-T",
    doi = "10.1103/PhysRevD.107.055013",
    journal = "Phys. Rev. D",
    volume = "107",
    number = "5",
    pages = "055013",
    year = "2023"
}

@article{Lawson:2019brd,
    author = "Lawson, Matthew and Millar, Alexander J. and Pancaldi, Matteo and Vitagliano, Edoardo and Wilczek, Frank",
    title = "{Tunable axion plasma haloscopes}",
    eprint = "1904.11872",
    archivePrefix = "arXiv",
    primaryClass = "hep-ph",
    reportNumber = "NORDITA-2019-038, MIT-CTP-5116, MPP-2019-83",
    doi = "10.1103/PhysRevLett.123.141802",
    journal = "Phys. Rev. Lett.",
    volume = "123",
    number = "14",
    pages = "141802",
    year = "2019"
}

@article{Chiles:2021gxk,
    author = "Chiles, Jeff and others",
    title = "{New Constraints on Dark Photon Dark Matter with Superconducting Nanowire Detectors in an Optical Haloscope}",
    eprint = "2110.01582",
    archivePrefix = "arXiv",
    primaryClass = "hep-ex",
    doi = "10.1103/PhysRevLett.128.231802",
    journal = "Phys. Rev. Lett.",
    volume = "128",
    number = "23",
    pages = "231802",
    year = "2022"
}

@article{Barbieri:2016vwg,
    author = "Barbieri, R. and Braggio, C. and Carugno, G. and Gallo, C. S. and Lombardi, A. and Ortolan, A. and Pengo, R. and Ruoso, G. and Speake, C. C.",
    title = "{Searching for galactic axions through magnetized media: the QUAX proposal}",
    eprint = "1606.02201",
    archivePrefix = "arXiv",
    primaryClass = "hep-ph",
    doi = "10.1016/j.dark.2017.01.003",
    journal = "Phys. Dark Univ.",
    volume = "15",
    pages = "135--141",
    year = "2017"
}

@article{QUAX:2023gop,
    author = "Di Vora, R. and others",
    collaboration = "QUAX",
    title = "{Search for galactic axions with a traveling wave parametric amplifier}",
    eprint = "2304.07505",
    archivePrefix = "arXiv",
    primaryClass = "hep-ex",
    reportNumber = "FERMILAB-PUB-23-229-SQMS-V",
    doi = "10.1103/PhysRevD.108.062005",
    journal = "Phys. Rev. D",
    volume = "108",
    number = "6",
    pages = "062005",
    year = "2023"
}

@article{QUAX:2024fut,
    author = "Rettaroli, A. and others",
    collaboration = "QUAX",
    title = "{Search for axion dark matter with the QUAX{\textendash}LNF tunable haloscope}",
    eprint = "2402.19063",
    archivePrefix = "arXiv",
    primaryClass = "physics.ins-det",
    reportNumber = "FERMILAB-PUB-24-0511-SQMS-V",
    doi = "10.1103/PhysRevD.110.022008",
    journal = "Phys. Rev. D",
    volume = "110",
    number = "2",
    pages = "022008",
    year = "2024"
}

@article{Choi:2003wr,
    author = "Choi, Ki-woon",
    title = "{A QCD axion from higher dimensional gauge field}",
    eprint = "hep-ph/0308024",
    archivePrefix = "arXiv",
    reportNumber = "KAIST-TH-2003-07",
    doi = "10.1103/PhysRevLett.92.101602",
    journal = "Phys. Rev. Lett.",
    volume = "92",
    pages = "101602",
    year = "2004"
}

@article{Craig:2024dnl,
    author = "Craig, Nathaniel and Kongsore, Marius",
    title = "{High-quality axions from higher-form symmetries in extra dimensions}",
    eprint = "2408.10295",
    archivePrefix = "arXiv",
    primaryClass = "hep-ph",
    doi = "10.1103/PhysRevD.111.015047",
    journal = "Phys. Rev. D",
    volume = "111",
    number = "1",
    pages = "015047",
    year = "2025"
}

@article{Mehta:2020kwu,
    author = "Mehta, Viraf M. and Demirtas, Mehmet and Long, Cody and Marsh, David J. E. and Mcallister, Liam and Stott, Matthew J.",
    title = "{Superradiance Exclusions in the Landscape of Type IIB String Theory}",
    eprint = "2011.08693",
    archivePrefix = "arXiv",
    primaryClass = "hep-th",
    reportNumber = "KCL-PH-TH/2020-77",
    month = "11",
    year = "2020"
}

@article{Gorghetto:2018ocs,
    author = "Gorghetto, Marco and Villadoro, Giovanni",
    title = "{Topological Susceptibility and QCD Axion Mass: QED and NNLO corrections}",
    eprint = "1812.01008",
    archivePrefix = "arXiv",
    primaryClass = "hep-ph",
    doi = "10.1007/JHEP03(2019)033",
    journal = "JHEP",
    volume = "03",
    pages = "033",
    year = "2019"
}

@article{Srednicki:1985xd,
    author = "Srednicki, Mark",
    title = "{Axion Couplings to Matter. 1. CP Conserving Parts}",
    reportNumber = "Print-85-0247 (UC,SANTA BARBARA)",
    doi = "10.1016/0550-3213(85)90054-9",
    journal = "Nucl. Phys. B",
    volume = "260",
    pages = "689--700",
    year = "1985"
}

@article{GrillidiCortona:2015jxo,
    author = "Grilli di Cortona, Giovanni and Hardy, Edward and Pardo Vega, Javier and Villadoro, Giovanni",
    title = "{The QCD axion, precisely}",
    eprint = "1511.02867",
    archivePrefix = "arXiv",
    primaryClass = "hep-ph",
    doi = "10.1007/JHEP01(2016)034",
    journal = "JHEP",
    volume = "01",
    pages = "034",
    year = "2016"
}

@article{DiLuzio:2016sbl,
    author = "Di Luzio, Luca and Mescia, Federico and Nardi, Enrico",
    title = "{Redefining the Axion Window}",
    eprint = "1610.07593",
    archivePrefix = "arXiv",
    primaryClass = "hep-ph",
    reportNumber = "IPPP-16-99",
    doi = "10.1103/PhysRevLett.118.031801",
    journal = "Phys. Rev. Lett.",
    volume = "118",
    number = "3",
    pages = "031801",
    year = "2017"
}

@article{DiLuzio:2017pfr,
    author = "Di Luzio, Luca and Mescia, Federico and Nardi, Enrico",
    title = "{Window for preferred axion models}",
    eprint = "1705.05370",
    archivePrefix = "arXiv",
    primaryClass = "hep-ph",
    reportNumber = "IPPP-17-41",
    doi = "10.1103/PhysRevD.96.075003",
    journal = "Phys. Rev. D",
    volume = "96",
    number = "7",
    pages = "075003",
    year = "2017"
}

@article{Choi:2017gpf,
    author = "Choi, Kiwoon and Im, Sang Hui and Park, Chan Beom and Yun, Seokhoon",
    title = "{Minimal Flavor Violation with Axion-like Particles}",
    eprint = "1708.00021",
    archivePrefix = "arXiv",
    primaryClass = "hep-ph",
    reportNumber = "CTPU-17-30, KIAS-P17058",
    doi = "10.1007/JHEP11(2017)070",
    journal = "JHEP",
    volume = "11",
    pages = "070",
    year = "2017"
}

@article{Chala:2020wvs,
    author = "Chala, Mikael and Guedes, Guilherme and Ramos, Maria and Santiago, Jose",
    title = "{Running in the ALPs}",
    eprint = "2012.09017",
    archivePrefix = "arXiv",
    primaryClass = "hep-ph",
    doi = "10.1140/epjc/s10052-021-08968-2",
    journal = "Eur. Phys. J. C",
    volume = "81",
    number = "2",
    pages = "181",
    year = "2021"
}

@article{Bauer:2020jbp,
    author = "Bauer, Martin and Neubert, Matthias and Renner, Sophie and Schnubel, Marvin and Thamm, Andrea",
    title = "{The Low-Energy Effective Theory of Axions and ALPs}",
    eprint = "2012.12272",
    archivePrefix = "arXiv",
    primaryClass = "hep-ph",
    reportNumber = "IPPP/20/69, MITP/20-070 SISSA 30/2020/FISI, ZH-TH-47/20",
    doi = "10.1007/JHEP04(2021)063",
    journal = "JHEP",
    volume = "04",
    pages = "063",
    year = "2021"
}

@article{Choi:2021kuy,
    author = "Choi, Kiwoon and Im, Sang Hui and Kim, Hee Jung and Seong, Hyeonseok",
    title = "{Precision axion physics with running axion couplings}",
    eprint = "2106.05816",
    archivePrefix = "arXiv",
    primaryClass = "hep-ph",
    reportNumber = "CTPU-PTC-21-26",
    doi = "10.1007/JHEP08(2021)058",
    journal = "JHEP",
    volume = "08",
    pages = "058",
    year = "2021"
}

@article{DiLuzio:2023tqe,
    author = "Di Luzio, Luca and Giannotti, Maurizio and Mescia, Federico and Nardi, Enrico and Okawa, Shohei and Piazza, Gioacchino",
    title = "{Running effects on QCD axion phenomenology}",
    eprint = "2305.11958",
    archivePrefix = "arXiv",
    primaryClass = "hep-ph",
    doi = "10.1103/PhysRevD.108.115004",
    journal = "Phys. Rev. D",
    volume = "108",
    number = "11",
    pages = "115004",
    year = "2023"
}

@article{DiLuzio:2022tyc,
    author = "Di Luzio, Luca and Mescia, Federico and Nardi, Enrico and Okawa, Shohei",
    title = "{Renormalization group effects in astrophobic axion models}",
    eprint = "2205.15326",
    archivePrefix = "arXiv",
    primaryClass = "hep-ph",
    doi = "10.1103/PhysRevD.106.055016",
    journal = "Phys. Rev. D",
    volume = "106",
    number = "5",
    pages = "055016",
    year = "2022"
}

@article{Dine:1982ah,
      author         = "Dine, Michael and Fischler, Willy",
      title          = "{The Not So Harmless Axion}",
      journal        = "Phys. Lett.",
      volume         = "120B",
      year           = "1983",
      pages          = "137-141",
      doi            = "10.1016/0370-2693(83)90639-1",
      reportNumber   = "UPR-0201T",
      SLACcitation   = "%%CITATION = PHLTA,120B,137;%%"
}

@article{Abbott:1982af,
      author         = "Abbott, L. F. and Sikivie, P.",
      title          = "{A Cosmological Bound on the Invisible Axion}",
      journal        = "Phys. Lett.",
      volume         = "120B",
      year           = "1983",
      pages          = "133-136",
      doi            = "10.1016/0370-2693(83)90638-X",
      reportNumber   = "PRINT-82-0695 (BRANDEIS)",
      SLACcitation   = "%%CITATION = PHLTA,120B,133;%%"
}

@article{Preskill:1982cy,
      author         = "Preskill, John and Wise, Mark B. and Wilczek, Frank",
      title          = "{Cosmology of the Invisible Axion}",
      journal        = "Phys. Lett.",
      volume         = "120B",
      year           = "1983",
      pages          = "127-132",
      doi            = "10.1016/0370-2693(83)90637-8",
      reportNumber   = "HUTP-82-A048, NSF-ITP-82-103",
      SLACcitation   = "%%CITATION = PHLTA,120B,127;%%"
}

@article{DiLuzio:2017ogq,
    author = "Di Luzio, Luca and Mescia, Federico and Nardi, Enrico and Panci, Paolo and Ziegler, Robert",
    title = "{Astrophobic Axions}",
    eprint = "1712.04940",
    archivePrefix = "arXiv",
    primaryClass = "hep-ph",
    reportNumber = "IPPP-17-102, CERN-TH-2017-256",
    doi = "10.1103/PhysRevLett.120.261803",
    journal = "Phys. Rev. Lett.",
    volume = "120",
    number = "26",
    pages = "261803",
    year = "2018"
}

@article{DiLuzio:2024xnt,
    author = "Di Luzio, Luca and Hoof, Sebastian and Marinissen, Coenraad and Plakkot, Vaisakh",
    title = "{Catalogues of cosmologically self-consistent hadronic QCD axion models}",
    eprint = "2412.17896",
    archivePrefix = "arXiv",
    primaryClass = "hep-ph",
    doi = "10.1088/1475-7516/2025/04/072",
    journal = "JCAP",
    volume = "04",
    pages = "072",
    year = "2025"
}

@article{Plakkot:2021xyx,
    author = "Plakkot, Vaisakh and Hoof, Sebastian",
    title = "{Anomaly ratio distributions of hadronic axion models with multiple heavy quarks}",
    eprint = "2107.12378",
    archivePrefix = "arXiv",
    primaryClass = "hep-ph",
    doi = "10.1103/PhysRevD.104.075017",
    journal = "Phys. Rev. D",
    volume = "104",
    number = "7",
    pages = "075017",
    year = "2021"
}

@article{Davis:1986xc,
    author = "Davis, Richard Lynn",
    title = "{Cosmic Axions from Cosmic Strings}",
    reportNumber = "SLAC-PUB-3895",
    doi = "10.1016/0370-2693(86)90300-X",
    journal = "Phys. Lett. B",
    volume = "180",
    pages = "225--230",
    year = "1986"
}

@article{Hiramatsu:2012gg,
    author = "Hiramatsu, Takashi and Kawasaki, Masahiro and Saikawa, Ken'ichi and Sekiguchi, Toyokazu",
    title = "{Production of dark matter axions from collapse of string-wall systems}",
    eprint = "1202.5851",
    archivePrefix = "arXiv",
    primaryClass = "hep-ph",
    reportNumber = "ICRR-REPORT-608-2011-25, IPMU12-0025, YITP-12-9",
    doi = "10.1103/PhysRevD.85.105020",
    journal = "Phys. Rev. D",
    volume = "85",
    pages = "105020",
    year = "2012",
    note = "[Erratum: Phys.Rev.D 86, 089902 (2012)]"
}

@article{Farina:2016tgd,
    author = "Farina, Marco and Pappadopulo, Duccio and Rompineve, Fabrizio and Tesi, Andrea",
    title = "{The photo-philic QCD axion}",
    eprint = "1611.09855",
    archivePrefix = "arXiv",
    primaryClass = "hep-ph",
    doi = "10.1007/JHEP01(2017)095",
    journal = "JHEP",
    volume = "01",
    pages = "095",
    year = "2017"
}

@article{Agrawal:2017cmd,
    author = "Agrawal, Prateek and Fan, JiJi and Reece, Matthew and Wang, Lian-Tao",
    title = "{Experimental Targets for Photon Couplings of the QCD Axion}",
    eprint = "1709.06085",
    archivePrefix = "arXiv",
    primaryClass = "hep-ph",
    doi = "10.1007/JHEP02(2018)006",
    journal = "JHEP",
    volume = "02",
    pages = "006",
    year = "2018"
}

@article{Darme:2020gyx,
    author = "Darm{\'e}, Luc and Di Luzio, Luca and Giannotti, Maurizio and Nardi, Enrico",
    title = "{Selective enhancement of the QCD axion couplings}",
    eprint = "2010.15846",
    archivePrefix = "arXiv",
    primaryClass = "hep-ph",
    reportNumber = "DESY-20-177, DESY 20-177",
    doi = "10.1103/PhysRevD.103.015034",
    journal = "Phys. Rev. D",
    volume = "103",
    number = "1",
    pages = "015034",
    year = "2021"
}

@article{Hook:2018jle,
    author = "Hook, Anson",
    title = "{Solving the Hierarchy Problem Discretely}",
    eprint = "1802.10093",
    archivePrefix = "arXiv",
    primaryClass = "hep-ph",
    doi = "10.1103/PhysRevLett.120.261802",
    journal = "Phys. Rev. Lett.",
    volume = "120",
    number = "26",
    pages = "261802",
    year = "2018"
}

@article{DiLuzio:2021pxd,
    author = "Di Luzio, Luca and Gavela, Belen and Quilez, Pablo and Ringwald, Andreas",
    title = "{An even lighter QCD axion}",
    eprint = "2102.00012",
    archivePrefix = "arXiv",
    primaryClass = "hep-ph",
    reportNumber = "DESY-21-010, DESY 21-010, IFT-UAM/CSIC-20-143, FTUAM-20-21",
    doi = "10.1007/JHEP05(2021)184",
    journal = "JHEP",
    volume = "05",
    pages = "184",
    year = "2021"
}

@article{DiLuzio:2021gos,
    author = "Di Luzio, Luca and Gavela, Belen and Quilez, Pablo and Ringwald, Andreas",
    title = "{Dark matter from an even lighter QCD axion: trapped misalignment}",
    eprint = "2102.01082",
    archivePrefix = "arXiv",
    primaryClass = "hep-ph",
    reportNumber = "DESY 21-011, DESY-21-011, IFT-UAM/CSIC-20-144, FTUAM-20-21",
    doi = "10.1088/1475-7516/2021/10/001",
    journal = "JCAP",
    volume = "10",
    pages = "001",
    year = "2021"
}

@article{Bjorkeroth:2019jtx,
    author = {Bj{\"o}rkeroth, Fredrik and Di Luzio, Luca and Mescia, Federico and Nardi, Enrico and Panci, Paolo and Ziegler, Robert},
    title = "{Axion-electron decoupling in nucleophobic axion models}",
    eprint = "1907.06575",
    archivePrefix = "arXiv",
    primaryClass = "hep-ph",
    reportNumber = "CERN-TH-2019-118, DESY-19-194",
    doi = "10.1103/PhysRevD.101.035027",
    journal = "Phys. Rev. D",
    volume = "101",
    number = "3",
    pages = "035027",
    year = "2020"
}

@article{Coughlan:1983ci,
    author = "Coughlan, G. D. and Fischler, W. and Kolb, Edward W. and Raby, S. and Ross, Graham G.",
    title = "{Cosmological Problems for the Polonyi Potential}",
    reportNumber = "LA-UR-83-1423",
    doi = "10.1016/0370-2693(83)91091-2",
    journal = "Phys. Lett. B",
    volume = "131",
    pages = "59--64",
    year = "1983"
}

@article{Banks:1993en,
    author = "Banks, Tom and Kaplan, David B. and Nelson, Ann E.",
    title = "{Cosmological implications of dynamical supersymmetry breaking}",
    eprint = "hep-ph/9308292",
    archivePrefix = "arXiv",
    reportNumber = "UCSD-PTH-93-26, RU-37",
    doi = "10.1103/PhysRevD.49.779",
    journal = "Phys. Rev. D",
    volume = "49",
    pages = "779--787",
    year = "1994"
}

@article{Choi:2010gm,
    author = "Choi, Kiwoon and Nilles, Hans Peter and Shin, Chang Sub and Trapletti, Michele",
    title = "{Sparticle Spectrum of Large Volume Compactification}",
    eprint = "1011.0999",
    archivePrefix = "arXiv",
    primaryClass = "hep-th",
    doi = "10.1007/JHEP02(2011)047",
    journal = "JHEP",
    volume = "02",
    pages = "047",
    year = "2011"
}

@article{Cicoli:2013cha,
    author = "Cicoli, Michele and Klevers, Denis and Krippendorf, Sven and Mayrhofer, Christoph and Quevedo, Fernando and Valandro, Roberto",
    title = "{Explicit de Sitter Flux Vacua for Global String Models with Chiral Matter}",
    eprint = "1312.0014",
    archivePrefix = "arXiv",
    primaryClass = "hep-th",
    doi = "10.1007/JHEP05(2014)001",
    journal = "JHEP",
    volume = "05",
    pages = "001",
    year = "2014"
}

@article{Allahverdi:2014ppa,
    author = "Allahverdi, Rouzbeh and Cicoli, Michele and Dutta, Bhaskar and Sinha, Kuver",
    title = "{Correlation between Dark Matter and Dark Radiation in String Compactifications}",
    eprint = "1401.4364",
    archivePrefix = "arXiv",
    primaryClass = "hep-ph",
    doi = "10.1088/1475-7516/2014/10/002",
    journal = "JCAP",
    volume = "10",
    pages = "002",
    year = "2014"
}

@article{Chauhan:2025rdj,
    author = "Chauhan, Aman and Cicoli, Michele and Krippendorf, Sven and Maharana, Anshuman and Piantadosi, Pellegrino and Schachner, Andreas",
    title = "{Deep observations of the Type IIB flux landscape}",
    eprint = "2501.03984",
    archivePrefix = "arXiv",
    primaryClass = "hep-th",
    doi = "10.1007/JHEP07(2025)271",
    journal = "JHEP",
    volume = "07",
    pages = "271",
    year = "2025"
}

@article{deCarlos:1993wie,
    author = "de Carlos, B. and Casas, J. A. and Quevedo, F. and Roulet, E.",
    title = "{Model independent properties and cosmological implications of the dilaton and moduli sectors of 4-d strings}",
    eprint = "hep-ph/9308325",
    archivePrefix = "arXiv",
    reportNumber = "CERN-TH-6958-93, NEIP-93-006, IEM-FT-75-93",
    doi = "10.1016/0370-2693(93)91538-X",
    journal = "Phys. Lett. B",
    volume = "318",
    pages = "447--456",
    year = "1993"
}

@article{Cicoli:2016chb,
    author = "Cicoli, Michele and Ciupke, David and de Alwis, Senarath and Muia, Francesco",
    title = "{$\alpha'$ Inflation: moduli stabilisation and observable tensors from higher derivatives}",
    eprint = "1607.01395",
    archivePrefix = "arXiv",
    primaryClass = "hep-th",
    reportNumber = "DESY-16-103",
    doi = "10.1007/JHEP09(2016)026",
    journal = "JHEP",
    volume = "09",
    pages = "026",
    year = "2016"
}

@article{Cicoli:2011qg,
    author = "Cicoli, Michele and Mayrhofer, Christoph and Valandro, Roberto",
    title = "{Moduli Stabilisation for Chiral Global Models}",
    eprint = "1110.3333",
    archivePrefix = "arXiv",
    primaryClass = "hep-th",
    reportNumber = "DESY-11-179, ZMP-HH-11-15",
    doi = "10.1007/JHEP02(2012)062",
    journal = "JHEP",
    volume = "02",
    pages = "062",
    year = "2012"
}

@article{Balasubramanian:2005zx,
    author = "Balasubramanian, Vijay and Berglund, Per and Conlon, Joseph P. and Quevedo, Fernando",
    title = "{Systematics of moduli stabilisation in Calabi-Yau flux compactifications}",
    eprint = "hep-th/0502058",
    archivePrefix = "arXiv",
    reportNumber = "DAMTP-2005-10, UNH-05-01, UPR-1109-T",
    doi = "10.1088/1126-6708/2005/03/007",
    journal = "JHEP",
    volume = "03",
    pages = "007",
    year = "2005"
}

@article{Cicoli:2024bxw,
    author = "Cicoli, Michele and Grassi, Antonella and Lacombe, Osmin and Pedro, Francisco G.",
    title = "{Chiral global embedding of Fibre Inflation with $ \overline{\textrm{D}3} $ uplift}",
    eprint = "2412.08723",
    archivePrefix = "arXiv",
    primaryClass = "hep-th",
    doi = "10.1007/JHEP06(2025)090",
    journal = "JHEP",
    volume = "06",
    pages = "090",
    year = "2025"
}

@article{Cicoli:2017axo,
    author = "Cicoli, Michele and Ciupke, David and Diaz, Victor A. and Guidetti, Veronica and Muia, Francesco and Shukla, Pramod",
    title = "{Chiral Global Embedding of Fibre Inflation Models}",
    eprint = "1709.01518",
    archivePrefix = "arXiv",
    primaryClass = "hep-th",
    doi = "10.1007/JHEP11(2017)207",
    journal = "JHEP",
    volume = "11",
    pages = "207",
    year = "2017"
}

@article{Cicoli:2016xae,
    author = "Cicoli, Michele and Muia, Francesco and Shukla, Pramod",
    title = "{Global Embedding of Fibre Inflation Models}",
    eprint = "1611.04612",
    archivePrefix = "arXiv",
    primaryClass = "hep-th",
    doi = "10.1007/JHEP11(2016)182",
    journal = "JHEP",
    volume = "11",
    pages = "182",
    year = "2016"
}

@article{Cicoli:2008va,
    author = "Cicoli, Michele and Conlon, Joseph P. and Quevedo, Fernando",
    title = "{General Analysis of LARGE Volume Scenarios with String Loop Moduli Stabilisation}",
    eprint = "0805.1029",
    archivePrefix = "arXiv",
    primaryClass = "hep-th",
    reportNumber = "DAMTP-2008-16",
    doi = "10.1088/1126-6708/2008/10/105",
    journal = "JHEP",
    volume = "10",
    pages = "105",
    year = "2008"
}

@article{Cicoli:2011yh,
    author = "Cicoli, Michele and Goodsell, Mark and Jaeckel, Joerg and Ringwald, Andreas",
    title = "{Testing String Vacua in the Lab: From a Hidden CMB to Dark Forces in Flux Compactifications}",
    eprint = "1103.3705",
    archivePrefix = "arXiv",
    primaryClass = "hep-th",
    reportNumber = "DESY-11-042, IPPP-11-13, DCTP-11-26",
    doi = "10.1007/JHEP07(2011)114",
    journal = "JHEP",
    volume = "07",
    pages = "114",
    year = "2011"
}

@article{Cicoli:2017zbx,
    author = "Cicoli, Michele and Diaz, Victor A. and Guidetti, Veronica and Rummel, Markus",
    title = "{The 3.5 keV Line from Stringy Axions}",
    eprint = "1707.02987",
    archivePrefix = "arXiv",
    primaryClass = "hep-th",
    doi = "10.1007/JHEP10(2017)192",
    journal = "JHEP",
    volume = "10",
    pages = "192",
    year = "2017"
}

@article{Broeckel:2021dpz,
    author = "Broeckel, Igor and Cicoli, Michele and Maharana, Anshuman and Singh, Kajal and Sinha, Kuver",
    title = "{Moduli stabilisation and the statistics of axion physics in the landscape}",
    eprint = "2105.02889",
    archivePrefix = "arXiv",
    primaryClass = "hep-th",
    doi = "10.1007/JHEP01(2022)191",
    journal = "JHEP",
    volume = "08",
    pages = "059",
    year = "2021",
    note = "[Addendum: JHEP 01, 191 (2022)]"
}

@article{Cicoli:2022fzy,
    author = "Cicoli, Michele and Hebecker, Arthur and Jaeckel, Joerg and Wittner, Manuel",
    title = "{Axions in string theory {\textemdash} slaying the Hydra of dark radiation}",
    eprint = "2203.08833",
    archivePrefix = "arXiv",
    primaryClass = "hep-th",
    doi = "10.1007/JHEP09(2022)198",
    journal = "JHEP",
    volume = "09",
    pages = "198",
    year = "2022"
}

@article{Conlon:2006tq,
    author = "Conlon, Joseph P.",
    title = "{The QCD axion and moduli stabilisation}",
    eprint = "hep-th/0602233",
    archivePrefix = "arXiv",
    reportNumber = "DAMTP-2006-17",
    doi = "10.1088/1126-6708/2006/05/078",
    journal = "JHEP",
    volume = "05",
    pages = "078",
    year = "2006"
}

@article{Cicoli:2012sz,
    author = "Cicoli, Michele and Goodsell, Mark and Ringwald, Andreas",
    title = "{The type IIB string axiverse and its low-energy phenomenology}",
    eprint = "1206.0819",
    archivePrefix = "arXiv",
    primaryClass = "hep-th",
    reportNumber = "DESY-12-058, CERN-PH-TH-2012-153",
    doi = "10.1007/JHEP10(2012)146",
    journal = "JHEP",
    volume = "10",
    pages = "146",
    year = "2012"
}

@article{Peccei:1977hh,
    author = "Peccei, R. D. and Quinn, Helen R.",
    title = "{CP Conservation in the Presence of Instantons}",
    reportNumber = "ITP-568-STANFORD",
    doi = "10.1103/PhysRevLett.38.1440",
    journal = "Phys. Rev. Lett.",
    volume = "38",
    pages = "1440--1443",
    year = "1977"
}

@article{Chigusa:2021mci,
    author = "Chigusa, So and Moroi, Takeo and Nakayama, Kazunori",
    title = "{Axion/hidden-photon dark matter conversion into condensed matter axion}",
    eprint = "2102.06179",
    archivePrefix = "arXiv",
    primaryClass = "hep-ph",
    reportNumber = "KEK-TH-2299",
    doi = "10.1007/JHEP08(2021)074",
    journal = "JHEP",
    volume = "08",
    pages = "074",
    year = "2021"
}

@article{Millar:2016cjp,
    author = "Millar, Alexander J. and Raffelt, Georg G. and Redondo, Javier and Steffen, Frank D.",
    title = "{Dielectric Haloscopes to Search for Axion Dark Matter: Theoretical Foundations}",
    eprint = "1612.07057",
    archivePrefix = "arXiv",
    primaryClass = "hep-ph",
    doi = "10.1088/1475-7516/2017/01/061",
    journal = "JCAP",
    volume = "01",
    pages = "061",
    year = "2017"
}

@article{BREAD:2021tpx,
    author = "Liu, Jesse and others",
    collaboration = "BREAD",
    title = "{Broadband Solenoidal Haloscope for Terahertz Axion Detection}",
    eprint = "2111.12103",
    archivePrefix = "arXiv",
    primaryClass = "physics.ins-det",
    reportNumber = "FERMILAB-PUB-21-694-AD-PPD-TD",
    doi = "10.1103/PhysRevLett.128.131801",
    journal = "Phys. Rev. Lett.",
    volume = "128",
    number = "13",
    pages = "131801",
    year = "2022"
}

@article{Bajjali:2023uis,
    author = "Bajjali, Fayez and others",
    title = "{First results from BRASS-p broadband searches for hidden photon dark matter}",
    eprint = "2306.05934",
    archivePrefix = "arXiv",
    primaryClass = "hep-ex",
    doi = "10.1088/1475-7516/2023/08/077",
    journal = "JCAP",
    volume = "08",
    pages = "077",
    year = "2023"
}

@article{DeMiguel:2020rpn,
    author = "De Miguel, Javier",
    title = "{A dark matter telescope probing the 6 to 60 GHz band}",
    eprint = "2003.06874",
    archivePrefix = "arXiv",
    primaryClass = "physics.ins-det",
    doi = "10.1088/1475-7516/2021/04/075",
    journal = "JCAP",
    volume = "04",
    pages = "075",
    year = "2021"
}

@article{DiLuzio:2020wdo,
    author = "Di Luzio, Luca and Giannotti, Maurizio and Nardi, Enrico and Visinelli, Luca",
    title = "{The landscape of QCD axion models}",
    eprint = "2003.01100",
    archivePrefix = "arXiv",
    primaryClass = "hep-ph",
    reportNumber = "DESY 20-036, DESY-20-036",
    doi = "10.1016/j.physrep.2020.06.002",
    journal = "Phys. Rept.",
    volume = "870",
    pages = "1--117",
    year = "2020"
}

@article{Quiskamp:2023ehr,
    author = "Quiskamp, Aaron and McAllister, Ben T. and Altin, Paul and Ivanov, Eugene N. and Goryachev, Maxim and Tobar, Michael E.",
    title = "{Exclusion of Axionlike-Particle Cogenesis Dark Matter in a Mass Window above 100{\,}{\,}{\ensuremath{\mu}}eV}",
    eprint = "2310.00904",
    archivePrefix = "arXiv",
    primaryClass = "hep-ex",
    doi = "10.1103/PhysRevLett.132.031601",
    journal = "Phys. Rev. Lett.",
    volume = "132",
    number = "3",
    pages = "031601",
    year = "2024"
}

@article{McAllister:2023ipr,
    author = "McAllister, Ben T. and Quiskamp, Aaron P. and Tobar, Michael E.",
    title = "{Tunable rectangular resonant cavities for axion haloscopes}",
    eprint = "2309.12098",
    archivePrefix = "arXiv",
    primaryClass = "physics.ins-det",
    doi = "10.1103/PhysRevD.109.015013",
    journal = "Phys. Rev. D",
    volume = "109",
    number = "1",
    pages = "015013",
    year = "2024"
}

@article{CAST:2024eil,
    author = {Altenm{\"u}ller, K. and others},
    collaboration = "CAST",
    title = "{New Upper Limit on the Axion-Photon Coupling with an Extended CAST Run with a Xe-Based Micromegas Detector}",
    eprint = "2406.16840",
    archivePrefix = "arXiv",
    primaryClass = "hep-ex",
    doi = "10.1103/PhysRevLett.133.221005",
    journal = "Phys. Rev. Lett.",
    volume = "133",
    number = "22",
    pages = "221005",
    year = "2024"
}

@article{Demirtas:2018akl,
    author = "Demirtas, Mehmet and Long, Cody and McAllister, Liam and Stillman, Mike",
    title = "{The Kreuzer-Skarke Axiverse}",
    eprint = "1808.01282",
    archivePrefix = "arXiv",
    primaryClass = "hep-th",
    doi = "10.1007/JHEP04(2020)138",
    journal = "JHEP",
    volume = "04",
    pages = "138",
    year = "2020"
}

@article{Demirtas:2022hqf,
    author = "Demirtas, Mehmet and Rios-Tascon, Andres and McAllister, Liam",
    title = "{CYTools: A Software Package for Analyzing Calabi-Yau Manifolds}",
    eprint = "2211.03823",
    archivePrefix = "arXiv",
    primaryClass = "hep-th",
    month = "11",
    year = "2022"
}

@article{Kreuzer:2000xy,
    author = "Kreuzer, Maximilian and Skarke, Harald",
    title = "{Complete classification of reflexive polyhedra in four-dimensions}",
    eprint = "hep-th/0002240",
    archivePrefix = "arXiv",
    reportNumber = "HUB-EP-00-13, TUW-00-07",
    doi = "10.4310/ATMP.2000.v4.n6.a2",
    journal = "Adv. Theor. Math. Phys.",
    volume = "4",
    pages = "1209--1230",
    year = "2000"
}

@article{Benabou:2025kgx,
    author = "Benabou, Joshua N. and Fraser, Katherine and Reig, Mario and Safdi, Benjamin R.",
    title = "{String theory and grand unification suggest a submicroelectronvolt QCD axion}",
    eprint = "2505.15884",
    archivePrefix = "arXiv",
    primaryClass = "hep-ph",
    doi = "10.1103/lthr-97lm",
    journal = "Phys. Rev. D",
    volume = "112",
    number = "6",
    pages = "066003",
    year = "2025"
}

@article{Fallon:2025lvn,
    author = "Fallon, Sebastian Vander Ploeg and Halverson, James and McAllister, Liam and Zhu, Yunhao",
    title = "{F-theory Axiverse}",
    eprint = "2511.20458",
    archivePrefix = "arXiv",
    primaryClass = "hep-th",
    month = "11",
    year = "2025"
}

@article{Demirtas:2021gsq,
    author = "Demirtas, Mehmet and Gendler, Naomi and Long, Cody and McAllister, Liam and Moritz, Jakob",
    title = "{PQ axiverse}",
    eprint = "2112.04503",
    archivePrefix = "arXiv",
    primaryClass = "hep-th",
    doi = "10.1007/JHEP06(2023)092",
    journal = "JHEP",
    volume = "06",
    pages = "092",
    year = "2023"
}

@article{Sheridan:2024vtt,
    author = "Sheridan, Elijah and Carta, Federico and Gendler, Naomi and Jain, Mudit and Marsh, David J. E. and McAllister, Liam and Righi, Nicole and Rogers, Keir K. and Schachner, Andreas",
    title = "{Fuzzy axions and associated relics}",
    eprint = "2412.12012",
    archivePrefix = "arXiv",
    primaryClass = "hep-th",
    reportNumber = "KCL-PH-TH/2024-75, KCL-PH-TH/2024-75",
    doi = "10.1007/JHEP09(2025)016",
    journal = "JHEP",
    volume = "09",
    pages = "016",
    year = "2025"
}

@article{Loladze:2025uvf,
    author = "Loladze, Vazha and Platschorre, Arthur and Reig, Mario",
    title = "{Higher axion strings}",
    eprint = "2503.18707",
    archivePrefix = "arXiv",
    primaryClass = "hep-ph",
    doi = "10.1007/JHEP08(2025)182",
    journal = "JHEP",
    volume = "08",
    pages = "182",
    year = "2025"
}

@article{Gendler:2024adn,
    author = "Gendler, Naomi and Marsh, David J. E.",
    title = "{Possible Implications of QCD Axion Dark Matter Constraints from Helioscopes and Haloscopes for the String Theory Landscape}",
    eprint = "2407.07143",
    archivePrefix = "arXiv",
    primaryClass = "hep-th",
    doi = "10.1103/PhysRevLett.134.081602",
    journal = "Phys. Rev. Lett.",
    volume = "134",
    number = "8",
    pages = "081602",
    year = "2025"
}

@article{Gendler:2023kjt,
    author = "Gendler, Naomi and Marsh, David J. E. and McAllister, Liam and Moritz, Jakob",
    title = "{Glimmers from the axiverse}",
    eprint = "2309.13145",
    archivePrefix = "arXiv",
    primaryClass = "hep-th",
    reportNumber = "KCL-PH-TH/2023-49",
    doi = "10.1088/1475-7516/2024/09/071",
    journal = "JCAP",
    volume = "09",
    pages = "071",
    year = "2024"
}

@article{Mehta:2021pwf,
    author = "Mehta, Viraf M. and Demirtas, Mehmet and Long, Cody and Marsh, David J. E. and McAllister, Liam and Stott, Matthew J.",
    title = "{Superradiance in string theory}",
    eprint = "2103.06812",
    archivePrefix = "arXiv",
    primaryClass = "hep-th",
    doi = "10.1088/1475-7516/2021/07/033",
    journal = "JCAP",
    volume = "07",
    pages = "033",
    year = "2021"
}

@article{Petrossian-Byrne:2025mto,
    author = "Petrossian-Byrne, Rudin and Villadoro, Giovanni",
    title = "{Open string axiverse}",
    eprint = "2503.16387",
    archivePrefix = "arXiv",
    primaryClass = "hep-ph",
    doi = "10.1007/JHEP07(2025)049",
    journal = "JHEP",
    volume = "07",
    pages = "049",
    year = "2025"
}

@article{Kakhidze:1990in,
    author = "Kakhidze, A. I. and Kolokolov, I. V.",
    title = "{Antiferromagnetic axions detector}",
    reportNumber = "IYF-90-131",
    journal = "Sov. Phys. JETP",
    volume = "72",
    pages = "598--600",
    year = "1991"
}

@article{PhysRevLett.111.017204,
  title = {Lifetimes of Antiferromagnetic Magnons in Two and Three Dimensions: Experiment, Theory, and Numerics},
  author = {Bayrakci, S. P. and Tennant, D. A. and Leininger, Ph. and Keller, T. and Gibson, M. C. R. and Wilson, S. D. and Birgeneau, R. J. and Keimer, B.},
  journal = {Phys. Rev. Lett.},
  volume = {111},
  issue = {1},
  pages = {017204},
  numpages = {5},
  year = {2013},
  month = {Jul},
  publisher = {American Physical Society},
  doi = {10.1103/PhysRevLett.111.017204},
  url = {https://link.aps.org/doi/10.1103/PhysRevLett.111.017204}
}

@article{Marsh:2018dlj,
    author = "Marsh, David J. E. and Fong, Kin-Chung and Lentz, Erik W. and Smejkal, Libo{\v{r}} and Ali, Mazhar N.",
    title = "{Proposal to Detect Dark Matter using Axionic Topological Antiferromagnets}",
    eprint = "1807.08810",
    archivePrefix = "arXiv",
    primaryClass = "hep-ph",
    doi = "10.1103/PhysRevLett.123.121601",
    journal = "Phys. Rev. Lett.",
    volume = "123",
    number = "12",
    pages = "121601",
    year = "2019"
}

@article{Schutte-Engel:2021bqm,
    author = {Sch{\"u}tte-Engel, Jan and Marsh, David J. E. and Millar, Alexander J. and Sekine, Akihiko and Chadha-Day, Francesca and Hoof, Sebastian and Ali, Mazhar N. and Fong, Kin-Chung and Hardy, Edward and {\v{S}}mejkal, Libor},
    title = "{Axion quasiparticles for axion dark matter detection}",
    eprint = "2102.05366",
    archivePrefix = "arXiv",
    primaryClass = "hep-ph",
    reportNumber = "IPPP/20/78, NORDITA-2021-007",
    doi = "10.1088/1475-7516/2021/08/066",
    journal = "JCAP",
    volume = "08",
    pages = "066",
    year = "2021"
}

@article{Arias-Aragon:2020qtn,
    author = "Arias-Arag{\'o}n, Fernando and D'eramo, Francesco and Ferreira, Ricardo Z. and Merlo, Luca and Notari, Alessio",
    title = "{Cosmic Imprints of XENON1T Axions}",
    eprint = "2007.06579",
    archivePrefix = "arXiv",
    primaryClass = "hep-ph",
    reportNumber = "FTUAM-20-12, IFT-UAM/CSIC-20-104",
    doi = "10.1088/1475-7516/2020/11/025",
    journal = "JCAP",
    volume = "11",
    pages = "025",
    year = "2020"
}

@article{DEramo:2021psx,
    author = "D'Eramo, Francesco and Hajkarim, Fazlollah and Yun, Seokhoon",
    title = "{Thermal Axion Production at Low Temperatures: A Smooth Treatment of the QCD Phase Transition}",
    eprint = "2108.04259",
    archivePrefix = "arXiv",
    primaryClass = "hep-ph",
    doi = "10.1103/PhysRevLett.128.152001",
    journal = "Phys. Rev. Lett.",
    volume = "128",
    number = "15",
    pages = "152001",
    year = "2022"
}

@article{DEramo:2021lgb,
    author = "D'Eramo, Francesco and Hajkarim, Fazlollah and Yun, Seokhoon",
    title = "{Thermal QCD Axions across Thresholds}",
    eprint = "2108.05371",
    archivePrefix = "arXiv",
    primaryClass = "hep-ph",
    doi = "10.1007/JHEP10(2021)224",
    journal = "JHEP",
    volume = "10",
    pages = "224",
    year = "2021"
}

@article{DEramo:2023nzt,
    author = "D'Eramo, Francesco and Hajkarim, Fazlollah and Lenoci, Alessandro",
    title = "{Dark radiation from the primordial thermal bath in momentum space}",
    eprint = "2311.04974",
    archivePrefix = "arXiv",
    primaryClass = "hep-ph",
    reportNumber = "DESY-23-177",
    doi = "10.1088/1475-7516/2024/03/009",
    journal = "JCAP",
    volume = "03",
    pages = "009",
    year = "2024"
}

@article{Graf:2010tv,
    author = "Graf, Peter and Steffen, Frank Daniel",
    title = "{Thermal axion production in the primordial quark-gluon plasma}",
    eprint = "1008.4528",
    archivePrefix = "arXiv",
    primaryClass = "hep-ph",
    reportNumber = "MPP-2010-20",
    doi = "10.1103/PhysRevD.83.075011",
    journal = "Phys. Rev. D",
    volume = "83",
    pages = "075011",
    year = "2011"
}

@article{Caloni:2024olo,
    author = "Caloni, Luca and Stengel, Patrick and Lattanzi, Massimiliano and Gerbino, Martina",
    title = "{Constraining UV freeze-in of light relics with current and next-generation CMB observations}",
    eprint = "2405.09449",
    archivePrefix = "arXiv",
    primaryClass = "astro-ph.CO",
    reportNumber = "CA21106",
    doi = "10.1088/1475-7516/2024/10/106",
    journal = "JCAP",
    volume = "10",
    pages = "106",
    year = "2024"
}

@article{Ferreira:2020bpb,
    author = "Ferreira, Ricardo Z. and Notari, Alessio and Rompineve, Fabrizio",
    title = "{Dine-Fischler-Srednicki-Zhitnitsky axion in the CMB}",
    eprint = "2012.06566",
    archivePrefix = "arXiv",
    primaryClass = "hep-ph",
    doi = "10.1103/PhysRevD.103.063524",
    journal = "Phys. Rev. D",
    volume = "103",
    number = "6",
    pages = "063524",
    year = "2021"
}

@article{DiLuzio:2021vjd,
    author = "Di Luzio, Luca and Martinelli, Guido and Piazza, Gioacchino",
    title = "{Breakdown of chiral perturbation theory for the axion hot dark matter bound}",
    eprint = "2101.10330",
    archivePrefix = "arXiv",
    primaryClass = "hep-ph",
    reportNumber = "DESY-21-012, DESY 21-012",
    doi = "10.1103/PhysRevLett.126.241801",
    journal = "Phys. Rev. Lett.",
    volume = "126",
    number = "24",
    pages = "241801",
    year = "2021"
}

@article{Planck:2018vyg,
    author = "Aghanim, N. and others",
    collaboration = "Planck",
    title = "{Planck 2018 results. VI. Cosmological parameters}",
    eprint = "1807.06209",
    archivePrefix = "arXiv",
    primaryClass = "astro-ph.CO",
    doi = "10.1051/0004-6361/201833910",
    journal = "Astron. Astrophys.",
    volume = "641",
    pages = "A6",
    year = "2020",
    note = "[Erratum: Astron.Astrophys. 652, C4 (2021)]"
}

@article{Masso:2002np,
    author = "Masso, Eduard and Rota, Francesc and Zsembinszki, Gabriel",
    title = "{On axion thermalization in the early universe}",
    eprint = "hep-ph/0203221",
    archivePrefix = "arXiv",
    reportNumber = "UAB-FT-522",
    doi = "10.1103/PhysRevD.66.023004",
    journal = "Phys. Rev. D",
    volume = "66",
    pages = "023004",
    year = "2002"
}

@article{DEramo:2014urw,
    author = "D'Eramo, Francesco and Hall, Lawrence J. and Pappadopulo, Duccio",
    title = "{Multiverse Dark Matter: SUSY or Axions}",
    eprint = "1409.5123",
    archivePrefix = "arXiv",
    primaryClass = "hep-ph",
    reportNumber = "UCB-PTH-14-35",
    doi = "10.1007/JHEP11(2014)108",
    journal = "JHEP",
    volume = "11",
    pages = "108",
    year = "2014"
}

@article{Chang:1993gm,
    author = "Chang, Sanghyeon and Choi, Kiwoon",
    title = "{Hadronic axion window and the big bang nucleosynthesis}",
    eprint = "hep-ph/9306216",
    archivePrefix = "arXiv",
    reportNumber = "SNUTP-93-11",
    doi = "10.1016/0370-2693(93)90656-3",
    journal = "Phys. Lett. B",
    volume = "316",
    pages = "51--56",
    year = "1993"
}

@article{Brust:2013ova,
    author = "Brust, Christopher and Kaplan, David E. and Walters, Matthew T.",
    title = "{New Light Species and the CMB}",
    eprint = "1303.5379",
    archivePrefix = "arXiv",
    primaryClass = "hep-ph",
    reportNumber = "UMD-PP-013-005",
    doi = "10.1007/JHEP12(2013)058",
    journal = "JHEP",
    volume = "12",
    pages = "058",
    year = "2013"
}

@article{SimonsObservatory:2018koc,
    author = "Ade, Peter and others",
    collaboration = "Simons Observatory",
    title = "{The Simons Observatory: Science goals and forecasts}",
    eprint = "1808.07445",
    archivePrefix = "arXiv",
    primaryClass = "astro-ph.CO",
    doi = "10.1088/1475-7516/2019/02/056",
    journal = "JCAP",
    volume = "02",
    pages = "056",
    year = "2019"
}

@article{CMB-S4:2022ght,
    author = "Abazajian, Kevork and others",
    collaboration = "CMB-S4",
    title = "{Snowmass 2021 CMB-S4 White Paper}",
    eprint = "2203.08024",
    archivePrefix = "arXiv",
    primaryClass = "astro-ph.CO",
    month = "3",
    year = "2022"
}

@article{DiLuzio:2022gsc,
    author = "Di Luzio, Luca and Martin Camalich, Jorge and Martinelli, Guido and Oller, Jos{\'e} Antonio and Piazza, Gioacchino",
    title = "{Axion-pion thermalization rate in unitarized NLO chiral perturbation theory}",
    eprint = "2211.05073",
    archivePrefix = "arXiv",
    primaryClass = "hep-ph",
    doi = "10.1103/PhysRevD.108.035025",
    journal = "Phys. Rev. D",
    volume = "108",
    number = "3",
    pages = "035025",
    year = "2023"
}

@article{Caloni:2022uya,
    author = "Caloni, Luca and Gerbino, Martina and Lattanzi, Massimiliano and Visinelli, Luca",
    title = "{Novel cosmological bounds on thermally-produced axion-like particles}",
    eprint = "2205.01637",
    archivePrefix = "arXiv",
    primaryClass = "astro-ph.CO",
    doi = "10.1088/1475-7516/2022/09/021",
    journal = "JCAP",
    volume = "09",
    pages = "021",
    year = "2022"
}

@article{Bianchini:2023ubu,
    author = "Bianchini, Federico and di Cortona, Giovanni Grilli and Valli, Mauro",
    title = "{QCD axion: Some like it hot}",
    eprint = "2310.08169",
    archivePrefix = "arXiv",
    primaryClass = "hep-ph",
    doi = "10.1103/PhysRevD.110.123527",
    journal = "Phys. Rev. D",
    volume = "110",
    number = "12",
    pages = "123527",
    year = "2024"
}

@article{Salvio:2013iaa,
    author = "Salvio, Alberto and Strumia, Alessandro and Xue, Wei",
    title = "{Thermal axion production}",
    eprint = "1310.6982",
    archivePrefix = "arXiv",
    primaryClass = "hep-ph",
    reportNumber = "FTUAM-13-29, IFT-UAM-CSIC-13-113",
    doi = "10.1088/1475-7516/2014/01/011",
    journal = "JCAP",
    volume = "01",
    pages = "011",
    year = "2014"
}

@article{Badziak:2024qjg,
    author = "Badziak, Marcin and Laletin, Maxim",
    title = "{Precise predictions for the QCD axion contribution to dark radiation with full phase-space evolution}",
    eprint = "2410.18186",
    archivePrefix = "arXiv",
    primaryClass = "hep-ph",
    doi = "10.1007/JHEP02(2025)108",
    journal = "JHEP",
    volume = "02",
    pages = "108",
    year = "2025"
}

@article{Badziak:2025mkt,
    author = "Badziak, Marcin and Gomu{\l}ka, Adam and Laletin, Maxim and Szafra{\'n}ski, Krzysztof",
    title = "{Improved cosmological constraints on axion-lepton interactions}",
    eprint = "2511.14864",
    archivePrefix = "arXiv",
    primaryClass = "hep-ph",
    month = "11",
    year = "2025"
}

@article{DEramo:2024jhn,
    author = "D'Eramo, Francesco and Lenoci, Alessandro",
    title = "{Back to the phase space: Thermal axion dark radiation via couplings to standard model fermions}",
    eprint = "2410.21253",
    archivePrefix = "arXiv",
    primaryClass = "hep-ph",
    doi = "10.1103/PhysRevD.110.116028",
    journal = "Phys. Rev. D",
    volume = "110",
    number = "11",
    pages = "116028",
    year = "2024"
}

@article{DEramo:2021usm,
    author = "D'Eramo, Francesco and Yun, Seokhoon",
    title = "{Flavor violating axions in the early Universe}",
    eprint = "2111.12108",
    archivePrefix = "arXiv",
    primaryClass = "hep-ph",
    doi = "10.1103/PhysRevD.105.075002",
    journal = "Phys. Rev. D",
    volume = "105",
    number = "7",
    pages = "075002",
    year = "2022"
}

@article{DEramo:2022nvb,
    author = "D'Eramo, Francesco and Di Valentino, Eleonora and Giar{\`e}, William and Hajkarim, Fazlollah and Melchiorri, Alessandro and Mena, Olga and Renzi, Fabrizio and Yun, Seokhoon",
    title = "{Cosmological bound on the QCD axion mass, redux}",
    eprint = "2205.07849",
    archivePrefix = "arXiv",
    primaryClass = "astro-ph.CO",
    doi = "10.1088/1475-7516/2022/09/022",
    journal = "JCAP",
    volume = "09",
    pages = "022",
    year = "2022"
}

@article{Green:2021hjh,
    author = "Green, Daniel and Guo, Yi and Wallisch, Benjamin",
    title = "{Cosmological implications of axion-matter couplings}",
    eprint = "2109.12088",
    archivePrefix = "arXiv",
    primaryClass = "astro-ph.CO",
    doi = "10.1088/1475-7516/2022/02/019",
    journal = "JCAP",
    volume = "02",
    number = "02",
    pages = "019",
    year = "2022"
}

@article{Arias-Aragon:2020shv,
    author = "Arias-Arag{\'o}n, Fernando and D'Eramo, Francesco and Ferreira, Ricardo Z. and Merlo, Luca and Notari, Alessio",
    title = "{Production of Thermal Axions across the ElectroWeak Phase Transition}",
    eprint = "2012.04736",
    archivePrefix = "arXiv",
    primaryClass = "hep-ph",
    doi = "10.1088/1475-7516/2021/03/090",
    journal = "JCAP",
    volume = "03",
    pages = "090",
    year = "2021"
}

@article{DEramo:2018vss,
    author = "D'Eramo, Francesco and Ferreira, Ricardo Z. and Notari, Alessio and Bernal, Jos{\'e} Luis",
    title = "{Hot Axions and the $H_0$ tension}",
    eprint = "1808.07430",
    archivePrefix = "arXiv",
    primaryClass = "hep-ph",
    doi = "10.1088/1475-7516/2018/11/014",
    journal = "JCAP",
    volume = "11",
    pages = "014",
    year = "2018"
}

@article{Ferreira:2018vjj,
    author = "Ferreira, Ricardo Z. and Notari, Alessio",
    title = "{Observable Windows for the QCD Axion Through the Number of Relativistic Species}",
    eprint = "1801.06090",
    archivePrefix = "arXiv",
    primaryClass = "hep-ph",
    doi = "10.1103/PhysRevLett.120.191301",
    journal = "Phys. Rev. Lett.",
    volume = "120",
    number = "19",
    pages = "191301",
    year = "2018"
}

@article{Baumann:2016wac,
    author = "Baumann, Daniel and Green, Daniel and Wallisch, Benjamin",
    title = "{New Target for Cosmic Axion Searches}",
    eprint = "1604.08614",
    archivePrefix = "arXiv",
    primaryClass = "astro-ph.CO",
    doi = "10.1103/PhysRevLett.117.171301",
    journal = "Phys. Rev. Lett.",
    volume = "117",
    number = "17",
    pages = "171301",
    year = "2016"
}

@article{Hofmann:1998pp,
    author = "Hofmann, Christoph P.",
    title = "{Spin wave scattering in the effective Lagrangian perspective}",
    eprint = "cond-mat/9805277",
    archivePrefix = "arXiv",
    reportNumber = "UCSD-PTH-98-15",
    doi = "10.1103/PhysRevB.60.388",
    journal = "Phys. Rev. B",
    volume = "60",
    pages = "388",
    year = "1999"
}

@article{Li:2009tca,
    author = "Li, Rundong and Wang, Jing and Qi, Xiaoliang and Zhang, Shou-Cheng",
    title = "{Dynamical Axion Field in Topological Magnetic Insulators}",
    eprint = "0908.1537",
    archivePrefix = "arXiv",
    primaryClass = "cond-mat.other",
    doi = "10.1038/nphys1534",
    journal = "Nature Phys.",
    volume = "6",
    pages = "284",
    year = "2010"
}

@article{Qiu:2025bbi,
    author = "Qiu, Jian-Xiang and others",
    title = "{Observation of the axion quasiparticle in 2D MnBi$_{2}$Te$_{4}$}",
    eprint = "2504.12572",
    archivePrefix = "arXiv",
    primaryClass = "cond-mat.mes-hall",
    doi = "10.1038/s41586-025-08862-x",
    journal = "Nature",
    volume = "641",
    number = "8061",
    pages = "62--69",
    year = "2025"
}

@article{Wilczek:1987mv,
    author = "Wilczek, Frank",
    title = "{Two Applications of Axion Electrodynamics}",
    reportNumber = "NSF-ITP-86-147",
    doi = "10.1103/PhysRevLett.58.1799",
    journal = "Phys. Rev. Lett.",
    volume = "58",
    pages = "1799",
    year = "1987"
}

@article{Peccei:1977ur,
    author = "Peccei, R. D. and Quinn, Helen R.",
    title = "{Constraints Imposed by CP Conservation in the Presence of Instantons}",
    reportNumber = "ITP-572-STANFORD",
    doi = "10.1103/PhysRevD.16.1791",
    journal = "Phys. Rev. D",
    volume = "16",
    pages = "1791--1797",
    year = "1977"
}

@article{Weinberg:1977ma,
    author = "Weinberg, Steven",
    title = "{A New Light Boson?}",
    reportNumber = "HUTP-77/A074",
    doi = "10.1103/PhysRevLett.40.223",
    journal = "Phys. Rev. Lett.",
    volume = "40",
    pages = "223--226",
    year = "1978"
}

@article{Wilczek:1977pj,
    author = "Wilczek, Frank",
    title = "{Problem of Strong  $P$  and  $T$  Invariance in the Presence of Instantons}",
    reportNumber = "Print-77-0939 (COLUMBIA)",
    doi = "10.1103/PhysRevLett.40.279",
    journal = "Phys. Rev. Lett.",
    volume = "40",
    pages = "279--282",
    year = "1978"
}

@article{Kim:1979if,
    author = "Kim, Jihn E.",
    title = "{Weak Interaction Singlet and Strong CP Invariance}",
    reportNumber = "UPR-0120T",
    doi = "10.1103/PhysRevLett.43.103",
    journal = "Phys. Rev. Lett.",
    volume = "43",
    pages = "103",
    year = "1979"
}

@article{Shifman:1979if,
    author = "Shifman, Mikhail A. and Vainshtein, A. I. and Zakharov, Valentin I.",
    title = "{Can Confinement Ensure Natural CP Invariance of Strong Interactions?}",
    reportNumber = "ITEP-64-1979",
    doi = "10.1016/0550-3213(80)90209-6",
    journal = "Nucl. Phys. B",
    volume = "166",
    pages = "493--506",
    year = "1980"
}

@article{Zhitnitsky:1980tq,
    author = "Zhitnitsky, A. R.",
    title = "{On Possible Suppression of the Axion Hadron Interactions. (In Russian)}",
    journal = "Sov. J. Nucl. Phys.",
    volume = "31",
    pages = "260",
    year = "1980"
}

@article{Dine:1981rt,
    author = "Dine, Michael and Fischler, Willy and Srednicki, Mark",
    title = "{A Simple Solution to the Strong CP Problem with a Harmless Axion}",
    reportNumber = "Print-81-0320 (IAS,PRINCETON)",
    doi = "10.1016/0370-2693(81)90590-6",
    journal = "Phys. Lett. B",
    volume = "104",
    pages = "199--202",
    year = "1981"
}

@article{Georgi:1981pu,
    author = "Georgi, Howard M. and Hall, Lawrence J. and Wise, Mark B.",
    title = "{Grand Unified Models With an Automatic {Peccei-Quinn} Symmetry}",
    reportNumber = "HUTP-81/A031",
    doi = "10.1016/0550-3213(81)90433-8",
    journal = "Nucl. Phys. B",
    volume = "192",
    pages = "409--416",
    year = "1981"
}

@article{Blumenhagen:2007sm,
    author = "Blumenhagen, Ralph and Moster, Sebastian and Plauschinn, Erik",
    title = "{Moduli Stabilisation versus Chirality for MSSM like Type IIB Orientifolds}",
    eprint = "0711.3389",
    archivePrefix = "arXiv",
    primaryClass = "hep-th",
    reportNumber = "MPP-2007-169",
    doi = "10.1088/1126-6708/2008/01/058",
    journal = "JHEP",
    volume = "01",
    pages = "058",
    year = "2008"
}

@article{Cicoli:2007xp,
    author = "Cicoli, Michele and Conlon, Joseph P. and Quevedo, Fernando",
    title = "{Systematics of String Loop Corrections in Type IIB Calabi-Yau Flux Compactifications}",
    eprint = "0708.1873",
    archivePrefix = "arXiv",
    primaryClass = "hep-th",
    reportNumber = "DAMTP-2007-75",
    doi = "10.1088/1126-6708/2008/01/052",
    journal = "JHEP",
    volume = "01",
    pages = "052",
    year = "2008"
}

@article{Hees:2018fpg,
    author = "Hees, Aur{\'e}lien and Minazzoli, Olivier and Savalle, Etienne and Stadnik, Yevgeny V. and Wolf, Peter",
    title = "{Violation of the equivalence principle from light scalar dark matter}",
    eprint = "1807.04512",
    archivePrefix = "arXiv",
    primaryClass = "gr-qc",
    doi = "10.1103/PhysRevD.98.064051",
    journal = "Phys. Rev. D",
    volume = "98",
    number = "6",
    pages = "064051",
    year = "2018"
}

@article{Conlon:2010ji,
    author = "Conlon, Joseph P. and Pedro, Francisco G.",
    title = "{Moduli Redefinitions and Moduli Stabilisation}",
    eprint = "1003.0388",
    archivePrefix = "arXiv",
    primaryClass = "hep-th",
    reportNumber = "OUTP-10-06P",
    doi = "10.1007/JHEP06(2010)082",
    journal = "JHEP",
    volume = "06",
    pages = "082",
    year = "2010"
}

@article{Aparicio:2014wxa,
    author = "Aparicio, Luis and Cicoli, Michele and Krippendorf, Sven and Maharana, Anshuman and Muia, Francesco and Quevedo, Fernando",
    title = "{Sequestered de Sitter String Scenarios: Soft-terms}",
    eprint = "1409.1931",
    archivePrefix = "arXiv",
    primaryClass = "hep-th",
    doi = "10.1007/JHEP11(2014)071",
    journal = "JHEP",
    volume = "11",
    pages = "071",
    year = "2014"
}

@article{Blumenhagen:2009gk,
    author = "Blumenhagen, R. and Conlon, J. P. and Krippendorf, S. and Moster, S. and Quevedo, F.",
    title = "{SUSY Breaking in Local String/F-Theory Models}",
    eprint = "0906.3297",
    archivePrefix = "arXiv",
    primaryClass = "hep-th",
    reportNumber = "MPP-2009-76, OUTP-09-14P, DAMTP-2009-48, CERN-PH-TH-2009-089",
    doi = "10.1088/1126-6708/2009/09/007",
    journal = "JHEP",
    volume = "09",
    pages = "007",
    year = "2009"
}

@article{Cicoli:2021dhg,
    author = "Cicoli, Michele and Etxebarria, I{\~n}aki Garc{\'\i}a and Quevedo, Fernando and Schachner, Andreas and Shukla, Pramod and Valandro, Roberto",
    title = "{The Standard Model quiver in de Sitter string compactifications}",
    eprint = "2106.11964",
    archivePrefix = "arXiv",
    primaryClass = "hep-th",
    doi = "10.1007/JHEP08(2021)109",
    journal = "JHEP",
    volume = "08",
    pages = "109",
    year = "2021"
}

@article{Cicoli:2017shd,
    author = "Cicoli, Michele and Garc{\`\i}a-Etxebarria, I{\~n}aki and Mayrhofer, Christoph and Quevedo, Fernando and Shukla, Pramod and Valandro, Roberto",
    title = "{Global Orientifolded Quivers with Inflation}",
    eprint = "1706.06128",
    archivePrefix = "arXiv",
    primaryClass = "hep-th",
    doi = "10.1007/JHEP11(2017)134",
    journal = "JHEP",
    volume = "11",
    pages = "134",
    year = "2017"
}

@article{Cicoli:2013mpa,
    author = "Cicoli, Michele and Krippendorf, Sven and Mayrhofer, Christoph and Quevedo, Fernando and Valandro, Roberto",
    title = "{D3/D7 Branes at Singularities: Constraints from Global Embedding and Moduli Stabilisation}",
    eprint = "1304.0022",
    archivePrefix = "arXiv",
    primaryClass = "hep-th",
    doi = "10.1007/JHEP07(2013)150",
    journal = "JHEP",
    volume = "07",
    pages = "150",
    year = "2013"
}

@article{Cicoli:2012vw,
    author = "Cicoli, Michele and Krippendorf, Sven and Mayrhofer, Christoph and Quevedo, Fernando and Valandro, Roberto",
    title = "{D-Branes at del Pezzo Singularities: Global Embedding and Moduli Stabilisation}",
    eprint = "1206.5237",
    archivePrefix = "arXiv",
    primaryClass = "hep-th",
    reportNumber = "DAMTP-2012-47, ZMP-HH-12-10",
    doi = "10.1007/JHEP09(2012)019",
    journal = "JHEP",
    volume = "09",
    pages = "019",
    year = "2012"
}

@article{Berezhiani:2000gh,
    author = "Berezhiani, Zurab and Gianfagna, Leonida and Giannotti, Maurizio",
    title = "{Strong CP problem and mirror world: The Weinberg-Wilczek axion revisited}",
    eprint = "hep-ph/0009290",
    archivePrefix = "arXiv",
    reportNumber = "DFAQ-TH-2000-04",
    doi = "10.1016/S0370-2693(00)01392-7",
    journal = "Phys. Lett. B",
    volume = "500",
    pages = "286--296",
    year = "2001"
}

@article{Rubakov:1997vp,
    author = "Rubakov, V. A.",
    title = "{Grand unification and heavy axion}",
    eprint = "hep-ph/9703409",
    archivePrefix = "arXiv",
    reportNumber = "INR-97-231",
    doi = "10.1134/1.567390",
    journal = "JETP Lett.",
    volume = "65",
    pages = "621--624",
    year = "1997"
}

@article{Dine:1986bg,
    author = "Dine, Michael and Seiberg, Nathan",
    title = "{String Theory and the Strong {CP} Problem}",
    reportNumber = "Print-86-0091 (CITY COLL.,N.Y.), CCNY-HEP-86/2",
    doi = "10.1016/0550-3213(86)90043-X",
    journal = "Nucl. Phys. B",
    volume = "273",
    pages = "109--124",
    year = "1986"
}

@article{Barr:1992qq,
    author = "Barr, Stephen M. and Seckel, D.",
    title = "{Planck scale corrections to axion models}",
    reportNumber = "BA-92-11",
    doi = "10.1103/PhysRevD.46.539",
    journal = "Phys. Rev. D",
    volume = "46",
    pages = "539--549",
    year = "1992"
}

@article{Kamionkowski:1992mf,
    author = "Kamionkowski, Marc and March-Russell, John",
    title = "{Planck scale physics and the Peccei-Quinn mechanism}",
    eprint = "hep-th/9202003",
    archivePrefix = "arXiv",
    reportNumber = "IASSNS-HEP-92-9, PUPT-92-1309",
    doi = "10.1016/0370-2693(92)90492-M",
    journal = "Phys. Lett. B",
    volume = "282",
    pages = "137--141",
    year = "1992"
}

@article{Holman:1992us,
    author = "Holman, Richard and Hsu, Stephen D. H. and Kephart, Thomas W. and Kolb, Edward W. and Watkins, Richard and Widrow, Lawrence M.",
    title = "{Solutions to the strong CP problem in a world with gravity}",
    eprint = "hep-ph/9203206",
    archivePrefix = "arXiv",
    reportNumber = "NSF-ITP-92-06, CMU-HEP92-05, FERMILAB-PUB-92-034-A, HUTP-92-A011, VAND-TH-92-2",
    doi = "10.1016/0370-2693(92)90491-L",
    journal = "Phys. Lett. B",
    volume = "282",
    pages = "132--136",
    year = "1992"
}

@article{Ghigna:1992iv,
    author = "Ghigna, S. and Lusignoli, Maurizio and Roncadelli, M.",
    title = "{Instability of the invisible axion}",
    reportNumber = "ROME-877-1992, FNT-T-92-12",
    doi = "10.1016/0370-2693(92)90019-Z",
    journal = "Phys. Lett. B",
    volume = "283",
    pages = "278--281",
    year = "1992"
}

@article{Randall:1992ut,
    author = "Randall, Lisa",
    title = "{Composite axion models and Planck scale physics}",
    reportNumber = "MIT-CTP-2074",
    doi = "10.1016/0370-2693(92)91928-3",
    journal = "Phys. Lett. B",
    volume = "284",
    pages = "77--80",
    year = "1992"
}

@article{Dobrescu:1996jp,
    author = "Dobrescu, Bogdan A.",
    title = "{The Strong CP problem versus Planck scale physics}",
    eprint = "hep-ph/9609221",
    archivePrefix = "arXiv",
    reportNumber = "BUHEP-96-30",
    doi = "10.1103/PhysRevD.55.5826",
    journal = "Phys. Rev. D",
    volume = "55",
    pages = "5826--5833",
    year = "1997"
}

@article{Babu:2002ic,
    author = "Babu, K. S. and Gogoladze, Ilia and Wang, Kai",
    title = "{Stabilizing the axion by discrete gauge symmetries}",
    eprint = "hep-ph/0212339",
    archivePrefix = "arXiv",
    reportNumber = "OSU-HEP-02-18",
    doi = "10.1016/S0370-2693(03)00411-8",
    journal = "Phys. Lett. B",
    volume = "560",
    pages = "214--222",
    year = "2003"
}

@article{Redi:2016esr,
    author = "Redi, Michele and Sato, Ryosuke",
    title = "{Composite Accidental Axions}",
    eprint = "1602.05427",
    archivePrefix = "arXiv",
    primaryClass = "hep-ph",
    doi = "10.1007/JHEP05(2016)104",
    journal = "JHEP",
    volume = "05",
    pages = "104",
    year = "2016"
}

@article{Fukuda:2017ylt,
    author = "Fukuda, Hajime and Ibe, Masahiro and Suzuki, Motoo and Yanagida, Tsutomu T.",
    title = "{A ''gauged'' $U(1)$ Peccei{\textendash}Quinn symmetry}",
    eprint = "1703.01112",
    archivePrefix = "arXiv",
    primaryClass = "hep-ph",
    reportNumber = "IPMU-17-0040",
    doi = "10.1016/j.physletb.2017.05.071",
    journal = "Phys. Lett. B",
    volume = "771",
    pages = "327--331",
    year = "2017"
}

@article{DiLuzio:2017tjx,
    author = "Di Luzio, Luca and Nardi, Enrico and Ubaldi, Lorenzo",
    title = "{Accidental Peccei-Quinn symmetry protected to arbitrary order}",
    eprint = "1704.01122",
    archivePrefix = "arXiv",
    primaryClass = "hep-ph",
    reportNumber = "IPPP-17-29",
    doi = "10.1103/PhysRevLett.119.011801",
    journal = "Phys. Rev. Lett.",
    volume = "119",
    number = "1",
    pages = "011801",
    year = "2017"
}

@article{Bonnefoy:2018ibr,
    author = "Bonnefoy, Quentin and Dudas, Emilian and Pokorski, Stefan",
    title = "{Axions in a highly protected gauge symmetry model}",
    eprint = "1804.01112",
    archivePrefix = "arXiv",
    primaryClass = "hep-ph",
    doi = "10.1140/epjc/s10052-018-6528-z",
    journal = "Eur. Phys. J. C",
    volume = "79",
    number = "1",
    pages = "31",
    year = "2019"
}

@article{Lillard:2018fdt,
    author = "Lillard, Benjamin and Tait, Tim M. P.",
    title = "{A High Quality Composite Axion}",
    eprint = "1811.03089",
    archivePrefix = "arXiv",
    primaryClass = "hep-ph",
    reportNumber = "UCI-HEP-TR-2018-12",
    doi = "10.1007/JHEP11(2018)199",
    journal = "JHEP",
    volume = "11",
    pages = "199",
    year = "2018"
}

@article{Gavela:2018paw,
    author = "Gavela, M. B. and Ibe, M. and Quilez, P. and Yanagida, T. T.",
    title = "{Automatic Peccei{\textendash}Quinn symmetry}",
    eprint = "1812.08174",
    archivePrefix = "arXiv",
    primaryClass = "hep-ph",
    reportNumber = "IPMU18-0205, IFT-UAM/CSIC-18-129, FTUAM-18-29",
    doi = "10.1140/epjc/s10052-019-7046-3",
    journal = "Eur. Phys. J. C",
    volume = "79",
    number = "6",
    pages = "542",
    year = "2019"
}

@article{Lee:2018yak,
    author = "Lee, Hye-Sung and Yin, Wen",
    title = "{Peccei-Quinn symmetry from a hidden gauge group structure}",
    eprint = "1811.04039",
    archivePrefix = "arXiv",
    primaryClass = "hep-ph",
    doi = "10.1103/PhysRevD.99.015041",
    journal = "Phys. Rev. D",
    volume = "99",
    number = "1",
    pages = "015041",
    year = "2019"
}

@article{Ardu:2020qmo,
    author = "Ardu, Marco and Di Luzio, Luca and Landini, Giacomo and Strumia, Alessandro and Teresi, Daniele and Wang, Jin-Wei",
    title = "{Axion quality from the (anti)symmetric of SU($ \mathcal{N} $)}",
    eprint = "2007.12663",
    archivePrefix = "arXiv",
    primaryClass = "hep-ph",
    reportNumber = "DESY 20-124, DESY-20-124",
    doi = "10.1007/JHEP11(2020)090",
    journal = "JHEP",
    volume = "11",
    pages = "090",
    year = "2020"
}

@article{DiLuzio:2020qio,
    author = "Di Luzio, Luca",
    title = "{Accidental SO(10) axion from gauged flavour}",
    eprint = "2008.09119",
    archivePrefix = "arXiv",
    primaryClass = "hep-ph",
    reportNumber = "DESY 20-133, DESY-20-133",
    doi = "10.1007/JHEP11(2020)074",
    journal = "JHEP",
    volume = "11",
    pages = "074",
    year = "2020"
}

@article{Yin:2020dfn,
    author = "Yin, Wen",
    title = "{Scale and quality of Peccei-Quinn symmetry and weak gravity conjectures}",
    eprint = "2007.13320",
    archivePrefix = "arXiv",
    primaryClass = "hep-ph",
    doi = "10.1007/JHEP10(2020)032",
    journal = "JHEP",
    volume = "10",
    pages = "032",
    year = "2020"
}

@article{Contino:2021ayn,
    author = "Contino, Roberto and Podo, Alessandro and Revello, Filippo",
    title = "{Chiral models of composite axions and accidental Peccei-Quinn symmetry}",
    eprint = "2112.09635",
    archivePrefix = "arXiv",
    primaryClass = "hep-ph",
    doi = "10.1007/JHEP04(2022)180",
    journal = "JHEP",
    volume = "04",
    pages = "180",
    year = "2022"
}

@article{DiLuzio:2025jhv,
    author = "Di Luzio, Luca and Landini, Giacomo and Mescia, Federico and Susi{\v{c}}, Vasja",
    title = "{High-quality Peccei-Quinn symmetry from the interplay of vertical and horizontal gauge symmetries}",
    eprint = "2503.16648",
    archivePrefix = "arXiv",
    primaryClass = "hep-ph",
    doi = "10.1140/epjc/s10052-025-15175-w",
    journal = "Eur. Phys. J. C",
    volume = "86",
    number = "1",
    pages = "5",
    year = "2026"
}

@article{Witten:1984dg,
    author = "Witten, Edward",
    title = "{Some Properties of O(32) Superstrings}",
    reportNumber = "Print-84-0838 (PRINCETON)",
    doi = "10.1016/0370-2693(84)90422-2",
    journal = "Phys. Lett. B",
    volume = "149",
    pages = "351--356",
    year = "1984"
}

@article{Eroncel:2022vjg,
    author = {Er{\"o}ncel, Cem and Sato, Ryosuke and Servant, Geraldine and S{\o}rensen, Philip},
    title = "{ALP dark matter from kinetic fragmentation: opening up the parameter window}",
    eprint = "2206.14259",
    archivePrefix = "arXiv",
    primaryClass = "hep-ph",
    reportNumber = "DESY 22-106, OU-HET-1148",
    doi = "10.1088/1475-7516/2022/10/053",
    journal = "JCAP",
    volume = "10",
    pages = "053",
    year = "2022"
}

@article{Brockway:1996yr,
    author = "Brockway, Jack W. and Carlson, Eric D. and Raffelt, Georg G.",
    title = "{SN 1987A gamma-ray limits on the conversion of pseudoscalars}",
    eprint = "astro-ph/9605197",
    archivePrefix = "arXiv",
    reportNumber = "MPI-PHT-96-42",
    doi = "10.1016/0370-2693(96)00778-2",
    journal = "Phys. Lett. B",
    volume = "383",
    pages = "439--443",
    year = "1996"
}

@article{Grifols:1996id,
    author = "Grifols, J. A. and Mass{\'o}, E. and Toldr{\`a}, R.",
    title = "{Gamma Rays from SN 1987A due to Pseudoscalar Conversion}",
    eprint = "astro-ph/9606028",
    archivePrefix = "arXiv",
    reportNumber = "UAB-FT-391",
    doi = "10.1103/PhysRevLett.77.2372",
    journal = "Phys. Rev. Lett.",
    volume = "77",
    pages = "2372--2375",
    year = "1996"
}

@article{Payez:2014xsa,
    author = "Payez, Alexandre and Evoli, Carmelo and Fischer, Tobias and Giannotti, Maurizio and Mirizzi, Alessandro and Ringwald, Andreas",
    title = "{Revisiting the SN1987A gamma-ray limit on ultralight axion-like particles}",
    eprint = "1410.3747",
    archivePrefix = "arXiv",
    primaryClass = "astro-ph.HE",
    reportNumber = "DESY-14-164",
    doi = "10.1088/1475-7516/2015/02/006",
    journal = "JCAP",
    volume = "02",
    pages = "006",
    year = "2015"
}

@article{Calore:2020tjw,
    author = "Calore, Francesca and Carenza, Pierluca and Giannotti, Maurizio and Jaeckel, Joerg and Mirizzi, Alessandro",
    title = "{Bounds on axionlike particles from the diffuse supernova flux}",
    eprint = "2008.11741",
    archivePrefix = "arXiv",
    primaryClass = "hep-ph",
    doi = "10.1103/PhysRevD.102.123005",
    journal = "Phys. Rev. D",
    volume = "102",
    number = "12",
    pages = "123005",
    year = "2020"
}

@article{Calore:2021hhn,
    author = "Calore, Francesca and Carenza, Pierluca and Eckner, Christopher and Fischer, Tobias and Giannotti, Maurizio and Jaeckel, Joerg and Kotake, Kei and Kuroda, Takami and Mirizzi, Alessandro and Sivo, Francesco",
    title = "{3D template-based Fermi-LAT constraints on the diffuse supernova axion-like particle background}",
    eprint = "2110.03679",
    archivePrefix = "arXiv",
    primaryClass = "astro-ph.HE",
    reportNumber = "LAPTH-038/21",
    doi = "10.1103/PhysRevD.105.063028",
    journal = "Phys. Rev. D",
    volume = "105",
    number = "6",
    pages = "063028",
    year = "2022"
}

@article{Hoof:2022xbe,
    author = "Hoof, Sebastian and Schulz, Lena",
    title = "{Updated constraints on axion-like particles from temporal information in supernova SN1987A gamma-ray data}",
    eprint = "2212.09764",
    archivePrefix = "arXiv",
    primaryClass = "hep-ph",
    reportNumber = "TTP22-072",
    doi = "10.1088/1475-7516/2023/03/054",
    journal = "JCAP",
    volume = "03",
    pages = "054",
    year = "2023",
    note = "For the associated data see 
    \href{https://github.com/sebhoof/snax}{https://github.com/sebhoof/snax}"
}

@article{Lella:2023bfb,
    author = "Lella, Alessandro and Carenza, Pierluca and Co', Giampaolo and Lucente, Giuseppe and Giannotti, Maurizio and Mirizzi, Alessandro and Rauscher, Thomas",
    title = "{Getting the most on supernova axions}",
    eprint = "2306.01048",
    archivePrefix = "arXiv",
    primaryClass = "hep-ph",
    doi = "10.1103/PhysRevD.109.023001",
    journal = "Phys. Rev. D",
    volume = "109",
    number = "2",
    pages = "023001",
    year = "2024"
}

@misc{AxionLimits,
  author       = {Ciaran O'Hare},
  title        = {cajohare/AxionLimits: AxionLimits},
  month        = jul,
  year         = 2020,
  publisher    = {Zenodo},
  version      = {v1.0},
  doi          = {10.5281/zenodo.3932430},
  howpublished = {\url{https://cajohare.github.io/AxionLimits/}}
}

@ARTICLE{1980SoPh...65...15F,
       author = {{Forrest}, D.~J. and {Chupp}, E.~L. and {Ryan}, J.~M. and {Cherry}, M.~L. and {Gleske}, I.~U. and {Reppin}, C. and {Pinkau}, K. and {Rieger}, E. and {Kanbach}, G. and {Kinzer}, R.~L. and {Share}, G. and {Johnson}, W.~N. and {Kurfess}, J.~D.},
        title = "{The gamma ray spectrometer for the Solar Maximum Mission.}",
      journal = {Sol. Phys.},
         year = 1980,
        month = feb,
       volume = {65},
       number = {1},
        pages = {15-23},
          doi = {10.1007/BF00151381},
       adsurl = {https://ui.adsabs.harvard.edu/abs/1980SoPh...65...15F}
}

@article{Chupp:1989kx,
    author = "Chupp, E. L. and Vestrand, W. T. and Reppin, C.",
    title = "{Experimental Limits on the Radiative Decay of {SN1987A} Neutrinos}",
    doi = "10.1103/PhysRevLett.62.505",
    journal = "Phys. Rev. Lett.",
    volume = "62",
    pages = "505--508",
    year = "1989"
}

@article{Oberauer:1993yr,
    author = "Oberauer, L. and Hagner, C. and Raffelt, G. and Rieger, E.",
    title = "{Supernova bounds on neutrino radiative decays}",
    doi = "10.1016/0927-6505(93)90004-W",
    journal = "Astropart. Phys.",
    volume = "1",
    pages = "377--386",
    year = "1993"
}

@ARTICLE{Graur+2017,
       author = {{Graur}, Or and {Bianco}, Federica B. and {Modjaz}, Maryam and {Shivvers}, Isaac and {Filippenko}, Alexei V. and {Li}, Weidong and {Smith}, Nathan},
        title = "{LOSS Revisited. II. The Relative Rates of Different Types of Supernovae Vary between Low- and High-mass Galaxies}",
      journal = {Astrophys. J.},
         year = 2017,
        month = mar,
       volume = {837},
       number = {2},
          eid = {121},
        pages = {121},
          doi = {10.3847/1538-4357/aa5eb7},
archivePrefix = {arXiv},
       eprint = {1609.02923},
 primaryClass = {astro-ph.HE},
       adsurl = {https://ui.adsabs.harvard.edu/abs/2017ApJ...837..121G}
}

@ARTICLE{2014ApJ...781...73D,
       author = {{de la Chevroti{\`e}re}, A. and {St-Louis}, N. and {Moffat}, A.~F.~J. and {MiMeS Collaboration}},
        title = "{Searching for Magnetic Fields in 11 Wolf-Rayet Stars: Analysis of Circular Polarization Measurements from ESPaDOnS}",
      journal = {Astrophys. J.},
         year = 2014,
        month = feb,
       volume = {781},
       number = {2},
          eid = {73},
        pages = {73},
          doi = {10.1088/0004-637X/781/2/73},
       adsurl = {https://ui.adsabs.harvard.edu/abs/2014ApJ...781...73D}
}

@ARTICLE{2016MNRAS.458.3381H,
       author = {{Hubrig}, S. and {Scholz}, K. and {Hamann}, W.-R. and {Sch{\"o}ller}, M. and {Ignace}, R. and {Ilyin}, I. and {Gayley}, K.~G. and {Oskinova}, L.~M.},
        title = "{Searching for a magnetic field in Wolf-Rayet stars using FORS 2 spectropolarimetry}",
      journal = {Mon. Not. R. Astron. Soc.},
         year = 2016,
        month = may,
       volume = {458},
       number = {3},
        pages = {3381-3393},
          doi = {10.1093/mnras/stw558},
archivePrefix = {arXiv},
       eprint = {1603.01441},
 primaryClass = {astro-ph.SR},
       adsurl = {https://ui.adsabs.harvard.edu/abs/2016MNRAS.458.3381H}
}

@article{Raffelt:1993ix,
    author = "Raffelt, Georg and Seckel, David",
    title = "{A selfconsistent approach to neutral current processes in supernova cores}",
    eprint = "astro-ph/9312019",
    archivePrefix = "arXiv",
    reportNumber = "MPI-PH-93-90, BA-93-43",
    doi = "10.1103/PhysRevD.52.1780",
    journal = "Phys. Rev. D",
    volume = "52",
    pages = "1780--1799",
    year = "1995"
}

@ARTICLE{2020MNRAS.499L.116H,
       author = {{Hubrig}, S. and {Sch{\"o}ller}, M. and {Cikota}, A. and {J{\"a}rvinen}, S.~P.},
        title = "{The search for magnetic fields in two Wolf-Rayet stars and the discovery of a variable magnetic field in WR 55}",
      journal = {Mon. Not. R. Astron. Soc.},
         year = 2020,
        month = dec,
       volume = {499},
       number = {1},
        pages = {L116-L120},
          doi = {10.1093/mnrasl/slaa170},
archivePrefix = {arXiv},
       eprint = {2010.00983},
 primaryClass = {astro-ph.SR},
       adsurl = {https://ui.adsabs.harvard.edu/abs/2020MNRAS.499L.116H}
}

@article{Keil:1996ju,
    author = "Keil, Wolfgang and Janka, Hans-Thomas and Schramm, David N. and Sigl, Gunter and Turner, Michael S. and Ellis, John R.",
    title = "{A Fresh look at axions and SN-1987A}",
    eprint = "astro-ph/9612222",
    archivePrefix = "arXiv",
    reportNumber = "FERMILAB-PUB-95-406-A, CERN-TH-96-112",
    doi = "10.1103/PhysRevD.56.2419",
    journal = "Phys. Rev. D",
    volume = "56",
    pages = "2419--2432",
    year = "1997"
}

@ARTICLE{2023MNRAS.524L..21J,
       author = {{J{\"a}rvinen}, S.~P. and {Hubrig}, S. and {Jayaraman}, R. and {Cikota}, A. and {Sch{\"o}ller}, M.},
        title = "{The magnetic, spectroscopic, and photometric variability of the Wolf-Rayet star WR 55}",
      journal = {Mon. Not. R. Astron. Soc.},
         year = 2023,
        month = sep,
       volume = {524},
       number = {1},
        pages = {L21-L25},
          doi = {10.1093/mnrasl/slad068},
archivePrefix = {arXiv},
       eprint = {2306.05038},
 primaryClass = {astro-ph.SR},
       adsurl = {https://ui.adsabs.harvard.edu/abs/2023MNRAS.524L..21J}
}

@article{Shenar:2023zoa,
    author = "Shenar, Tomer and others",
    title = "{A massive helium star with a sufficiently strong magnetic field to form a magnetar}",
    eprint = "2308.08591",
    archivePrefix = "arXiv",
    primaryClass = "astro-ph.SR",
    doi = "10.1126/science.ade3293",
    journal = "Science",
    volume = "381",
    number = "6659",
    pages = "761--765",
    year = "2023"
}

@article{Tauris:2015xra,
    author = "Tauris, Thomas M. and Langer, Norbert and Podsiadlowski, Philipp",
    title = "{Ultra-stripped supernovae: progenitors and fate}",
    eprint = "1505.00270",
    archivePrefix = "arXiv",
    primaryClass = "astro-ph.SR",
    doi = "10.1093/mnras/stv990",
    journal = "Mon. Not. R. Astron. Soc.",
    volume = "451",
    number = "2",
    pages = "2123--2144",
    year = "2015"
}

@article{Kleiser:2011ms,
    author = "Kleiser, Io K. W. and others",
    title = "{Peculiar Type II Supernovae from Blue Supergiants}",
    eprint = "1101.1298",
    archivePrefix = "arXiv",
    primaryClass = "astro-ph.CO",
    doi = "10.1111/j.1365-2966.2011.18708.x",
    journal = "Mon. Not. R. Astron. Soc.",
    volume = "415",
    pages = "372",
    year = "2011"
}

@article{Pessi:2025wht,
    author = "Pessi, T. and Desai, D. D. and Prieto, J. L. and Kochanek, C. S. and Shappee, B. J. and Anderson, J. P. and Beacom, J. F. and Dong, Subo and Stanek, K. Z. and Thompson, T. A.",
    title = "{Supernova rates and luminosity functions from ASAS-SN II: 2014--2017 core-collapse supernovae and their subtypes}",
    eprint = "2508.10985",
    archivePrefix = "arXiv",
    primaryClass = "astro-ph.HE",
    doi = "10.1051/0004-6361/202556799",
    month = "8",
    year = "2025"
}

@article{LIGOScientific:2017vwq,
    author = "Abbott, B. P. and others",
    collaboration = "LIGO Scientific, Virgo",
    title = "{GW170817: Observation of Gravitational Waves from a Binary Neutron Star Inspiral}",
    eprint = "1710.05832",
    archivePrefix = "arXiv",
    primaryClass = "gr-qc",
    reportNumber = "LIGO-P170817",
    doi = "10.1103/PhysRevLett.119.161101",
    journal = "Phys. Rev. Lett.",
    volume = "119",
    number = "16",
    pages = "161101",
    year = "2017"
}

@article{LIGOScientific:2017ync,
    author = "Abbott, B. P. and others",
    collaboration = "LIGO Scientific, Virgo, Fermi GBM, INTEGRAL, IceCube, AstroSat Cadmium Zinc Telluride Imager Team, IPN, Insight-Hxmt, ANTARES, Swift, AGILE Team, 1M2H Team, Dark Energy Camera GW-EM, DES, DLT40, GRAWITA, Fermi-LAT, ATCA, ASKAP, Las Cumbres Observatory Group, OzGrav, DWF (Deeper Wider Faster Program), AST3, CAASTRO, VINROUGE, MASTER, J-GEM, GROWTH, JAGWAR, CaltechNRAO, TTU-NRAO, NuSTAR, Pan-STARRS, MAXI Team, TZAC Consortium, KU, Nordic Optical Telescope, ePESSTO, GROND, Texas Tech University, SALT Group, TOROS, BOOTES, MWA, CALET, IKI-GW Follow-up, H.E.S.S., LOFAR, LWA, HAWC, Pierre Auger, ALMA, Euro VLBI Team, Pi of Sky, Chandra Team at McGill University, DFN, ATLAS Telescopes, High Time Resolution Universe Survey, RIMAS, RATIR, SKA South Africa/MeerKAT",
    title = "{Multi-messenger Observations of a Binary Neutron Star Merger}",
    eprint = "1710.05833",
    archivePrefix = "arXiv",
    primaryClass = "astro-ph.HE",
    reportNumber = "LIGO-P1700294, VIR-0802A-17, FERMILAB-PUB-17-478-A-AE-CD",
    doi = "10.3847/2041-8213/aa91c9",
    journal = "Astrophys. J. Lett.",
    volume = "848",
    number = "2",
    pages = "L12",
    year = "2017"
}

@article{LIGOScientific:2017zic,
    author = "Abbott, B. P. and others",
    collaboration = "LIGO Scientific, Virgo, Fermi-GBM, INTEGRAL",
    title = "{Gravitational Waves and Gamma-rays from a Binary Neutron Star Merger: GW170817 and GRB 170817A}",
    eprint = "1710.05834",
    archivePrefix = "arXiv",
    primaryClass = "astro-ph.HE",
    reportNumber = "LIGO-P1700308",
    doi = "10.3847/2041-8213/aa920c",
    journal = "Astrophys. J. Lett.",
    volume = "848",
    number = "2",
    pages = "L13",
    year = "2017"
}

@article{LIGOScientific:2018hze,
    author = "Abbott, B. P. and others",
    collaboration = "LIGO Scientific, Virgo",
    title = "{Properties of the binary neutron star merger GW170817}",
    eprint = "1805.11579",
    archivePrefix = "arXiv",
    primaryClass = "gr-qc",
    doi = "10.1103/PhysRevX.9.011001",
    journal = "Phys. Rev. X",
    volume = "9",
    number = "1",
    pages = "011001",
    year = "2019"
}

@ARTICLE{Ma+2025,
       author = {{Ma}, Xiaoran and {Wang}, Xiaofeng and {Mo}, Jun and {Howell}, D. Andrew and {Pellegrino}, Craig and {Zhang}, Jujia and {Yan}, Shengyu and {Arcavi}, Iair and {Chen}, Zhihao and {Farah}, Joseph and {Padilla Gonzalez}, Estefania and {Guo}, Fangzhou and {Hiramatsu}, Daichi and {Li}, Gaici and {Lin}, Han and {Liu}, Jialian and {McCully}, Curtis and {Newsome}, Megan and {Sai}, Hanna and {Terreran}, Giacomo and {Xiang}, Danfeng and {Zhang}, Xinhan and {Zhang}, Tianmeng},
        title = "{Supernovae at distances $< 40$~Mpc: I. Catalogues and fractions of supernovae in a complete sample}",
      journal = {Astron. Astrophys.},
         year = 2025,
        month = jun,
       volume = {698},
          eid = {A305},
        pages = {A305},
          doi = {10.1051/0004-6361/202452684},
archivePrefix = {arXiv},
       eprint = {2504.04393},
 primaryClass = {astro-ph.HE},
       adsurl = {https://ui.adsabs.harvard.edu/abs/2025A&A...698A.305M}
}

@article{Eroncel:2022efc,
    author = {Er{\"o}ncel, Cem and Servant, G{\'e}raldine},
    title = "{ALP dark matter mini-clusters from kinetic fragmentation}",
    eprint = "2207.10111",
    archivePrefix = "arXiv",
    primaryClass = "hep-ph",
    reportNumber = "DESY 22-115",
    doi = "10.1088/1475-7516/2023/01/009",
    journal = "JCAP",
    volume = "01",
    pages = "009",
    year = "2023"
}

@article{Fasiello:2025ngh,
    author = "Fasiello, Matteo and Lizarraga, Joanes and Papageorgiou, Alexandros and Urio, Ander",
    title = "{Kinetic Fragmentation of the QCD Axion on the Lattice}",
    eprint = "2507.01822",
    archivePrefix = "arXiv",
    primaryClass = "astro-ph.CO",
    month = "7",
    year = "2025"
}

@article{Arvanitaki:2019rax,
    author = "Arvanitaki, Asimina and Dimopoulos, Savas and Galanis, Marios and Lehner, Luis and Thompson, Jedidiah O. and Van Tilburg, Ken",
    title = "{Large-misalignment mechanism for the formation of compact axion structures: Signatures from the QCD axion to fuzzy dark matter}",
    eprint = "1909.11665",
    archivePrefix = "arXiv",
    primaryClass = "astro-ph.CO",
    doi = "10.1103/PhysRevD.101.083014",
    journal = "Phys. Rev. D",
    volume = "101",
    number = "8",
    pages = "083014",
    year = "2020"
}

@article{Manzari:2024jns,
    author = "Manzari, Claudio Andrea and Park, Yujin and Safdi, Benjamin R. and Savoray, Inbar",
    title = "{Supernova Axions Convert to Gamma Rays in Magnetic Fields of Progenitor Stars}",
    eprint = "2405.19393",
    archivePrefix = "arXiv",
    primaryClass = "hep-ph",
    doi = "10.1103/PhysRevLett.133.211002",
    journal = "Phys. Rev. Lett.",
    volume = "133",
    number = "21",
    pages = "211002",
    year = "2024"
}

@article{Fiorillo:2025gnd,
    author = "Fiorillo, Damiano F. G. and Gil Muyor, {\'A}ngel and Janka, Hans-Thomas and Raffelt, Georg G. and Vitagliano, Edoardo",
    title = "{Axion-photon conversion in transient compact stars: Systematics, constraints, and opportunities}",
    eprint = "2509.13322",
    archivePrefix = "arXiv",
    primaryClass = "hep-ph",
    month = "9",
    year = "2025"
}

@article{Buschmann:2021juv,
    author = "Buschmann, Malte and Dessert, Christopher and Foster, Joshua W. and Long, Andrew J. and Safdi, Benjamin R.",
    title = "{Upper Limit on the QCD Axion Mass from Isolated Neutron Star Cooling}",
    eprint = "2111.09892",
    archivePrefix = "arXiv",
    primaryClass = "hep-ph",
    doi = "10.1103/PhysRevLett.128.091102",
    journal = "Phys. Rev. Lett.",
    volume = "128",
    number = "9",
    pages = "091102",
    year = "2022"
}

@article{Fiorillo:2025zzx,
    author = "Fiorillo, Damiano F. G. and Lella, Alessandro and O'Hare, Ciaran A. J. and Vitagliano, Edoardo",
    title = "{Leading Bounds on Micrometer to Picometer Fifth Forces from Neutron Star Cooling}",
    eprint = "2506.19906",
    archivePrefix = "arXiv",
    primaryClass = "hep-ph",
    reportNumber = "BARI-TH/776-25",
    doi = "10.1103/tlqz-713s",
    journal = "Phys. Rev. Lett.",
    volume = "135",
    number = "21",
    pages = "211003",
    year = "2025"
}

@article{Raffelt:1987yt,
  author         = "Raffelt, G. and Seckel, D.",
  title          = "{Bounds on Exotic Particle Interactions from SN 1987A}",
  journal        = "Phys. Rev. Lett.",
  volume         = "60",
  year           = "1988",
  pages          = "1793",
  doi            = "10.1103/PhysRevLett.60.1793"
}

@article{Ellis:1987pk,
  author         = "Ellis, J. R. and Olive, K. A.",
  title          = "{Constraints on light particles from supernova SN 1987A}",
  journal        = "Phys. Lett. B",
  volume         = "193",
  year           = "1987",
  pages          = "525",
  doi            = "10.1016/0370-2693(87)91710-2"
}

@article{Turner:1987by,
  author         = "Turner, M. S.",
  title          = "{Axions from SN 1987A}",
  journal        = "Phys. Rev. Lett.",
  volume         = "60",
  year           = "1988",
  pages          = "1797",
  doi            = "10.1103/PhysRevLett.60.1797"
}

@article{Mayle:1987as,
  author         = "Mayle, R. and Wilson, J. R. and Ellis, J. R. and Olive, K. A. and Schramm, D. N. and Steigman, G.",
  title          = "{Constraints on Axions from SN 1987A}",
  journal        = "Phys. Lett. B",
  volume         = "203",
  year           = "1988",
  pages          = "188",
  doi            = "10.1016/0370-2693(88)91595-X"
}

@article{Mayle:1989yx,
  author         = "Mayle, R. and Wilson, J. R. and Ellis, J. R. and Olive, K. A. and Schramm, D. N. and Steigman, G.",
  title          = "{Updated constraints on axions from SN1987A}",
  journal        = "Phys. Lett. B",
  volume         = "219",
  year           = "1989",
  pages          = "515",
  doi            = "10.1016/0370-2693(89)91104-0"
}

@article{Brinkmann:1988vi,
  author         = "Brinkmann, R. P. and Turner, M. S.",
  title          = "{Numerical Rates for Nucleon-Nucleon Axion Bremsstrahlung}",
  journal        = "Phys. Rev. D",
  volume         = "38",
  year           = "1988",
  pages          = "2338",
  doi            = "10.1103/PhysRevD.38.2338"
}

@article{Burrows:1988ah,
  author         = "Burrows, A. and Turner, M. S. and Brinkmann, R. P.",
  title          = "{Axions and SN 1987A}",
  journal        = "Phys. Rev. D",
  volume         = "39",
  year           = "1989",
  pages          = "1020",
  doi            = "10.1103/PhysRevD.39.1020"
}

@article{Burrows:1990pk,
  author         = "Burrows, A. and Ressell, M. T. and Turner, M. S.",
  title          = "{Axions and SN1987A: Axion trapping}",
  journal        = "Phys. Rev. D",
  volume         = "42",
  year           = "1990",
  pages          = "3297",
  doi            = "10.1103/PhysRevD.42.3297"
}

@article{Janka:1995ir,
  author         = "Janka, H.-T. and Keil, W. and Raffelt, G. and Seckel, D.",
  title          = "{Nucleon spin fluctuations and the supernova emission of neutrinos and axions}",
  journal        = "Phys. Rev. Lett.",
  volume         = "76",
  year           = "1996",
  pages          = "2621",
  doi            = "10.1103/PhysRevLett.76.2621",
  eprint         = "astro-ph/9507023"
}

@article{Turner:1991ax,
  author         = "Turner, M. S.",
  title          = "{Dirac neutrinos and SN 1987A}",
  journal        = "Phys. Rev. D",
  volume         = "45",
  year           = "1992",
  pages          = "1066",
  doi            = "10.1103/PhysRevD.45.1066"
}

@article{Carenza:2020cis,
  author         = "Carenza, P. and Fore, B. and Giannotti, M. and Mirizzi, A. and Reddy, S.",
  title          = "{Enhanced Supernova Axion Emission and its Implications}",
  journal        = "Phys. Rev. Lett.",
  volume         = "126",
  year           = "2021",
  pages          = "071102",
  doi            = "10.1103/PhysRevLett.126.071102",
  eprint         = "2010.02943",
  archivePrefix  = "arXiv",
  primaryClass   = "hep-ph"
}

@article{Fischer:2021jfm,
  author         = "Fischer, T. and Carenza, P. and Fore, B. and Giannotti, M. and Mirizzi, A. and Reddy, S.",
  title          = "{Observable signatures of enhanced axion emission from protoneutron stars}",
  journal        = "Phys. Rev. D",
  volume         = "104",
  year           = "2021",
  pages          = "103012",
  doi            = "10.1103/PhysRevD.104.103012",
  eprint         = "2108.13726",
  archivePrefix  = "arXiv",
  primaryClass   = "hep-ph"
}

@article{Raffelt:1991pw,
  author         = "Raffelt, G. and Seckel, D.",
  title          = "{Multiple scattering suppression of the bremsstrahlung emission of neutrinos and axions in supernovae}",
  journal        = "Phys. Rev. Lett.",
  volume         = "67",
  year           = "1991",
  pages          = "2605",
  doi            = "10.1103/PhysRevLett.67.2605"
}

@article{Candon:2025sdm,
    author = "Cand{\'o}n, Francisco R. and Fiorillo, Damiano F. G. and Gil Muyor, {\'A}ngel and Janka, Hans-Thomas and Raffelt, Georg G. and Vitagliano, Edoardo",
    title = "{Stripped-Envelope Supernovae for QCD Axion Detection}",
    eprint = "2511.13815",
    archivePrefix = "arXiv",
    primaryClass = "hep-ph",
    month = "11",
    year = "2025"
}

@article{Raffelt:1998pa,
  author         = "Raffelt, G. and Sigl, G.",
  title          = "{Numerical toy model calculation of the nucleon spin autocorrelation function in a supernova core}",
  journal        = "Phys. Rev. D",
  volume         = "60",
  year           = "1999",
  pages          = "023001",
  doi            = "10.1103/PhysRevD.60.023001",
  eprint         = "hep-ph/9808476",
  archivePrefix  = "arXiv"
}

@article{Sedrakian:2000kc,
  author         = "Sedrakian, A. and Dieperink, A. E. L.",
  title          = "{Coherent neutrino radiation in supernovae at two loops}",
  journal        = "Phys. Rev. D",
  volume         = "62",
  year           = "2000",
  pages          = "083002",
  doi            = "10.1103/PhysRevD.62.083002",
  eprint         = "astro-ph/0002228",
  archivePrefix  = "arXiv"
}

@article{vanDalen:2003zw,
  author         = "van Dalen, E. N. E. and Dieperink, A. E. L. and Tjon, J. A.",
  title          = "{Neutrino emission in neutron stars}",
  journal        = "Phys. Rev. C",
  volume         = "67",
  year           = "2003",
  pages          = "065807",
  doi            = "10.1103/PhysRevC.67.065807",
  eprint         = "nucl-th/0303037",
  archivePrefix  = "arXiv"
}

@article{Lykasov:2008yz,
  author         = "Lykasov, G. I. and Pethick, C. J. and Schwenk, A.",
  title          = "{Unified approach to structure factors and neutrino processes in nucleon matter}",
  journal        = "Phys. Rev. C",
  volume         = "78",
  year           = "2008",
  pages          = "045803",
  doi            = "10.1103/PhysRevC.78.045803",
  eprint         = "0808.0330",
  archivePrefix  = "arXiv",
  primaryClass   = "nucl-th"
}

@article{Carenza:2019pxu,
  author         = "Carenza, P. and Fischer, T. and Giannotti, M. and Guo, G. and Martínez-Pinedo, G. and Mirizzi, A.",
  title          = "{Improved axion emissivity from a supernova via nucleon-nucleon bremsstrahlung}",
  journal        = "JCAP",
  volume         = "10",
  year           = "2019",
  pages          = "016",
  doi            = "10.1088/1475-7516/2019/10/016",
  eprint         = "1906.11844",
  archivePrefix  = "arXiv",
  primaryClass   = "hep-ph",
  note           = "Erratum: JCAP 05 (2020) E01, doi:10.1088/1475-7516/2020/05/E01"
}

@article{ericson1988pions,
  title={Pions and nuclei},
  author={Ericson, Torleif and Weise, Wolfram},
  year={1988}
}

@article{Fore:2023gwv,
    author = "Fore, Bryce and Kaiser, Norbert and Reddy, Sanjay and Warrington, Neill C.",
    title = "{Mass of charged pions in neutron-star matter}",
    eprint = "2301.07226",
    archivePrefix = "arXiv",
    primaryClass = "nucl-th",
    doi = "10.1103/PhysRevC.110.025803",
    journal = "Phys. Rev. C",
    volume = "110",
    number = "2",
    pages = "025803",
    year = "2024"
}

@article{Fore:2019wib,
    author = "Fore, Bryce and Reddy, Sanjay",
    title = "{Pions in hot dense matter and their astrophysical implications}",
    eprint = "1911.02632",
    archivePrefix = "arXiv",
    primaryClass = "astro-ph.HE",
    reportNumber = "INT-PUB-19-046",
    doi = "10.1103/PhysRevC.101.035809",
    journal = "Phys. Rev. C",
    volume = "101",
    number = "3",
    pages = "035809",
    year = "2020"
}

@article{Shternin:2018dcn,
    author = "Shternin, P. S. and Baldo, M. and Haensel, P.",
    title = "{In-medium enhancement of the modified Urca neutrino reaction rates}",
    eprint = "1807.06569",
    archivePrefix = "arXiv",
    primaryClass = "astro-ph.HE",
    doi = "10.1016/j.physletb.2018.09.035",
    journal = "Phys. Lett. B",
    volume = "786",
    pages = "28--34",
    year = "2018"
}

@article{Schwenk:2003bc,
    author = "Schwenk, Achim and Friman, Bengt",
    title = "{Polarization contributions to the spin dependence of the effective interaction in neutron matter}",
    eprint = "nucl-th/0307089",
    archivePrefix = "arXiv",
    doi = "10.1103/PhysRevLett.92.082501",
    journal = "Phys. Rev. Lett.",
    volume = "92",
    pages = "082501",
    year = "2004"
}

@article{Schwenk:2003pj,
    author = "Schwenk, Achim and Jaikumar, Prashanth and Gale, Charles",
    title = "{Neutrino bremsstrahlung in neutron matter from effective nuclear interactions}",
    eprint = "nucl-th/0309072",
    archivePrefix = "arXiv",
    doi = "10.1016/j.physletb.2004.01.036",
    journal = "Phys. Lett. B",
    volume = "584",
    pages = "241--250",
    year = "2004"
}

@article{ericson1989axion,
  title={Axion emission from SN1987A. Nuclear physics constraints},
  author={Ericson, Torleif EO and Mathiot, J-F},
  journal={Physics Letters B},
  volume={219},
  number={4},
  pages={507--514},
  year={1989},
  publisher={Elsevier}
}

@article{Ho:2022oaw,
    author = "Ho, Shu-Yu and Kim, Jongkuk and Ko, Pyungwon and Park, Jae-hyeon",
    title = "{Supernova axion emissivity with {\ensuremath{\Delta}}(1232) resonance in heavy baryon chiral perturbation theory}",
    eprint = "2212.01155",
    archivePrefix = "arXiv",
    primaryClass = "hep-ph",
    reportNumber = "KIAS-P22082",
    doi = "10.1103/PhysRevD.107.075002",
    journal = "Phys. Rev. D",
    volume = "107",
    number = "7",
    pages = "075002",
    year = "2023"
}

@article{Choi:2021ign,
    author = "Choi, Kiwoon and Kim, Hee Jung and Seong, Hyeonseok and Shin, Chang Sub",
    title = "{Axion emission from supernova with axion-pion-nucleon contact interaction}",
    eprint = "2110.01972",
    archivePrefix = "arXiv",
    primaryClass = "hep-ph",
    reportNumber = "CTPU-PTC-21-35",
    doi = "10.1007/JHEP02(2022)143",
    journal = "JHEP",
    volume = "02",
    pages = "143",
    year = "2022"
}

@article{Chang:2018rso,
  author         = "Chang, J. H. and Essig, R. and McDermott, S. D.",
  title          = "{Supernova 1987A Constraints on Sub-GeV Dark Sectors, Millicharged Particles, the QCD Axion, and an Axion-like Particle}",
  journal        = "JHEP",
  volume         = "09",
  year           = "2018",
  pages          = "051",
  doi            = "10.1007/JHEP09(2018)051",
  eprint         = "1803.00993",
  archivePrefix  = "arXiv",
  primaryClass   = "hep-ph"
}

@article{Fiorillo:2025yzf,
    author = "Fiorillo, Damiano F. G. and Pitik, Tetyana and Vitagliano, Edoardo",
    title = "{Energy Transfer by Feebly Interacting Particles in Supernovae: The Trapping Regime}",
    eprint = "2503.13653",
    archivePrefix = "arXiv",
    primaryClass = "hep-ph",
    doi = "10.1103/cz94-dqxt",
    journal = "Phys. Rev. Lett.",
    volume = "135",
    number = "7",
    pages = "071005",
    year = "2025"
}

@article{Caputo:2022rca,
    author = "Caputo, Andrea and Raffelt, Georg and Vitagliano, Edoardo",
    title = "{Radiative transfer in stars by feebly interacting bosons}",
    eprint = "2204.11862",
    archivePrefix = "arXiv",
    primaryClass = "astro-ph.SR",
    doi = "10.1088/1475-7516/2022/08/045",
    journal = "JCAP",
    volume = "08",
    number = "08",
    pages = "045",
    year = "2022"
}

@article{Fiorillo:2023frv,
    author = "Fiorillo, Damiano F. G. and Heinlein, Malte and Janka, Hans-Thomas and Raffelt, Georg and Vitagliano, Edoardo and Bollig, Robert",
    title = "{Supernova simulations confront SN 1987A neutrinos}",
    eprint = "2308.01403",
    archivePrefix = "arXiv",
    primaryClass = "astro-ph.HE",
    doi = "10.1103/PhysRevD.108.083040",
    journal = "Phys. Rev. D",
    volume = "108",
    number = "8",
    pages = "083040",
    year = "2023"
}

@article{Raffelt:2006cw,
    author = "Raffelt, Georg G.",
    editor = "Kuster, Markus and Raffelt, Georg and Beltran, Berta",
    title = "{Astrophysical axion bounds}",
    eprint = "hep-ph/0611350",
    archivePrefix = "arXiv",
    reportNumber = "MPP-2006-172",
    doi = "10.1007/978-3-540-73518-2_3",
    journal = "Lect. Notes Phys.",
    volume = "741",
    pages = "51--71",
    year = "2008"
}

@article{Hanhart:2000ae,
  author         = "Hanhart, C. and Phillips, D. R. and Reddy, S.",
  title          = "{Neutrino and axion emissivities of neutron stars from nucleon-nucleon scattering data}",
  journal        = "Phys. Lett. B",
  volume         = "499",
  year           = "2001",
  pages          = "9--15",
  doi            = "10.1016/S0370-2693(00)01382-4",
  eprint         = "astro-ph/0003445"
}

@article{Iwamoto:1984ir,
    author = "Iwamoto, N.",
    title = "{Axion Emission from Neutron Stars}",
    doi = "10.1103/PhysRevLett.53.1198",
    journal = "Phys. Rev. Lett.",
    volume = "53",
    pages = "1198--1201",
    year = "1984"
}

@article{Carenza:2024ehj,
    author = "Carenza, Pierluca and Giannotti, Maurizio and Isern, Jordi and Mirizzi, Alessandro and Straniero, Oscar",
    title = "{Axion astrophysics}",
    eprint = "2411.02492",
    archivePrefix = "arXiv",
    primaryClass = "hep-ph",
    reportNumber = "BARI-TH/66-24",
    doi = "10.1016/j.physrep.2025.02.002",
    journal = "Phys. Rept.",
    volume = "1117",
    pages = "1--102",
    year = "2025"
}

@article{Caputo:2024oqc,
    author = "Caputo, Andrea and Raffelt, Georg",
    title = "{Astrophysical Axion Bounds: The 2024 Edition}",
    eprint = "2401.13728",
    archivePrefix = "arXiv",
    primaryClass = "hep-ph",
    reportNumber = "MPP-2024-13, CERN-TH-2024-013",
    doi = "10.22323/1.454.0041",
    journal = "PoS",
    volume = "COSMICWISPers",
    pages = "041",
    year = "2024"
}

@article{Lecce:2025dbz,
    author = "Lecce, Francesca and Lella, Alessandro and Lucente, Giuseppe and Vijayan, Vimal and Bauswein, Andreas and Giannotti, Maurizio and Mirizzi, Alessandro",
    title = "{Probing axionlike particles with multimessenger observations of neutron star mergers}",
    eprint = "2504.02032",
    archivePrefix = "arXiv",
    primaryClass = "hep-ph",
    reportNumber = "BARI-TH/773-25",
    doi = "10.1103/krf3-lm4s",
    journal = "Phys. Rev. D",
    volume = "112",
    number = "2",
    pages = "023001",
    year = "2025"
}

@article{Fiorillo:2022piv,
    author = "Fiorillo, Damiano F. G. and Iocco, Fabio",
    title = "{Axions from neutron star mergers}",
    doi = "10.1103/PhysRevD.105.123007",
    journal = "Phys. Rev. D",
    volume = "105",
    number = "12",
    pages = "123007",
    year = "2022"
}

@article{CAST_He4_new,
  title = {New solar axion search using the CERN Axion Solar Telescope with $^{4}\mathrm{He}$ filling},
  author = {Arik, M. and Aune, S. and Barth, K. and Belov, A. and Br\"auninger, H. and Bremer, J. and Burwitz, V. and Cantatore, G. and Carmona, J. M. and Cetin, S. A. and Collar, J. I. and Da Riva, E. and Dafni, T. and Davenport, M. and Dermenev, A. and Eleftheriadis, C. and Elias, N. and Fanourakis, G. and Ferrer-Ribas, E. and Gal\'an, J. and Garc\'{\i}a, J. A. and Gardikiotis, A. and Garza, J. G. and Gazis, E. N. and Geralis, T. and Georgiopoulou, E. and Giomataris, I. and Gninenko, S. and G\'omez Marzoa, M. and Hasinoff, M. D. and Hoffmann, D. H. H. and Iguaz, F. J. and Irastorza, I. G. and Jacoby, J. and Jakov\ifmmode \check{c}\else \v{c}\fi{}i\ifmmode \acute{c}\else \'{c}\fi{}, K. and Karuza, M. and Kavuk, M. and Kr\ifmmode \check{c}\else \v{c}\fi{}mar, M. and Kuster, M. and Laki\ifmmode \acute{c}\else \'{c}\fi{}, B. and Laurent, J. M. and Liolios, A. and Ljubi\ifmmode \check{c}\else \v{c}\fi{}i\ifmmode \acute{c}\else \'{c}\fi{}, A. and Luz\'on, G. and Neff, S. and Niinikoski, T. and Nordt, A. and Ortega, I. and Papaevangelou, T. and Pivovaroff, M. J. and Raffelt, G. and Rodr\'{\i}guez, A. and Rosu, M. and Ruz, J. and Savvidis, I. and Shilon, I. and Solanki, S. K. and Stewart, L. and Tom\'as, A. and Vafeiadis, T. and Villar, J. and Vogel, J. K. and Yildiz, S. C. and Zioutas, K.},
  collaboration = {CAST Collaboration},
  journal = {Phys. Rev. D},
  volume = {92},
  issue = {2},
  pages = {021101},
  numpages = {6},
  year = {2015},
  month = {Jul},
  publisher = {American Physical Society},
  doi = {10.1103/PhysRevD.92.021101},
  url = {https://link.aps.org/doi/10.1103/PhysRevD.92.021101}
}

@article{CAST_He3_new,
  title = {Search for Solar Axions by the CERN Axion Solar Telescope with $^{3}\mathrm{He}$ Buffer Gas: Closing the Hot Dark Matter Gap},
  author = {Arik, M. and Aune, S. and Barth, K. and Belov, A. and Borghi, S. and Br\"auninger, H. and Cantatore, G. and Carmona, J. M. and Cetin, S. A. and Collar, J. I. and Da Riva, E. and Dafni, T. and Davenport, M. and Eleftheriadis, C. and Elias, N. and Fanourakis, G. and Ferrer-Ribas, E. and Friedrich, P. and Gal\'an, J. and Garc\'{\i}a, J. A. and Gardikiotis, A. and Garza, J. G. and Gazis, E. N. and Geralis, T. and Georgiopoulou, E. and Giomataris, I. and Gninenko, S. and G\'omez, H. and G\'omez Marzoa, M. and Gruber, E. and Guth\"orl, T. and Hartmann, R. and Hauf, S. and Haug, F. and Hasinoff, M. D. and Hoffmann, D. H. H. and Iguaz, F. J. and Irastorza, I. G. and Jacoby, J. and Jakov\ifmmode \check{c}\else \v{c}\fi{}i\ifmmode \acute{c}\else \'{c}\fi{}, K. and Karuza, M. and K\"onigsmann, K. and Kotthaus, R. and Kr\ifmmode \check{c}\else \v{c}\fi{}mar, M. and Kuster, M. and Laki\ifmmode \acute{c}\else \'{c}\fi{}, B. and Lang, P. M. and Laurent, J. M. and Liolios, A. and Ljubi\ifmmode \check{c}\else \v{c}\fi{}i\ifmmode \acute{c}\else \'{c}\fi{}, A. and Luz\'on, G. and Neff, S. and Niinikoski, T. and Nordt, A. and Papaevangelou, T. and Pivovaroff, M. J. and Raffelt, G. and Riege, H. and Rodr\'{\i}guez, A. and Rosu, M. and Ruz, J. and Savvidis, I. and Shilon, I. and Silva, P. S. and Solanki, S. K. and Stewart, L. and Tom\'as, A. and Tsagri, M. and van Bibber, K. and Vafeiadis, T. and Villar, J. and Vogel, J. K. and Yildiz, S. C. and Zioutas, K.},
  collaboration = {CAST Collaboration},
  journal = {Phys. Rev. Lett.},
  volume = {112},
  issue = {9},
  pages = {091302},
  numpages = {6},
  year = {2014},
  month = {Mar},
  publisher = {American Physical Society},
  doi = {10.1103/PhysRevLett.112.091302},
  url = {https://link.aps.org/doi/10.1103/PhysRevLett.112.091302}
}

@article{CAST_He3,
  title = {Search for Sub-eV Mass Solar Axions by the CERN Axion Solar Telescope with $^{3}\mathrm{He}$ Buffer Gas},
  author = {Arik, M. and Aune, S. and Barth, K. and Belov, A. and Borghi, S. and Br\"auninger, H. and Cantatore, G. and Carmona, J. M. and Cetin, S. A. and Collar, J. I. and Dafni, T. and Davenport, M. and Eleftheriadis, C. and Elias, N. and Ezer, C. and Fanourakis, G. and Ferrer-Ribas, E. and Friedrich, P. and Gal\'an, J. and Garc\'{\i}a, J. A. and Gardikiotis, A. and Gazis, E. N. and Geralis, T. and Giomataris, I. and Gninenko, S. and G\'omez, H. and Gruber, E. and Guth\"orl, T. and Hartmann, R. and Haug, F. and Hasinoff, M. D. and Hoffmann, D. H. H. and Iguaz, F. J. and Irastorza, I. G. and Jacoby, J. and Jakov\ifmmode \check{c}\else \v{c}\fi{}i\ifmmode \acute{c}\else \'{c}\fi{}, K. and Karuza, M. and K\"onigsmann, K. and Kotthaus, R. and Kr\ifmmode \check{c}\else \v{c}\fi{}mar, M. and Kuster, M. and Laki\ifmmode \acute{c}\else \'{c}\fi{}, B. and Laurent, J. M. and Liolios, A. and Ljubi\ifmmode \check{c}\else \v{c}\fi{}i\ifmmode \acute{c}\else \'{c}\fi{}, A. and Lozza, V. and Lutz, G. and Luz\'on, G. and Morales, J. and Niinikoski, T. and Nordt, A. and Papaevangelou, T. and Pivovaroff, M. J. and Raffelt, G. and Rashba, T. and Riege, H. and Rodr\'{\i}guez, A. and Rosu, M. and Ruz, J. and Savvidis, I. and Silva, P. S. and Solanki, S. K. and Stewart, L. and Tom\'as, A. and Tsagri, M. and van Bibber, K. and Vafeiadis, T. and Villar, J. A. and Vogel, J. K. and Yildiz, S. C. and Zioutas, K.},
  collaboration = {CAST Collaboration},
  journal = {Phys. Rev. Lett.},
  volume = {107},
  issue = {26},
  pages = {261302},
  numpages = {5},
  year = {2011},
  month = {Dec},
  publisher = {American Physical Society},
  doi = {10.1103/PhysRevLett.107.261302},
  url = {https://link.aps.org/doi/10.1103/PhysRevLett.107.261302}
}

@article{VanBibber_Raffelt,
  title = {Design for a practical laboratory detector for solar axions},
  author = {van Bibber, K. and McIntyre, P. M. and Morris, D. E. and Raffelt, G. G.},
  journal = {Phys. Rev. D},
  volume = {39},
  issue = {8},
  pages = {2089--2099},
  numpages = {0},
  year = {1989},
  month = {Apr},
  publisher = {American Physical Society},
  doi = {10.1103/PhysRevD.39.2089},
  url = {https://link.aps.org/doi/10.1103/PhysRevD.39.2089}
}

@article{CAST_TPC,
doi = {10.1088/1367-2630/9/6/171},
url = {https://doi.org/10.1088/1367-2630/9/6/171},
year = {2007},
month = {jun},
publisher = {},
volume = {9},
number = {6},
pages = {171},
author = {Autiero, D and Beltrán, B and Carmona, J M and Cebrián, S and Chesi, E and Davenport, M and Delattre, M and Di Lella, L and Formenti, F and Irastorza, I G and Gómez, H and Hasinoff, M and Lakić, B and Luzón, G and Morales, J and Musa, L and Ortiz, A and Placci, A and Rodrigurez, A and Ruz, J and Villar, J A and Zioutas, K},
title = {The CAST time projection chamber},
journal = {New Journal of Physics}
}

@article{Notari:2022ffe,
    author = "Notari, Alessio and Rompineve, Fabrizio and Villadoro, Giovanni",
    title = "{Improved Hot Dark Matter Bound on the QCD Axion}",
    eprint = "2211.03799",
    archivePrefix = "arXiv",
    primaryClass = "hep-ph",
    reportNumber = "CERN-TH-2022-165",
    doi = "10.1103/PhysRevLett.131.011004",
    journal = "Phys. Rev. Lett.",
    volume = "131",
    number = "1",
    pages = "011004",
    year = "2023"
}

@article{Bouzoud:2024bom,
    author = "Bouzoud, Killian and Ghiglieri, Jacopo",
    title = "{Thermal axion production at hard and soft momenta}",
    eprint = "2404.06113",
    archivePrefix = "arXiv",
    primaryClass = "hep-ph",
    doi = "10.1007/JHEP01(2025)163",
    journal = "JHEP",
    volume = "01",
    pages = "163",
    year = "2025"
}

@article{CAST_MM,
doi = {10.1088/1367-2630/9/6/170},
url = {https://doi.org/10.1088/1367-2630/9/6/170},
year = {2007},
month = {jun},
publisher = {},
volume = {9},
number = {6},
pages = {170},
author = {Abbon, P and Andriamonje, S and Aune, S and Dafni, T and Davenport, M and Delagnes, E and de Oliveira, R and Fanourakis, G and Ferrer Ribas, E and Franz, J and Geralis, T and Giganon, A and Gros, M and Giomataris, Y and Irastorza, I G and Kousouris, K and Morales, J and Papaevangelou, T and Ruz, J and Zachariadou, K and Zioutas, K},
title = {The Micromegas detector of the CAST experiment},
journal = {New Journal of Physics}
}

@article{CAST_CCD,
doi = {10.1088/1367-2630/9/6/169},
url = {https://doi.org/10.1088/1367-2630/9/6/169},
year = {2007},
month = {jun},
publisher = {},
volume = {9},
number = {6},
pages = {169},
author = {Kuster, M and Bräuninger, H and Cebrián, S and Davenport, M and Eleftheriadis, C and Englhauser, J and Fischer, H and Franz, J and Friedrich, P and Hartmann, R and Heinsius, F H and Hoffmann, D H H and Hoffmeister, G and Joux, J N and Kang, D and Königsmann, K and Kotthaus, R and Papaevangelou, T and Lasseur, C and Lippitsch, A and Lutz, G and Morales, J and Rodríguez, A and Strüder, L and Vogel, J and Zioutas},
title = {The x-ray telescope of CAST},
journal = {New Journal of Physics}
}

@article{CAST_GRIDPix,
title = {A GridPix-based X-ray detector for the CAST experiment},
journal = {Nuclear Instruments and Methods in Physics Research Section A: Accelerators, Spectrometers, Detectors and Associated Equipment},
volume = {867},
pages = {101-107},
year = {2017},
issn = {0168-9002},
doi = {https://doi.org/10.1016/j.nima.2017.04.007},
url = {https://www.sciencedirect.com/science/article/pii/S0168900217304400},
author = {C. Krieger and J. Kaminski and M. Lupberger and K. Desch}
}

@book{Raffelt:1996wa,
    author = "Raffelt, G. G.",
    title = "{Stars as laboratories for fundamental physics}: {The astrophysics of neutrinos, axions, and other weakly interacting particles}",
    isbn = "978-0-226-70272-8",
    month = "5",
    year = "1996"
}

@Article{BabyIAXO_raytracing,
author={Ahyoune, S.
and Altenm{\"u}ller, K.
and Antol{\'i}n, I.
and Basso, S.
and Brun, P.
and Cand{\'o}n, F. R.
and Castel, J. F.
and Cebri{\'a}n, S.
and Chouhan, D.
and Della Ceca, R.
and Cervera-Cort{\'e}s, M.
and Chernov, V.
and Civitani, M. M.
and Cogollos, C.
and Costa, E.
and Cotroneo, V.
and Dafn{\'i}, T.
and Derbin, A.
and Desch, K.
and D{\'i}az-Mart{\'i}n, M. C.
and D{\'i}az-Morcillo, A.
and D{\'i}ez-Ib{\'a}{\~{n}}ez, D.
and Diez Pardos, C.
and Dinter, M.
and D{\"o}brich, B.
and Drachnev, I.
and Dudarev, A.
and Ezquerro, A.
and Fabiani, S.
and Ferrer-Ribas, E.
and Finelli, F.
and Fleck, I.
and Gal{\'a}n, J.
and Galanti, G.
and Galaverni, M.
and Garc{\'i}a, J. A.
and Garc{\'i}a-Barcel{\'o}, J. M.
and Gastaldo, L.
and Giannotti, M.
and Giganon, A.
and Goblin, C.
and Goyal, N.
and Gu, Y.
and Hagge, L.
and Helary, L.
and Hengstler, D.
and Heuchel, D.
and Hoof, S.
and Iglesias-Marzoa, R.
and Iguaz, F. J.
and I{\~{n}}iguez, C.
and Irastorza, I. G.
and Jakov{\v{c}}i{\'{c}}, K.
and K{\"a}fer, D.
and Kaminski, J.
and Karstensen, S.
and Law, M.
and Lindner, A.
and Loidl, M.
and Loiseau, C.
and L{\'o}pez-Alegre, G.
and Lozano-Guerrero, A.
and Lubsandorzhiev, B.
and Luz{\'o}n, G.
and Manthos, I.
and Margalejo, C.
and Mar{\'i}n-Franch, A.
and Marqu{\'e}s, J.
and Marutzky, F.
and Menneglier, C.
and Mentink, M.
and Mertens, S.
and Miralda-Escud{\'e}, J.
and Mirallas, H.
and Muleri, F.
and Muratova, V.
and Navarro-Madrid, J. R.
and Navick, X. F.
and Nikolopoulos, K.
and Notari, A.
and Nozik, A.
and Obis, L.
and Ortiz-de-Sol{\'o}rzano, A.
and O'Shea, T.
and von Oy, J.
and Pareschi, G.
and Papaevangelou, T.
and Perez, K.
and P{\'e}rez, O.
and Picatoste, E.
and Pivovaroff, M. J.
and Porr{\'o}n, J.
and Puyuelo, M. J.
and Quintana, A.
and Redondo, J.
and Reuther, D.
and Ringwald, A.
and Rodrigues, M.
and Rubini, A.
and Rueda-Teruel, S.
and Rueda-Teruel, F.
and Ruiz-Ch{\'o}liz, E.
and Ruz, J.
and Schaffran, J.
and Schiffer, T.
and Schmidt, S.
and Schneekloth, U.
and Sch{\"o}nfeld, L.
and Schott, M.
and Segui, L.
and Singh, U. R.
and Soffitta, P.
and Spiga, D.
and Stern, M.
and Straniero, O.
and Tavecchio, F.
and Unzhakov, E.
and Ushakov, N. A.
and Vecchi, G.
and Vogel, J. K.
and Voronin, D. M.
and Ward, R.
and Weltman, A.
and Wiesinger, C.
and Wolf, R.
and Yanes-D{\'i}az, A.
and Yu, Y.
and collaboration, IAXO},
title={An accurate solar axions ray-tracing response of BabyIAXO},
journal={Journal of High Energy Physics},
year={2025},
month={Feb},
day={25},
volume={2025},
number={2},
pages={159},
issn={1029-8479},
doi={10.1007/JHEP02(2025)159},
url={https://doi.org/10.1007/JHEP02(2025)159}
}

@article{Biljana_PRD,
  title = {Weighing the solar axion},
  author = {Dafni, Theopisti and O'Hare, Ciaran A. J. and Laki\ifmmode \acute{c}\else \'{c}\fi{}, Biljana and Gal\'an, Javier and Iguaz, Francisco J. and Irastorza, Igor G. and Jakov\ifmmode \check{c}\else \v{c}\fi{}i\ifmmode \acute{c}\else \'{c}\fi{}, Kre\ifmmode \check{s}\else \v{s}\fi{}imir and Luz\'on, Gloria and Redondo, Javier and Ruiz Ch\'oliz, Elisa},
  journal = {Phys. Rev. D},
  volume = {99},
  issue = {3},
  pages = {035037},
  numpages = {13},
  year = {2019},
  month = {Feb},
  publisher = {American Physical Society},
  doi = {10.1103/PhysRevD.99.035037},
  url = {https://link.aps.org/doi/10.1103/PhysRevD.99.035037}
}

@INPROCEEDINGS{JRuz_IEEE_mass,
  author={Ruz, J.},
  booktitle={2011 IEEE Nuclear Science Symposium Conference Record}, 
  title={Determination of effective axion masses in the helium-3 buffer of CAST}, 
  year={2011},
  volume={},
  number={},
  pages={135-140},
  doi={10.1109/NSSMIC.2011.6154466}}

@Article{BabyIAXO_Conceptual,
author={Abeln, A.
and Altenm{\"u}ller, K.
and Arguedas Cuendis, S.
and Armengaud, E.
and Atti{\'e}, D.
and Aune, S.
and Basso, S.
and Berg{\'e}, L.
and Biasuzzi, B.
and Borges De Sousa, P. T. C.
and Brun, P.
and Bykovskiy, N.
and Calvet, D.
and Carmona, J. M.
and Castel, J. F.
and Cebri{\'a}n, S.
and Chernov, V.
and Christensen, F. E.
and Civitani, M. M.
and Cogollos, C.
and Dafn{\'i}, T.
and Derbin, A.
and Desch, K.
and D{\'i}ez, D.
and Dinter, M.
and D{\"o}brich, B.
and Drachnev, I.
and Dudarev, A.
and Dumoulin, L.
and Ferreira, D. D. M.
and Ferrer-Ribas, E.
and Fleck, I.
and Gal{\'a}n, J.
and Gasc{\'o}n, D.
and Gastaldo, L.
and Giannotti, M.
and Giomataris, Y.
and Giuliani, A.
and Gninenko, S.
and Golm, J.
and Golubev, N.
and Hagge, L.
and Hahn, J.
and Hailey, C. J.
and Hengstler, D.
and Henriksen, P. L.
and Houdy, T.
and Iglesias-Marzoa, R.
and Iguaz, F. J.
and Irastorza, I. G.
and I{\~{n}}iguez, C.
and Jakov{\v{c}}i{\'{c}}, K.
and Kaminski, J.
and Kanoute, B.
and Karstensen, S.
and Kravchuk, L.
and Laki{\'{c}}, B.
and Lasserre, T.
and Laurent, P.
and Limousin, O.
and Lindner, A.
and Loidl, M.
and Lomskaya, I.
and L{\'o}pez-Alegre, G.
and Lubsandorzhiev, B.
and Ludwig, K.
and Luz{\'o}n, G.
and Malbrunot, C.
and Margalejo, C.
and Marin-Franch, A.
and Marnieros, S.
and Marutzky, F.
and Mauricio, J.
and Menesguen, Y.
and Mentink, M.
and Mertens, S.
and Mescia, F.
and Miralda-Escud{\'e}, J.
and Mirallas, H.
and Mols, J. P.
and Muratova, V.
and Navick, X. F.
and Nones, C.
and Notari, A.
and Nozik, A.
and Obis, L.
and Oriol, C.
and Orsini, F.
and Ortiz de Sol{\'o}rzano, A.
and Oster, S.
and Pais Da Silva, H. P.
and Pantuev, V.
and Papaevangelou, T.
and Pareschi, G.
and Perez, K.
and P{\'e}rez, O.
and Picatoste, E.
and Pivovaroff, M. J.
and Poda, D. V.
and Redondo, J.
and Ringwald, A.
and Rodrigues, M.
and Rueda-Teruel, F.
and Rueda-Teruel, S.
and Ruiz-Choliz, E.
and Ruz, J.
and Saemann, E. O.
and Salvado, J.
and Schiffer, T.
and Schmidt, S.
and Schneekloth, U.
and Schott, M.
and Segui, L.
and Tavecchio, F.
and ten Kate, H. H. J.
and Tkachev, I.
and Troitsky, S.
and Unger, D.
and Unzhakov, E.
and Ushakov, N.
and Vogel, J. K.
and Voronin, D.
and Weltman, A.
and Werthenbach, U.
and Wuensch, W.
and Yanes-D{\'i}az, A.
and collaboration, The IAXO},
title={Conceptual design of BabyIAXO, the intermediate stage towards the International Axion Observatory},
journal={Journal of High Energy Physics},
year={2021},
month={May},
day={17},
volume={2021},
number={5},
pages={137},
issn={1029-8479},
doi={10.1007/JHEP05(2021)137},
url={https://doi.org/10.1007/JHEP05(2021)137}
}

@article{IAXO_PhysPot,
doi = {10.1088/1475-7516/2019/06/047},
url = {https://doi.org/10.1088/1475-7516/2019/06/047},
year = {2019},
month = {jun},
publisher = {},
volume = {2019},
number = {06},
pages = {047},
author = {Armengaud, E. and Attié, D. and Basso, S. and Brun, P. and Bykovskiy, N. and Carmona, J.M. and Castel, J.F. and Cebrián, S. and Cicoli, M. and Civitani, M. and Cogollos, C. and Conlon, J.P. and Costa, D. and Dafni, T. and Daido, R. and Derbin, A.V. and Descalle, M.A. and Desch, K. and Dratchnev, I.S. and Döbrich, B. and Dudarev, A. and Ferrer-Ribas, E. and Fleck, I. and Galán, J. and Galanti, G. and Garrido, L. and Gascon, D. and Gastaldo, L. and Germani, C. and Ghisellini, G. and Giannotti, M. and Giomataris, I. and Gninenko, S. and Golubev, N. and Graciani, R. and Irastorza, I.G. and Jakovčić, K. and Kaminski, J. and Krčmar, M. and Krieger, C. and Lakić, B. and Lasserre, T. and Laurent, P. and Limousin, O. and Lindner, A. and Lomskaya, I. and Lubsandorzhiev, B. and Luzón, G. and Marsh, M. C. D. and Margalejo, C. and Mescia, F. and Meyer, M. and Miralda-Escudé, J. and Mirallas, H. and Muratova, V.N. and Navick, X.F. and Nones, C. and Notari, A. and Nozik, A. and de Solórzano, A. Ortiz and Pantuev, V. and Papaevangelou, T. and Pareschi, G. and Perez, K. and Picatoste, E. and Pivovaroff, M.J. and Redondo, J. and Ringwald, A. and Roncadelli, M. and Ruiz-Chóliz, E. and Ruz, J. and Saikawa, K. and Salvadó, J. and Samperiz, M.P. and Schiffer, T. and Schmidt, S. and Schneekloth, U. and Schott, M. and Silva, H. and Tagliaferri, G. and Takahashi, F. and Tavecchio, F. and Kate, H. ten and Tkachev, I. and Troitsky, S. and Unzhakov, E. and Vedrine, P. and Vogel, J.K. and Weinsheimer, C. and Weltman, A. and Yin, W.},
title = {Physics potential of the International Axion Observatory (IAXO)},
journal = {Journal of Cosmology and Astroparticle Physics}
}

@article{zhang2020large,
  title={Large dynamical axion field in topological antiferromagnetic insulator Mn2Bi2Te5},
  author={Zhang, Jinlong and Wang, Dinghui and Shi, Minji and Zhu, Tongshuai and Zhang, Haijun and Wang, Jing},
  journal={Chinese Physics Letters},
  volume={37},
  number={7},
  pages={077304},
  year={2020},
  publisher={IOP Publishing}
}

@article{qi2008topological,
  title={Topological field theory of time-reversal invariant insulators},
  author={Qi, Xiao-Liang and Hughes, Taylor L and Zhang, Shou-Cheng},
  journal={Physical Review B—Condensed Matter and Materials Physics},
  volume={78},
  number={19},
  pages={195424},
  year={2008},
  publisher={APS}
}

@article{essin2009magnetoelectric,
  title={Magnetoelectric polarizability and axion electrodynamics in crystalline insulators},
  author={Essin, Andrew M and Moore, Joel E and Vanderbilt, David},
  journal={Physical review letters},
  volume={102},
  number={14},
  pages={146805},
  year={2009},
  publisher={APS}
}

@article{nenno2020axion,
  title={Axion physics in condensed-matter systems},
  author={Nenno, Dennis M and Garcia, Christina AC and Gooth, Johannes and Felser, Claudia and Narang, Prineha},
  journal={Nature Reviews Physics},
  volume={2},
  number={12},
  pages={682--696},
  year={2020},
  publisher={Nature Publishing Group UK London}
}

@article{sekine2021axion,
  title={Axion electrodynamics in topological materials},
  author={Sekine, Akihiko and Nomura, Kentaro},
  journal={Journal of Applied Physics},
  volume={129},
  number={14},
  year={2021},
  publisher={AIP Publishing}
}

@article{otrokov2019prediction,
  title={Prediction and observation of an antiferromagnetic topological insulator},
  author={Otrokov, Mikhail M and Klimovskikh, Ilya I and Bentmann, Hendrik and Estyunin, D and Zeugner, Alexander and Aliev, Ziya S and Ga{\ss}, Sebastian and Wolter, AUB and Koroleva, AV and Shikin, Alexander M and others},
  journal={Nature},
  volume={576},
  number={7787},
  pages={416--422},
  year={2019},
  publisher={Nature Publishing Group UK London}
}

@article{Raffelt:1990yz,
    author = "Raffelt, Georg G.",
    title = "{Astrophysical methods to constrain axions and other novel particle phenomena}",
    reportNumber = "MPI-PAE-PTH-29-90",
    doi = "10.1016/0370-1573(90)90054-6",
    journal = "Phys. Rept.",
    volume = "198",
    pages = "1--113",
    year = "1990"
}

@article{Alekseev:1988gp,
    author = "Alekseev, E. N. and Alekseeva, L. N. and Krivosheina, I. V. and Volchenko, V. I.",
    title = "{Detection of the Neutrino Signal From {SN1987A} in the {LMC} Using the Inr Baksan Underground Scintillation Telescope}",
    doi = "10.1016/0370-2693(88)91651-6",
    journal = "Phys. Lett. B",
    volume = "205",
    pages = "209--214",
    year = "1988"
}

@article{Cavan-Piton:2024ayu,
    author = "Cavan-Piton, Mael and Guadagnoli, Diego and Oertel, Micaela and Seong, Hyeonseok and Vittorio, Ludovico",
    title = "{Axion Emission from Strange Matter in Core-Collapse SNe}",
    eprint = "2401.10979",
    archivePrefix = "arXiv",
    primaryClass = "hep-ph",
    reportNumber = "LAPTH-006/24, DESY-24-011",
    doi = "10.1103/PhysRevLett.133.121002",
    journal = "Phys. Rev. Lett.",
    volume = "133",
    number = "12",
    pages = "121002",
    year = "2024"
}

@article{Lucente:2022vuo,
    author = "Lucente, Giuseppe and Mastrototaro, Leonardo and Carenza, Pierluca and Di Luzio, Luca and Giannotti, Maurizio and Mirizzi, Alessandro",
    title = "{Axion signatures from supernova explosions through the nucleon electric-dipole portal}",
    eprint = "2203.15812",
    archivePrefix = "arXiv",
    primaryClass = "hep-ph",
    doi = "10.1103/PhysRevD.105.123020",
    journal = "Phys. Rev. D",
    volume = "105",
    number = "12",
    pages = "123020",
    year = "2022"
}

@article{Hyper-Kamiokande:2018ofw,
    author = "Abe, K. and others",
    collaboration = "Hyper-Kamiokande",
    title = "{Hyper-Kamiokande Design Report}",
    eprint = "1805.04163",
    archivePrefix = "arXiv",
    primaryClass = "physics.ins-det",
    month = "5",
    year = "2018"
}

@article{Georgi:1986df,
    author = "Georgi, Howard and Kaplan, David B. and Randall, Lisa",
    title = "{Manifesting the Invisible Axion at Low-energies}",
    reportNumber = "HUTP-86/A004",
    doi = "10.1016/0370-2693(86)90688-X",
    journal = "Phys. Lett. B",
    volume = "169",
    pages = "73--78",
    year = "1986"
}

@article{Hirata:1988ad,
    author = "Hirata, K. S. and others",
    title = "{Observation in the Kamiokande-II Detector of the Neutrino Burst from Supernova SN 1987a}",
    doi = "10.1103/PhysRevD.38.448",
    journal = "Phys. Rev. D",
    volume = "38",
    pages = "448--458",
    year = "1988"
}

@article{Alekseev:1987ej,
    author = "Alekseev, E. N. and Alekseeva, L. N. and Volchenko, V. I. and Krivosheina, I. V.",
    editor = "Tran Thanh Van, J.",
    title = "{Possible Detection of a Neutrino Signal on 23 February 1987 at the Baksan Underground Scintillation Telescope of the Institute of Nuclear Research}",
    journal = "JETP Lett.",
    volume = "45",
    pages = "589--592",
    year = "1987"
}

@article{Cavan-Piton:2025nsj,
    author = "Cavan-Piton, Mael and Guadagnoli, Diego and Iohner, Axel and Fernandez-Menendez, Pablo and Vittorio, Ludovico",
    title = "{Probing the general axion-nucleon interaction in water Cherenkov experiments}",
    eprint = "2503.17490",
    archivePrefix = "arXiv",
    primaryClass = "hep-ph",
    reportNumber = "LAPTH-011/25",
    doi = "10.1007/JHEP07(2025)070",
    journal = "JHEP",
    volume = "07",
    pages = "070",
    year = "2025"
}

@article{Kamiokande-II:1987idp,
    author = "Hirata, K. and others",
    editor = "Wali, K. C.",
    collaboration = "Kamiokande-II",
    title = "{Observation of a Neutrino Burst from the Supernova SN 1987a}",
    reportNumber = "UT-ICEPP-87-01, UPR-142E",
    doi = "10.1103/PhysRevLett.58.1490",
    journal = "Phys. Rev. Lett.",
    volume = "58",
    pages = "1490--1493",
    year = "1987"
}

@article{IMB:1988suc,
    author = "Bratton, C. B. and others",
    collaboration = "IMB",
    title = "{Angular Distribution of Events From Sn1987a}",
    reportNumber = "UM-PDK-88-1",
    doi = "10.1103/PhysRevD.37.3361",
    journal = "Phys. Rev. D",
    volume = "37",
    pages = "3361",
    year = "1988"
}

@article{Bionta:1987qt,
    author = "Bionta, R. M. and others",
    title = "{Observation of a Neutrino Burst in Coincidence with Supernova SN 1987a in the Large Magellanic Cloud}",
    reportNumber = "UCI-NEUTRINO-87-10",
    doi = "10.1103/PhysRevLett.58.1494",
    journal = "Phys. Rev. Lett.",
    volume = "58",
    pages = "1494",
    year = "1987"
}

@article{Abe:2011ts,
    author = "Abe, K. and others",
    title = "{Letter of Intent: The Hyper-Kamiokande Experiment --- Detector Design and Physics Potential ---}",
    eprint = "1109.3262",
    archivePrefix = "arXiv",
    primaryClass = "hep-ex",
    month = "9",
    year = "2011"
}

@article{Hempel:2009mc,
    author = "Hempel, Matthias and Schaffner-Bielich, Jurgen",
    title = "{Statistical Model for a Complete Supernova Equation of State}",
    eprint = "0911.4073",
    archivePrefix = "arXiv",
    primaryClass = "nucl-th",
    doi = "10.1016/j.nuclphysa.2010.02.010",
    journal = "Nucl. Phys. A",
    volume = "837",
    pages = "210--254",
    year = "2010"
}

@article{Saikawa:2024bta,
    author = "Saikawa, Ken'ichi and Redondo, Javier and Vaquero, Alejandro and Kaltschmidt, Mathieu",
    title = "{Spectrum of global string networks and the axion dark matter mass}",
    eprint = "2401.17253",
    archivePrefix = "arXiv",
    primaryClass = "hep-ph",
    reportNumber = "KANAZAWA-24-02, MPP-2024-18",
    doi = "10.1088/1475-7516/2024/10/043",
    journal = "JCAP",
    volume = "10",
    pages = "043",
    year = "2024"
}
\bibliographystyle{JHEP}

\end{document}